\documentclass[aip,reprint,twocolumn,superscriptaddress,showkeys, nofootinbib,notitlepage,amsmath,amssymb,floatfix,preprintnumbers]{revtex4-1}

\usepackage{caption,subcaption}
\usepackage{graphicx}
\usepackage{amsmath,amssymb,float}
\usepackage{natbib}
\usepackage{soul}
\usepackage{xcolor}
\sethlcolor{yellow} 
\usepackage{bm}
\usepackage{array}
\usepackage{hyperref}
\usepackage[labelformat=simple]{subcaption}                                           
\def\intgammaxi{{\int\xi\Pi d\xi}}
\def\intgammaeta{{\int\eta\Pi d\eta}}
\begin{document}
\title{Nemchinov-Dyson Solutions of the Two-Dimensional Axisymmetric Inviscid Compressible Flow Equations}
\author{Jesse F.~Giron}
\email{jgiron@lanl.gov}
\email{jfgiron@asu.edu}
\affiliation{Applied Physics, Los Alamos National Laboratory, P.O. Box 1663, MS T082, Los Alamos, New Mexico, USA 87545}
\affiliation{Department of Physics, Box 871504, Arizona State University, Tempe, Arizona, USA  85287-1504}
\author{Scott  D.~Ramsey}
\email{ramsey@lanl.gov}
\affiliation{Applied Physics, Los Alamos National Laboratory, P.O. Box 1663, MS T082, Los Alamos, New Mexico, USA 87545}
\author{Roy S.~Baty}
\email{rbaty@lanl.gov}
\affiliation{Applied Physics, Los Alamos National Laboratory, P.O. Box 1663, MS T082, Los Alamos, New Mexico, USA 87545}
\preprint{LA-UR 20-27826}
\date{November, 2020}
\begin{abstract}
We investigate the two-dimensional ($2$D) inviscid compressible flow equations in axisymmetric coordinates, constrained by an ideal gas equation of state (EOS). Beginning with the assumption that the $2$D velocity field is space-time separable and linearly variable in each corresponding spatial coordinate, we proceed to derive an infinite family of elliptic or hyperbolic, uniformly expanding or contracting ``gas cloud'' solutions. Construction of specific example solutions belonging to this family is dependent on the solution of a system of nonlinear, coupled, second-order ordinary differential equations, and the prescription of an additional physical process of interest (e.g., uniform temperature or uniform entropy flow). The physical and computational implications of these solutions as pertaining to quantitative code verification or model qualification studies are discussed in some detail. 
\end{abstract}
\maketitle
%
%
%
%%%%%%%%%%%%%%%%%%%%%%%%%%%%%%%%%%%%%%%%%%%%%%%%%%
%%%%%%%%%%%%%%%%%%%%%%%%%%%%%%%%%%%%%%%%%%%%%%%%%%
\section{Introduction}\label{sec:intro}
%%%%%%%%%%%%%%%%%%%%%%%%%%%%%%%%%%%%%%%%%%%%%%%%%%
%%%%%%%%%%%%%%%%%%%%%%%%%%%%%%%%%%%%%%%%%%%%%%%%%%
%
%
%
A classical family of self-similar solutions of the one-dimensional (1D) inviscid compressible flow (Euler) equations for an ideal gas involves the ``unsteady motion of a gas when the velocity is proportional to distance from the center of symmetry," as originally investigated by Sedov~\cite{sedov} (see also Zel'dovich {\it et al}~\cite{zeldovich_raizer}, Cantwell~\cite{cantwell}, and Atzeni and Meyer-ter-Vehn~\cite{atzeni2004physics}). As discussed in rigorous detail by Sedov~\cite{sedov}, these ``linear velocity'' solutions are intimately connected to a variety of other important inviscid compressible flow patterns, including

\begin{quote}
``...the problem of propagation of a detonation wave in a medium with variable density ... the problem of an intense explosion ... and the problem of an intense point explosion in a medium with variable initial density..."
\end{quote}
and, perhaps disseminated most widely, the adiabatic expansion of gas clouds~\cite{stanyukovich2016unsteady,sachdev2016shock}. These solutions and their generalizations have found extensive practical applications in the fields of laser and plasma physics as shown by Motz~\cite{motz1979physics}, Pert~\cite{pert_1980,pert1987use,pert_1989}, Hunter and London~\cite{hunter1988multidimensional}, and Anisimov et al.~\cite{anisimov1993gas,anisimov1996analytical} to name only a few, astrophysical modeling (e.g., the expansion of supernova remnants~\cite{stanyukovich2016unsteady} or stellar collapse processes~\cite{guo2020continued}), superfluid physics~\cite{kuznetsov2020expansion}, the evaluation of inertial confinement fusion (ICF) concepts~\cite{Kidder_1974_isentropic,Kidder_1974_laser,Kidder_1976,hora,hora_pfirsch,motz1979physics,coggeshall1986lie,coggeshall1992group,coggeshall1991analytic,atzeni2004physics,krauser}, and many other areas of physics besides. In addition to their physical applications, some of the Kidder~\cite{Kidder_1974_isentropic,Kidder_1974_laser,Kidder_1976} and Coggeshall~\cite{coggeshall1986lie,coggeshall1992group,coggeshall1991analytic} solutions have also more recently been exercised as test problems for the quantitative verification of inviscid compressible flow codes~\cite{morgan2014lagrangian,morgan2015point,burton2015reduction,burton2018compatible}. 

In all of these contexts, and consistent with the nature of both the model verification and model qualification processes as defined by Oberkampf {\it et al.}~\cite{oberkampf}, establishing the practical utility of mathematical models (as exemplified by their sufficient fidelity and predictive capability within applications of interest) demands an iterative process wherein those models are refined and improved as necessary. This notion in turn often motivates an ever-accelerating need for new surrogate problems to be used in conjunction with, for example, computational science codes of relevance to any of the aforementioned physical applications.

Against this broader backdrop, the $1$D inviscid Euler equations for an ideal gas thus represent a natural starting point for the exploration of a wider variety of flow scenarios with relevance to the aforementioned applications. Along these lines, possible modifications of the 1D inviscid Euler equations include but are not necessarily limited to the inclusion of non-ideal material constitutive laws, multi-fluid representations, charged particle transport phenomena, reaction-transport processes, relativistic effects, gravitational, electromagnetic, or thermal radiation field coupling, or higher-fidelity geometric effects such as two and three-dimensional ($2$D and $3$D, respectively) representations in various coordinate systems. Among these and many other possible choices, multi-dimensional generalizations of the $1$D linear velocity flows form the basis of this study.

An essential entry point into higher dimensional (i.e., $2$D or $3$D) fluid flows is Chandrasekhar's~\cite{chandrasekhar_1967} summary and thorough codification of the celebrated ``ellipsoidal figures of equilibrium'' as originally formulated at the dawn of fluid mechanics by luminaries including Newton, Maclaurin, Dirichlet, and others. Intimately related to these scenarios is the extensive body of literature pertaining to ellipsoidal gas clouds; see, for example, Ovsyannikov~\cite{ovsyannikovhydro}, Nemchinov~\cite{nemchinov_1965}, Anisimov and Lysikov~\cite{anisimov_1970}, Dyson~\cite{dyson_original,dyson_spinning}, Hara \textit{et al}~\cite{hara}, Tarasova~\cite{tarasova}, Shieh~\citep{shieh}, Rogers {\it et al}~\citep{Rogers_2011}, Gaffet~\cite{gaffet_1996,gaffet_1999,gaffet_2000,gaffet_2001_axis,gaffet_2001_liouville, gaffet_2005,gaffet_2010}, and Bogoyavlensky~\cite{bogoyavlenskymethods}, to name only a few references. All of these studies feature inherently $2$D or $3$D fluid bodies that are contracting, expanding, or rotating under various ancillary assumptions. Many of these scenarios also feature the linear velocity assumption, which despite its simplicity maintains a profound physical significance. As noted by Gaffet\cite{gaffet_1996},
\begin{quote}
``The physical motivation for considering the simplifying assumption … on the form of the velocity field, lies essentially in the fact that the basic kinematical quantity, the deformation tensor … is then uniformly distributed throughout space. That assumption … may be viewed as a natural generalization of the rigid flows that obtain when the uniform value of the deformation tensor vanishes...''
\end{quote}
and by Shieh~\citep{shieh},
\begin{quote}
``Those who are not familiar with this field often get a misleading impression ... that only trivial results can follow from such a simplifying assumption. ... Now, the assumption ... adds internal vortex motion as well as the pulsation of the semiaxes into the study of the problem. Furthermore, [it] contains the interactions of these types of motion. Even with the aid of modern computers, these interactions are not yet fully explored...''
\end{quote}
thus further motivating this study.

Within the tremendous body of work pertaining to ellipsoidal gas cloud motion, the seminal generalizations to $2$D and $3$D geometries of the $1$D linear velocity solutions appear to have been originated independently by Ovsyannikov~\cite{ovsyannikovhydro} and Dyson~\cite{dyson_original,dyson_spinning}. Both these models feature compressible fluid ellipsoids in the absence of any dissipative or otherwise ancillary effects, and so are associated with the $2$D or $3$D inviscid Euler equations for an ideal gas. As noted by Dyson~\cite{dyson_original}, the objective of these formulations is to find ``... a model which will describe the free expansion of a non-spherical cloud of gas into a vacuum.'' Commensurate with this objective, both Nemchinov~\cite{nemchinov_1965} (specializing Ovsyannikov’s~\cite{ovsyannikovhydro} more general results) and Dyson~\cite{dyson_original} proceed under the additional assumptions of
\begin{enumerate}
\item A $3$D Cartesian coordinate system, 
\item The gas cloud motion is irrotational,
\item The gas cloud expands by a uniform change of scale in each spatial coordinate (i.e., the linear velocity assumption), 
\item The expansion proceeds with spatially uniform temperature, leading to a Gaussian form of the cloud mass density distribution (Nemchinov~\cite{nemchinov_1965} also presents a case where this assumption is replaced with a quadratic temperature distribution),
\end{enumerate}
the aggregate result of which is the establishment of a fluid flow scenario that will hereafter be referred to as the ($3$D Cartesian) ``Nemchinov-Dyson problem'' for the sake of brevity, and also for some degree of consistency with the existing literature on the subject\footnote{An equally legitimate moniker for this scenario is the ($3$D Cartesian) ``irrotational Ovsyannikov-Dyson problem,'' in light of Ovsyannikov's seminal contributions to the more general scenario featuring rotation as well as expansion. Originally unaware of Ovsyannikov's earlier contributions, Dyson independently treated both the rotational case and its irrotational sub-case.}.

Motivated by the notion that ``in practice we are concerned only with axially symmetric expansions,'' Dyson~\cite{dyson_original} provides some numerical solutions for a $2$D axisymmetric form of his model. Nemchinov~\cite{nemchinov_1965} proceeds similarly, perhaps motivated by the fact that key behaviors and conclusions pertaining to the relevant motions are readily extracted from the slightly simpler model. In any event, deeper analytical studies of the Nemchinov-Dyson problem are provided by both Anisimov and Lysikov~\cite{anisimov_1970} and Gaffet~\cite{gaffet_1996,gaffet_1999}; see also Bogoyavlensky~\cite{bogoyavlenskymethods}. In particular, Anisimov and Lysikov~\cite{anisimov_1970} show for an ideal gas without inner degrees of freedom the Nemchinov-Dyson problem may be solved analytically in terms of elliptic integral functions of the third kind. Subsequent analytical studies along the same lines are provided by Hunter and London~\cite{hunter1988multidimensional} (see also references to H. Hietarinta appearing therein) and Gaffet~\cite{gaffet_1996,gaffet_1999}: all of these also feature the key assumption of a monoatomic polytropic gas, thus leading to full reductions and solutions of the Nemchinov-Dyson problem in terms of quadratures (of which the elliptic integral representations are a special case). 

Also exemplified by Dyson~\cite{dyson_spinning}, the notions of symmetries and associated conserved quantities play an important role in the construction of both Anisimov and Lysikov's~\cite{anisimov_1970}, Hunter and London's~\cite{hunter1988multidimensional}, and Gaffet's~\cite{gaffet_1996,gaffet_1999} analytical solutions. This correspondence also explicitly appears in related studies by Coggeshall~\cite{coggeshall1986lie,coggeshall1992group,coggeshall1991analytic}, who encodes the salient mathematics in the theory of invariance under groups of continuous point transformations (Lie groups). Using this systematic group-theoretic or symmetry analysis formalism, Coggeshall~\cite{coggeshall1992group,coggeshall_94_hydro} derives numerous new analytical solutions to the $2$D and $3$D inviscid Euler equations, including rotational and irrotational linear velocity instantiations in various coordinate systems, and other solutions featuring shock waves. Some of these analytical results bear close resemblance to solutions of the Nemchinov-Dyson problem established through other, but related means.

Inside of this voluminous amount of work performed ot date on the subject of expanding ellipsoidal gas clouds, our work seeks to address a variety of finer points appearing to arise in the existing literature with somewhat less frequency:
\begin{itemize}
\item In addition to the usual isothermal or Gaussian density instantiations, and the uniform entropy and parabolic temperature solutions, exploration of some alternate solution archetypes arising from a degree of freedom associated with all linear velocity solutions first recognized by Sedov~\cite{sedov}. 
\item In addition to the usual expansion scenarios, exploration of implosion or cumulation scenarios as discussed briefly by Bogoyavlensky~\cite{bogoyavlenskymethods}. Some scenarios of this type have been investigated in great detail especially in $1$D geometries, and have relevance to the aforementioned stellar formation or ICF processes.
\item In addition to the usual ellipsoidal gas cloud configurations, exploration of scenarios featuring cone, funnel, or otherwise hyperbolic-shaped gas clouds as discussed briefly by Bogoyavlensky~\cite{bogoyavlenskymethods}. While not exactly the ``funnel''-like motions appearing in the related rotational solutions, even irrotational solutions of this type may have relevance to solar flare processes. 
\item The provision of substantiating evidence in the interest of resolving a certain discrepancy between the works of Nemchinov~\cite{nemchinov_1965} and Hunter and London~\cite{hunter1988multidimensional}.
\end{itemize}
For consistency with many of the established results along these same lines, and cognizant of Dyson's~\cite{dyson_original} motivations as previously noted, these outcomes will be realized exlusively within the $2$D axisymmetric coordinate system.

The motivation behind this work is thus to leverage the existing developments in Nemchinov-Dyson and multi-dimensional Coggeshall~\cite{coggeshall1986lie,coggeshall1992group,coggeshall1991analytic} problems to obtain and analyze a variety of analytical or semi-analytical solutions in the $2$D axisymmetric coordinate system. Any new solutions derived in this program of study will therefore be available for further analysis from the standpoint of symmetry analysis theory in the style of, for example, McHardy \textit{et al.}~\cite{mchardy2019group}), or for integration within code verification or model qualification practices for the specific assessment of explicitly $2$D axisymmetric inviscid compressible flow solvers or codes (as opposed to their representations in other coordinate systems). 

In support of these goals, Sec.~\ref{sec:math_mod} provides an overview of the relevant mathematical model, including certain assumptions and results surrounding the assumed multi-dimensional geometry and equation of state constitutive law. A formalized definition of a generalized Nemchinov-Dyson problem for use throughout the remainder of this study is provided in Sec.~\ref{sec:dyson}, followed by derivation and analysis of some possible solution archetypes. Several detailed example solutions obtained via this formalism are presented in Sec.~\ref{subsec:examples}. Finally, we conclude and provide recommendations for future study in Sec.~\ref{sec:conclusion}.
%
%
%
%%%%%%%%%%%%%%%%%%%%%%%%%%%%%%%%%%%%%%%%%%%%%%%%%%
%%%%%%%%%%%%%%%%%%%%%%%%%%%%%%%%%%%%%%%%%%%%%%%%%%
\section{Mathematical Model}\label{sec:math_mod}
%%%%%%%%%%%%%%%%%%%%%%%%%%%%%%%%%%%%%%%%%%%%%%%%%%
%%%%%%%%%%%%%%%%%%%%%%%%%%%%%%%%%%%%%%%%%%%%%%%%%%
%
%
%
As shown by many authors (for example, Harlow and Amsden~\cite{harlow1971fluid}), the inviscid compressible flow (Euler) equations in a general coordinate system are as follows:
\begin{eqnarray}
\frac{\partial\rho}{\partial t}+\left(\vec{u}\cdot\vec{\nabla}\right)\rho + \rho \left(\vec{\nabla}\cdot\vec{u}\right)&=&0,\label{cons_mass}\\
\frac{\partial\vec{u}}{\partial t}+\left(\vec{u}\cdot\vec{\nabla}\right)\vec{u} +\frac{1}{\rho}\vec{\nabla}P&=&0,\label{cons_mom}\\
\frac{\partial E}{\partial t}+\left(\vec{u}\cdot\vec{\nabla}\right) E + \frac{1}{\rho}\vec{\nabla}\cdot\left(P\vec{u}\right)&=&0,\label{cons_enegy_E}
\end{eqnarray}
where the mass density $\rho\left(\vec{r},t\right)$, bulk flow velocity vector $\vec{u}\left(\vec{r},t\right)$, pressure $P\left(\vec{r},t\right)$, and total energy per unit mass $E\left(\vec{r},t\right)$ are functions of the position vector $\vec{r}$ and time $t$. The total energy per unit mass may be further decomposed into the specific internal energy $I\left(\vec{r},t\right)$ (SIE; internal energy per unit mass) and specific kinetic energy, that is
\begin{eqnarray}\label{fund_energy}
E\left(\vec{r},t\right) = I\left(\vec{r},t\right) + \frac{1}{2} \vec{u}\left(\vec{r},t\right)\cdot \vec{u}\left(\vec{r},t\right).
\end{eqnarray}
Conservation of mass, momentum and energy are reperesented by Eqs. (\ref{cons_mass})-(\ref{cons_enegy_E}), respectively. Equation (\ref{cons_enegy_E}) may be rewritten, using Eqs. (\ref{cons_mass}), (\ref{cons_mom}), and (\ref{fund_energy}), as 
\begin{eqnarray}\label{energy_relation_sub}
\frac{\partial I}{\partial t}+\left(\vec{u}\cdot\vec{\nabla}\right)I -\frac{P}{\rho^2}\left[\frac{\partial \rho}{\partial t}+\left(\vec{u}\cdot\vec{\nabla}\right)\rho\right]=0.
\end{eqnarray}
Equation~(\ref{energy_relation_sub}) may be further reduced using the fundemental thermodynamic relation~\cite{mandl1988statistical,bowley1999introductory,adkins1983equilibrium,landau2013statistical,zemansky1966basic} between $\rho$, $P$, $I$, the fluid temperature $T$, and the fluid entropy $S$; that is,
\begin{equation}\label{law}
dI= TdS +\frac{P}{\rho^2}d\rho.
\end{equation}
Using the chain rule and Eq. (\ref{law}), Eq. (\ref{energy_relation_sub}) becomes
\begin{equation}\label{isentropic_flow}
\frac{\partial S}{\partial t}+\left(\vec{u}\cdot\vec{\nabla}\right)S=0,
\end{equation}
also known as the equation for isentropic flow. Equation (\ref{isentropic_flow}) is expected to result from Eqs. (\ref{cons_mass})-(\ref{cons_enegy_E}) since they do not feature dissipative processes such as viscosity or heat conduction. Moreover, if the fluid entropy $S$ is assumed to be a function of the fluid density $\rho$ and pressure $P$, then Eq. (\ref{isentropic_flow}) may be expanded to yield
\begin{eqnarray}
\frac{\partial S}{\partial \rho}\biggr\rvert_P\left[\frac{\partial\rho}{\partial t}+\left(\vec{u}\cdot\vec{\nabla}\right)\rho\right]+\frac{\partial S}{\partial P}\biggr\rvert_\rho\left[\frac{\partial P}{\partial t}+\left(\vec{u}\cdot\vec{\nabla}\right)P \right] =0.\nonumber\\\label{isentropic_chain_rule}
\end{eqnarray}
Substituting Eq. (\ref{cons_mass}) into Eq. (\ref{isentropic_chain_rule}), we find
\begin{equation}\label{cons_energy_P}
\frac{\partial P}{\partial t}+\left(\vec{u}\cdot\vec{\nabla}\right)P + K_S\left(\vec{\nabla}\cdot \vec{u}\right)=0,
\end{equation}
where $K_S\left(\rho,P\right)$ is the adiabatic bulk modulus, which is defined by
\begin{equation}\label{abm_def}
K_S\left(\rho,P\right) \equiv -\rho\frac{\frac{\partial S}{\partial \rho}\biggr\rvert_P}{\frac{\partial S}{\partial P}\biggr\rvert_\rho},
\end{equation}
or as shown by Axford~\cite{axford},
\begin{equation}\label{bulkmod}
K_S(\rho,P) = \frac{P}{\rho}\frac{\partial P}{\partial I}\biggr\rvert_\rho + \rho \frac{\partial P}{\partial \rho}\biggr\rvert_I.
\end{equation}
The adiabatic bulk modulus appears only in the total energy (or entropy) conservation relation given by Eq.~(\ref{cons_energy_P}), and is a measure of the fluid's resistance to uniform, constant entropy compression. It is also related to the local fluid sound speed $c$ by
\begin{equation}\label{sound}
K_S = \rho c^2 .
\end{equation} 
%
%
%
%%%%%%%%%%%%%%%%%%%%%%%%%%%%%%%%%%%%%%%%%%%%%%%%%%
%%%%%%%%%%%%%%%%%%%%%%%%%%%%%%%%%%%%%%%%%%%%%%%%%%
\subsection{Axisymmetric Coordinate System}\label{subsec:axisym_coordinates}
%%%%%%%%%%%%%%%%%%%%%%%%%%%%%%%%%%%%%%%%%%%%%%%%%%
%%%%%%%%%%%%%%%%%%%%%%%%%%%%%%%%%%%%%%%%%%%%%%%%%%
%
%
%
\begin{figure}[t]
        \includegraphics[scale=0.45]{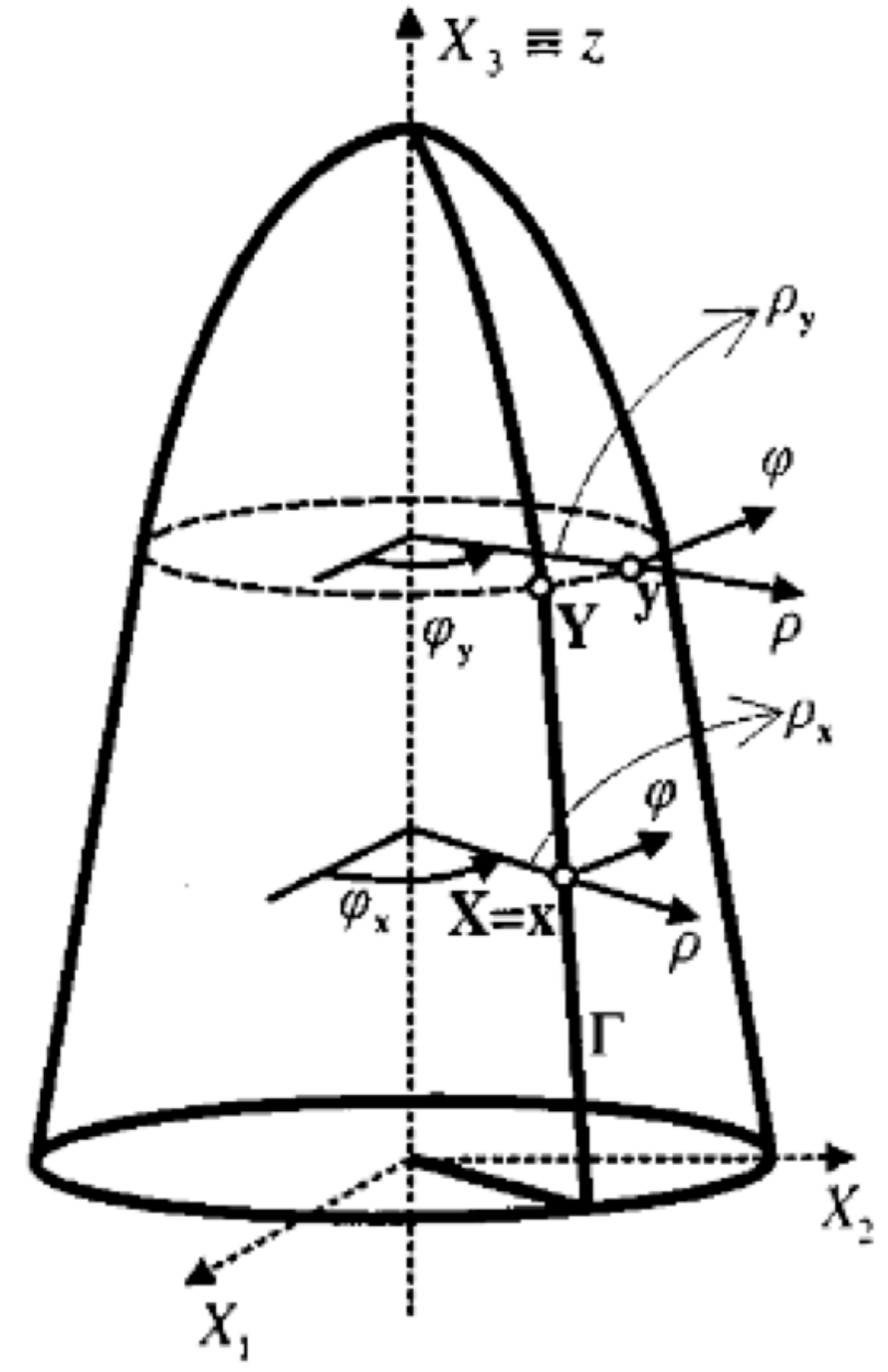}
        \caption{The 2D axisymmetric coordinate system, as found in Ref.~\cite{TsinopoulosStephanosV1999Aabe}. The object is rotated about the $z$-axis where the edge of the object is represented by the curve $\Gamma$. For the analysis studied in Eqs. (\ref{cons_mass_cyl})-(\ref{cons_energy_P_cyl}), $\rho$ in this diagram corresponds to the spatial coordinate $r$.}
\label{axisymmetric_coordinate_system}
\end{figure}

Further simplification of Eqs.~(\ref{cons_mass}), (\ref{cons_mom}), and (\ref{cons_energy_P}) is affected by selection of a spatial coordinate system, through which the various vector operators appearing in Eqs.~(\ref{cons_mass}), (\ref{cons_mom}), and (\ref{cons_energy_P}) may be resolved. Of particular interest to this work is the axisymmetric coordinate system, as depicted in Fig.~\ref{axisymmetric_coordinate_system}; this 2D $\left(r,z\right)$ geometry represents a natural bridge between 1D spherical and 3D Cartesian or spherical geometries, in that it allows for the existince of spherical, ellipsoidal, and other shapes in a 2D, axially symmetric setting.

Using the Lam\'{e} coefficient formalism~\cite{spiegel1959schaum}, in axisymmetric coordinates the various operators appearing in Eqs.~(\ref{cons_mass}), (\ref{cons_mom}), and (\ref{cons_energy_P}) are resolved as, for an arbitrary function $\varphi\left(r,z\right)$ (whose unit vectors are $\hat{e}_r$ and $\hat{e}_z$) and vector field $\vec{A}\left(r,z\right)$,
\begin{eqnarray}
\vec{\nabla}\varphi\left(r,z\right) &=& \frac{\partial\varphi}{\partial r}\hat{e}_r +\frac{\partial\varphi}{\partial z}\hat{e}_z\label{grad_2d} , \\
\vec{\nabla}\cdot\vec{A}\left(r,z\right)&=& \frac{1}{r} \left[\frac{\partial}{\partial r}\left(rA_r\right)+\frac{\partial}{\partial z}\left(rA_z\right)\right]\label{div_2d} ,
\end{eqnarray}
so that Eqs.~(\ref{cons_mass}), (\ref{cons_mom}), and (\ref{cons_energy_P}) become, respectively,
\begin{eqnarray}
\frac{\partial \rho}{\partial t} + u_r\frac{\partial \rho}{\partial r}+u_z\frac{\partial \rho}{\partial z}+\rho\left[\frac{\partial u_r}{\partial r}+\frac{\partial u_z}{\partial z}+ \frac{u_r}{r}\right]&=&0,\label{cons_mass_cyl}\;\;\;\;\;\;\;\;\\
\frac{\partial u_r}{\partial t} + u_r\frac{\partial u_r}{\partial r}+u_z\frac{\partial u_r}{\partial z} + \frac{1}{\rho}\frac{\partial P}{\partial r} &=& 0,\label{cons_mom_ur_cyl}\\
\frac{\partial u_z}{\partial t} + u_r\frac{\partial u_z}{\partial r}+u_z\frac{\partial u_z}{\partial z} + \frac{1}{\rho}\frac{\partial P}{\partial z} &=& 0,\label{cons_mom_uz_cyl}\\
\frac{\partial P}{\partial t} + u_r\frac{\partial P}{\partial r}+u_z\frac{\partial P}{\partial z}+K_S \left[\frac{\partial u_r}{\partial r}+\frac{\partial u_z}{\partial z}+ \frac{u_r}{r}\right]&=&0,\label{cons_energy_P_cyl}
\end{eqnarray}
where $u_r$ and $u_z$ denote, respectively, the $r$ and $z$ components of the bulk velocity field.
%
%
%
%%%%%%%%%%%%%%%%%%%%%%%%%%%%%%%%%%%%%%%%%%%%%%%%%%
%%%%%%%%%%%%%%%%%%%%%%%%%%%%%%%%%%%%%%%%%%%%%%%%%%
\subsection{Thermodynamic Considerations and the Equation of State}\label{subsec:Thermodynamics_EOS}
%%%%%%%%%%%%%%%%%%%%%%%%%%%%%%%%%%%%%%%%%%%%%%%%%%
%%%%%%%%%%%%%%%%%%%%%%%%%%%%%%%%%%%%%%%%%%%%%%%%%%
%
%
%
As written, Eqs.~(\ref{cons_mass_cyl})-(\ref{cons_energy_P_cyl}) are a system of four partial differential equations (PDEs) in the four unknowns $\rho$, $u_r$, $u_z$, and $P$. Solution of this system may be attempted under prescription of the functional form in $\rho$ and $P$ of the adiabatic bulk modulus $K_S$ appearing in the energy conservation relation. As suggested throughout Sec.~\ref{sec:math_mod}, the adiabatic bulk modulus itself is intimately related to the equation of state (EOS) closure model associated with a fluid archetype under consideration. In particular, when the fluid EOS assumes a form given by
\begin{equation}\label{eos_gen}
P = \mathcal{P}\left(\rho,I\right) \,
\end{equation}
for an arbitrary function $\mathcal{P}$ of the indicated arguments, the corresponding adiabatic bulk modulus may be calculated using Eq.~(\ref{bulkmod}), and the associated entropy form $S = \mathcal{S}\left(\rho,P\right)$ of the EOS (where $\mathcal{S}$ is another arbitrary function of the indicated arguments) may then be calculated using Eq.~(\ref{abm_def}).

One of the simplest closure models that may be assumed in the context of Eqs.~(\ref{cons_mass}), (\ref{cons_mom}), and (\ref{cons_energy_P}) is the ideal gas EOS, which is representative of a wide variety of relatively simple gases (e.g., monoatomic, diatomic, or other gases with relatively simple structure and associated internal degrees of freedom). The ideal gas EOS is given by
\begin{equation}\label{ideal_gas_eos}
P = \left(\gamma-1\right)\rho I,
\end{equation}
where the constant adiabatic index $\gamma>1$ is representative of the internal atomic or molecular degrees of freedom within the finer gas structure, and may be defined as
\begin{equation}\label{gamma_def}
\gamma \equiv \frac{c_P}{c_V} ,
\end{equation}
where $c_V$ and $c_P$ are the (constant) specific heat capacities (i.e., heat capacities per unit mass) of the gas at constant volume and pressure, respectively, such that
\begin{equation}\label{calor_perfect}
I = c_V T ,
\end{equation}
that is, the gas is also assumed to be calorically perfect. With Eqs.~(\ref{bulkmod}) and (\ref{ideal_gas_eos}), we find the adiabatic bulk modulus for an ideal gas is simply
\begin{eqnarray}\label{bulk_mod_ideal_gas_sub}
K_S&=& \gamma P, 
\end{eqnarray}
so that Eq.~(\ref{abm_def}) becomes 
\begin{equation}\label{abm_def_perfect}
\gamma P = -\rho\frac{\frac{\partial S}{\partial \rho}\biggr\rvert_P}{\frac{\partial S}{\partial P}\biggr\rvert_\rho},
\end{equation}
which may be solved using the method of characteristics to yield the associated form of the entropy as
\begin{equation}\label{entropy_ideal}
S = \mathcal{S}\left(P \rho^{-\gamma}\right) ,
\end{equation}
where $\mathcal{S}$ is an arbitrary function of the indicated argument; for simplicity, $\mathcal{S}$ will be taken as uniform throughout the remainder of this work, such that, without loss of generality,
\begin{equation}\label{entropy_ideal_easy}
S = P \rho^{-\gamma}.
\end{equation}

Using Eq.~(\ref{bulk_mod_ideal_gas_sub}), Eqs.~(\ref{cons_mass_cyl})-(\ref{cons_mom_uz_cyl}) remain unchanged while Eq.~(\ref{cons_energy_P_cyl}) finally becomes
\begin{equation}
\frac{\partial P}{\partial t} + u_r\frac{\partial P}{\partial r}+u_z\frac{\partial P}{\partial z}+\gamma P \left[\frac{\partial u_r}{\partial r}+\frac{\partial u_z}{\partial z}+ \frac{u_r}{r}\right]=0. \label{cons_energy_P_cyl_ideal}
\end{equation}
Equations~(\ref{cons_mass_cyl})-(\ref{cons_mom_uz_cyl}) and (\ref{cons_energy_P_cyl_ideal}) are the invsicid Euler equations that will be used throughout the remainder of this study.
%
%%%%%%%%%%%%%%%%%%%%%%%%%%%%%%%%%%%%%%%%%%%%%%%%%%
%%%%%%%%%%%%%%%%%%%%%%%%%%%%%%%%%%%%%%%%%%%%%%%%%%
\section{The Nemchinov-Dyson Problem}
\label{sec:dyson}
%%%%%%%%%%%%%%%%%%%%%%%%%%%%%%%%%%%%%%%%%%%%%%%%%%
%%%%%%%%%%%%%%%%%%%%%%%%%%%%%%%%%%%%%%%%%%%%%%%%%%
%
%
%
%
\begin{figure}[t]
        \includegraphics[scale=0.45]{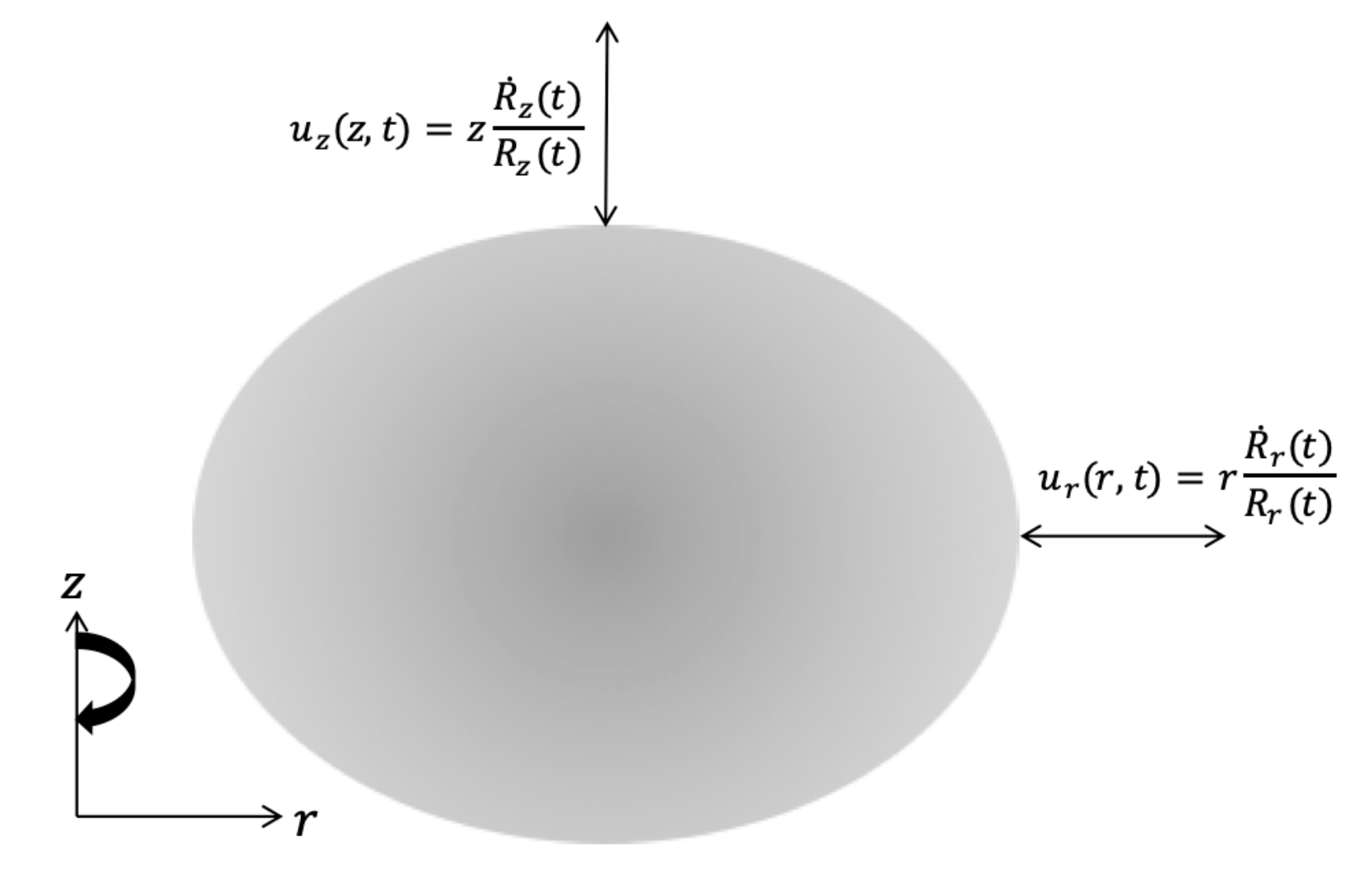}
        \caption{Notional depiction of the Nemchinov-Dyson problem. The ideal gas cloud (depicted here as an ellipse for ease of illustration) has an arbitrary eccentricity and interior state distribution at $t=0$, with Eqs.~(\ref{ur_assump}) and (\ref{uz_assump}) representing the flow velocities throughout [where the scale velocities $\dot{R}_r$ and $\dot{R}_z$ may take on either positive or negative values]; otherwise there is no angular velocity associated with this configuration. The $z$-axis is the axis of rotation as shown by the curved arrow.}
\label{nemchinov_dyson_depiction}
\end{figure}
As depicted in Fig.~\ref{nemchinov_dyson_depiction} and discussed extensively throughout Sec.~\ref{sec:intro}, a Nemchinov-Dyson solution of Eqs.~(\ref{cons_mass_cyl})-(\ref{cons_mom_uz_cyl}) and (\ref{cons_energy_P_cyl_ideal}) may be constructed by assuming separable, linear proportionalities between each of the featured flow velocity components and their associated spatial coordinates. In axisymmetric geometry, this homogeneity assumption proceeds according to
\begin{eqnarray}
u_r &=& r\frac{\dot{R_r}\left(t\right)}{R_r\left(t\right)},\label{ur_assump}\\
u_z &=& z\frac{\dot{R_z}\left(t\right)}{R_z\left(t\right)},\label{uz_assump}
\end{eqnarray}
where the ``scale radii'' $R_r > 0$ and $R_z > 0$ are functions of time to be determined, and the ``scale velocities'' $\dot{R}_q\left(q \in r,z\right)$ are defined by
\begin{equation}
\dot{R}_q \equiv \frac{dR_q}{dt}, \label{scale_velocities}
\end{equation}
such that the overdots denote time differentiation.

Substituting Eqs.~(\ref{ur_assump}) and (\ref{uz_assump}) into Eq.~(\ref{cons_mass_cyl}) yields
\begin{eqnarray}\label{mass_linvel_subbed}
\frac{\partial \rho}{\partial t}+r\frac{\dot{R_r}}{R_r}\frac{\partial \rho}{\partial r}+ z\frac{\dot{R_z}}{R_z}\frac{\partial \rho}{\partial z}+ \rho\left(2\frac{\dot{R}_r}{R_r}+\frac{\dot{R}_z }{R_z}\right)&=&0.
\end{eqnarray}
Using method of characteristics, we then find
\begin{eqnarray}\label{chars}
\frac{dt}{1}=\frac{dr}{\frac{\dot{R}_r}{R_r}r}=\frac{dz}{\frac{\dot{R}_z}{R_z}z}=\frac{d\rho}{-\rho\left(2\frac{\dot{R}_r}{R_r}+\frac{\dot{R}_z}{R_z}\right)},
\end{eqnarray}
or
\begin{eqnarray}
\frac{\dot{R}_r}{R_r}dt&=&\frac{dr}{r},\nonumber\\
\Rightarrow \xi(r,t) &=& \frac{r}{R_r},\label{xi_sim}
\end{eqnarray}
where the Lagrangian coordinate $\xi\left(r,t\right)$ is a similarity variable, in terms of which Eqs.~(\ref{cons_mass_cyl})-(\ref{cons_mom_uz_cyl}) and (\ref{cons_energy_P_cyl_ideal}) may be reduced. Also from Eq.~(\ref{chars}),
\begin{eqnarray}
\frac{\dot{R}_z}{R_z}dt&=&\frac{dz}{z},\nonumber\\
\Rightarrow \eta(z,t) &=& \frac{z}{R_z},\label{eta_sim}
\end{eqnarray}
where the Lagrangian coordinate $\eta\left(r,t\right)$ is another similarity variable, in terms of which Eqs.~(\ref{cons_mass_cyl})-(\ref{cons_mom_uz_cyl}) and (\ref{cons_energy_P_cyl_ideal}) may be reduced. Finally, we solve for $d\rho$ and $dt$ in Eq.~(\ref{chars}) and find
\begin{eqnarray}
\frac{d\rho}{\rho}&=&-\left(2\frac{\dot{R}_r}{R_r}+ \frac{\dot{R}_z}{R_z}\right)dt,\nonumber\\
\Rightarrow \rho\left(r,z,t\right)&=&\frac{1}{R_r^2R_z}\Pi\left(\xi,\eta\right),\label{rho_dyson_solution}
\end{eqnarray}
thus yielding a solution for the Nemchinov-Dyson density $\rho$ in terms of both the scale radii and $\Pi$, which is an arbitrary function of the arguments $\xi$ and $\eta$. 

We now substitute Eqs.~(\ref{ur_assump}), (\ref{uz_assump}), and (\ref{xi_sim})-(\ref{rho_dyson_solution}) into Eqs.~(\ref{cons_mom_ur_cyl}) and (\ref{cons_mom_uz_cyl}) and find
\begin{eqnarray}
\ddot{R}_r\xi + \frac{R_r^2R_z}{R_r}\frac{1}{\Pi\left(\xi,\eta\right)}\frac{\partial P}{\partial \xi}&=&0,\\
\ddot{R}_z\eta + \frac{R_r^2R_z}{R_z}\frac{1}{\Pi\left(\xi,\eta\right)}\frac{\partial P}{\partial \eta}&=&0,
\end{eqnarray}
respectively. Solving for the pressure $P$ in both of the above equations then yields
\begin{eqnarray}
P\left(r,z,t\right) &=& -\frac{\ddot{R}_r}{R_rR_z}\int \xi \Pi d\xi ,\label{def_gamma_1}\\
P\left(r,z,t\right) &=& -\frac{\ddot{R}_z}{R_r^2}\int \eta \Pi d\eta .\label{def_gamma_2}
\end{eqnarray}
Furthermore, substituting Eqs.~(\ref{ur_assump}), (\ref{uz_assump}), and (\ref{xi_sim})-(\ref{rho_dyson_solution}) into Eq.~(\ref{cons_energy_P_cyl_ideal}) yields a third equation for the pressure $P$, namely,
\begin{equation}\label{eng_linvel_subbed}
\frac{\partial P}{\partial t}+r\frac{\dot{R_r}}{R_r}\frac{\partial P}{\partial r}+ z\frac{\dot{R_z}}{R_z}\frac{\partial P}{\partial z}+ \gamma P \left(2\frac{\dot{R}_r}{R_r}+\frac{\dot{R}_z }{R_z}\right)=0 ,
\end{equation}
which, using the same method of characteristics procedure as used to solve Eq.~(\ref{mass_linvel_subbed}), has a solution given by
\begin{eqnarray}
P\left(r,z,t\right) = \frac{1}{\left(R_r^2R_z\right)^\gamma}\beta\left(\xi,\eta\right),\label{def_gamma_3}
\end{eqnarray}
where $\xi$ and $\eta$ retain their previous definitions and, like the function $\Pi$ appearing in the Nemchinov-Dyson density solution, $\beta$ is an arbitrary function of the the arguments $\xi$ and $\eta$.

Since on the grounds of physical realism the pressure $P$ must be a single-valued function, Eqs.~(\ref{def_gamma_1}), (\ref{def_gamma_2}), and (\ref{def_gamma_3}) yield the equivalences
\begin{eqnarray}
\frac{1}{\left(R_r^2R_z\right)^\gamma}\beta
&=&-\frac{\ddot{R}_z}{R_r^2}\int \eta \Pi d\eta \nonumber \\
&=&-\frac{\ddot{R}_r}{R_rR_z}\int \xi \Pi d\xi ,
\end{eqnarray}
or
\begin{eqnarray}
-\frac{R_r^{1-2\gamma}R_z^{1-\gamma}}{\ddot{R}_r}&=&\frac{1}{\beta}\intgammaxi ,\label{kappa_1_setup}\\
-\frac{R_r^{2-2\gamma}R_z^{-\gamma}}{\ddot{R}_z}&=&\frac{1}{\beta}\intgammaeta ,\label{kappa_2_setup}\\
\frac{R_r\ddot{R}_r}{R_z\ddot{R}_z}&=& \frac{\int\eta\Pi d\eta}{\int\xi\Pi d\xi}. \label{kappa_3_setup}
\end{eqnarray}
The left-hand side of each of Eqs.~(\ref{kappa_1_setup})-(\ref{kappa_3_setup}) depends only on $t$, while their right-hand sides depend not only on $t$, but also $r$ and $z$ (as parameterized through $\xi$ and $\eta$). As such, one possible means of satisfying Eqs.~(\ref{kappa_1_setup})-(\ref{kappa_3_setup}) is to enforce
\begin{eqnarray}
-\frac{R_r^{1-2\gamma}R_z^{1-\gamma}}{\ddot{R}_r}&=&\kappa_1=\frac{1}{\beta}\intgammaxi ,\label{kappa_1}\\
-\frac{R_r^{2-2\gamma}R_z^{-\gamma}}{\ddot{R}_z}&=& \kappa_2=\frac{1}{\beta}\intgammaeta ,\label{kappa_2}\\
\frac{R_r\ddot{R}_r}{R_z\ddot{R}_z}&=& \kappa_3 = \frac{\int\eta\Pi d\eta}{\int\xi\Pi d\xi}, \label{kappa_3}
\end{eqnarray}
where the $\kappa_i$ $\left(i \in 1,2,3\right)$ are constants. Substituting Eqs.~(\ref{kappa_1}) and (\ref{kappa_2}) into Eq.~(\ref{kappa_3}), we immediately find that the $\kappa_i$ must satisfy the constraint
\begin{equation}\label{kappas_relation}
\kappa_3 = \frac{\kappa_2}{\kappa_1}.
\end{equation}

First analyzing the right-hand equalities appearing in Eqs.~(\ref{kappa_1})-(\ref{kappa_3}), trivial rearrangements reveal various properties the otherwise arbitrary functions $\Pi$ and $\beta$ must feature:
\begin{eqnarray}
\kappa_1 \beta &=& \intgammaxi, \label{kappa_1_triv}\\
\kappa_2 \beta &=& \intgammaeta, \label{kappa_2_triv}\\
\kappa_3 \int\xi\Pi d\xi &=& \int\eta\Pi d\eta, \label{kappa_3_triv}
\end{eqnarray}
or, differentiating Eq.~(\ref{kappa_3_triv}) with respect to both $\xi$ and $\eta$ gives
\begin{equation}\label{kappa_3_deriv}
\kappa_3 \xi \frac{\partial \Pi}{\partial \eta} = \eta \frac{\partial \Pi}{\partial \xi} ,
\end{equation}
which may be solved for the function $\Pi$ appearing in Eq.~(\ref{rho_dyson_solution}) for the density $\rho$ using the method of characteristics to yield
\begin{equation}
\Pi\left(\xi,\eta\right) = \Pi\left(\frac{\zeta^2}{2}\right) ,
\label{pi_solution}
\end{equation}
such that $\Pi$ is revealed to be an arbitrary function only of the coordinate $\zeta$ defined by
\begin{equation}
\zeta^2 \equiv \kappa_3 \xi^2 + \eta^2 .
\label{zeta_def}
\end{equation}
With Eq.~(\ref{pi_solution}), Eq.~(\ref{rho_dyson_solution}) then becomes
\begin{eqnarray}
\rho\left(r,z,t\right)&=&\frac{1}{R_r^2 R_z}\Pi\left(\zeta \right).
\label{rho_dyson_solution_zeta}
\end{eqnarray}
Moreover, with Eqs.~(\ref{pi_solution}) and (\ref{zeta_def}), the integrals appearing in Eqs.~(\ref{def_gamma_1}) and (\ref{def_gamma_2}) for the pressure $P$ become
\begin{eqnarray}
\int \xi \Pi d\xi &=& \frac{\Gamma\left(\zeta\right)}{\kappa_3} \label{def_gamma_1_zeta} , \\
\int \eta \Pi d\eta &=& \Gamma\left(\zeta\right) \label{def_gamma_2_zeta} , 
\end{eqnarray}
where
\begin{equation}
\Gamma\left(\zeta\right) \equiv \int \zeta \Pi d\zeta ,
\label{big_gamma_def}
\end{equation}
such that both of Eqs.~(\ref{def_gamma_1_zeta}) and (\ref{def_gamma_2_zeta}) are guarananteed to give equivalent results in light of Eq.~(\ref{kappa_3_triv}); that is, Eqs.~(\ref{def_gamma_1}) and (\ref{def_gamma_2}) become
\begin{eqnarray}
P\left(r,z,t\right) &=& -\frac{\ddot{R}_r}{\kappa_3 R_r R_z}
\Gamma\left(\zeta\right) , \nonumber \\
                         &=& -\frac{\ddot{R}_z}{R_r^2}
\Gamma\left(\zeta\right) .
\label{p_dyson_solution_zeta}
\end{eqnarray}
which are equivalent with Eq.~(\ref{kappa_3}) taken into consideration.

In addition, with Eqs.~(\ref{ideal_gas_eos}), (\ref{rho_dyson_solution_zeta}), and (\ref{p_dyson_solution_zeta}), the SIE associated with the axisymmetric Nemchinov-Dyson solution is given by,
\begin{eqnarray}
I\left(r,z,t \right)
&=&-\frac{R_r \ddot{R}_r}{\kappa_3 \left(\gamma-1\right)}
\Upsilon\left(\zeta\right) \nonumber \\
&=&-\frac{R_z \ddot{R}_z}{\left(\gamma-1\right)}
\Upsilon\left(\zeta\right) ,
\label{sie_dyson_solution_zeta}
\end{eqnarray}
where
\begin{equation}
\Upsilon\left(\zeta\right) \equiv \frac{\int \zeta \Pi d\zeta}{\Pi} ,
\label{upsilon_def}
\end{equation}
such that both representations of Eq.~(\ref{sie_dyson_solution_zeta}) are again guaranteed to give equivalent results in light of Eq.~(\ref{kappa_3}). Likewise, and also with Eqs.~(\ref{entropy_ideal_easy}), (\ref{rho_dyson_solution}), and (\ref{p_dyson_solution_zeta}), the entropy associated with the axisymmetric Nemchinov-Dyson solution is given by 
\begin{eqnarray}
S\left(r,z,t\right)
&& = -\frac{\ddot{R}_r R_r^{2\gamma-1} R_z^{\gamma-1}}{\kappa_3} 
\Sigma\left(\zeta\right) \nonumber \\
&& = -\ddot{R}_z R_r^{2\gamma-2} R_z^{\gamma} 
\Sigma\left(\zeta\right)  ,
\label{entropy_dyson_solution_zeta}
\end{eqnarray}
where
\begin{equation}
\Sigma\left(\zeta\right) \equiv \Pi^{-\gamma} \int \zeta \Pi d\zeta ,
\label{sigma_def}
\end{equation}
such that both representations of Eq.~(\ref{entropy_dyson_solution_zeta}) are again guaranteed to give equivalent results in in light of Eq.~(\ref{kappa_3}).
 
Finally, turning to the left-hand equalities appearing in Eqs.~(\ref{kappa_1})-(\ref{kappa_3}). With Eqs.~(\ref{kappa_3}) and (\ref{kappas_relation}), Eqs.~(\ref{kappa_1}) and (\ref{kappa_2}) are revealed to be redundant. As such,
\begin{eqnarray}
-\frac{R_r^{1-2\gamma} R_z^{1-\gamma}}{\ddot{R}_r} &=& \kappa_1 , \label{ode_1_gen} \\
-\frac{R_r \ddot{R}_r}{R_z \ddot{R}_z} &=& \kappa_3 \label{ode_2_gen} ,
\end{eqnarray}
represent the two salient coupled, second-order, nonlinear ordinary differential equations (ODEs) in the scale radii $R_r$ and $R_z$; a solution of these ODEs thus resolves the time-dependence appearing in Eqs.~(\ref{rho_dyson_solution_zeta}), (\ref{p_dyson_solution_zeta}), (\ref{sie_dyson_solution_zeta}), and (\ref{entropy_dyson_solution_zeta}).

The Nemchinov-Dyson solution of Eqs.~(\ref{cons_mass_cyl})-(\ref{cons_mom_uz_cyl}) and (\ref{cons_energy_P_cyl_ideal}) is thus comprised of the $r$ and $z$ velocity components $u_r$ and $u_z$, density $\rho$, pressure $P$, SIE $I$, and entropy $S$ relations given by Eqs.~(\ref{ur_assump}), (\ref{uz_assump}), (\ref{rho_dyson_solution_zeta}), (\ref{p_dyson_solution_zeta}), (\ref{sie_dyson_solution_zeta}), and (\ref{entropy_dyson_solution_zeta}), respectively. Each of these flow variables features two principal components:
\begin{enumerate}
\item Time-dependence parameterized exclusively by the scale radii $R_r$ and $R_z$: these functions must satisfy the coupled, second-order, nonlinear ODE system given by Eqs.~(\ref{ode_1_gen}) and (\ref{ode_2_gen}). Some representative solutions of these ODEs are provided in Sec.~\ref{sec:scale_radius_sols}.
\item Spatial dependence parameterized exclusively by the function $\Pi$: in turn, $\Pi$ is a function only of the coordinate $\zeta$, which is itself defined in terms of the similarity variables $\xi$ and $\eta$ (and hence $r$ and $z$) via Eqs.~(\ref{xi_sim}), (\ref{eta_sim}), and (\ref{zeta_def}). Otherwise, the function $\Pi$ (and the related functions $\beta$, $\Gamma$, $\Upsilon$, and $\Sigma$)\footnote{Though the arbitrary function $\beta$ is related to $\Pi$, it is no longer needed in this study due to its relation to $\Gamma$ via Eqs. (\ref{kappa_1})-(\ref{kappa_3}).} is arbitrary. Some representative choices of $\Pi$ (and their attendant physical motivations) are provided in Sec.~\ref{sec:pi_sols}.
\end{enumerate}
%
%%%%%%%%%%%%%%%%%%%%%%%%%%%%%%%%%%%%%%%%%%%%%%%%%%
%%%%%%%%%%%%%%%%%%%%%%%%%%%%%%%%%%%%%%%%%%%%%%%%%%
\subsection{Solution Sets for $R_r$ and $R_z$}
\label{sec:scale_radius_sols}
%%%%%%%%%%%%%%%%%%%%%%%%%%%%%%%%%%%%%%%%%%%%%%%%%%
%%%%%%%%%%%%%%%%%%%%%%%%%%%%%%%%%%%%%%%%%%%%%%%%%%
%
To attempt solution of Eqs.~(\ref{ode_1_gen}) and (\ref{ode_2_gen}), we first use Eq.~(\ref{kappas_relation}) to write Eq.~(\ref{ode_2_gen}) as
\begin{eqnarray}
\frac{R_r \ddot{R}_r}{R_z \ddot{R}_z} &=& \frac{\kappa_2}{\kappa_1} \label{ode_2} .
\end{eqnarray}
Solution of Eqs.~(\ref{ode_1_gen}) and (\ref{ode_2}) requires the introduction of four initial conditions. Without loss of generality, these initial conditions may be expressed at $t = 0$ as
\begin{eqnarray}
R_r\left(t = 0\right) &=& R_{r,0}, \label{Rr_zero} \\
R_z\left(t = 0\right) &=& R_{z,0}, \label{Rz_zero} \\
\dot{R}_r\left(t = 0\right) &=& \dot{R}_{r,0}, \label{Rr_dot_zero} \\
\dot{R}_z\left(t = 0\right) &=& \dot{R}_{z,0}, \label{Rz_dot_zero}
\end{eqnarray}
where the constants $R_{r,0} > 0$ and $R_{z,0} > 0$, and the constants $\dot{R}_{r,0}$ and $\dot{R}_{z,0}$ are otherwise unconstrained (i.e., each may be positive, negative, or zero). In addition, Eqs.~(\ref{Rr_zero})-(\ref{Rz_dot_zero}) further suggest that the free constants $\kappa_1$ and $\kappa_2$ appearing in Eqs.~(\ref{ode_1_gen}) and (\ref{ode_2}), respectively, may be written as, by evaluating Eqs.~(\ref{ode_1_gen}) and (\ref{ode_2}) themselves at $t = 0$,
\begin{eqnarray}
\kappa_1 &=& -\frac{R_{r,0}^{1-2\gamma} R_{z,0}^{1-\gamma}}{\ddot{R}_{r,0}} , \label{kappa_1_def} \\
\kappa_2 &=& -\frac{R_{r,0}^{2-2\gamma} R_{z,0}^{-\gamma}}{\ddot{R}_{z,0}}, \label{kappa_2_def}
\end{eqnarray}
respectively, where $\ddot{R}_{r,0}$ and $\ddot{R}_{z,0}$ are the second derivatives of $R_r$ and $R_z$ evaluated at $t = 0$; as with the first derivatives these constants are unconstrained aside from being non-zero (in the interest of discarding any trivial solutions of Eqs.~(\ref{ode_1_gen}) and (\ref{ode_2})). As such, Eqs.~(\ref{kappa_1_def}) and (\ref{kappa_2_def}) reveal that constants $\kappa_1$ and $\kappa_2$ appearing in Eqs.~(\ref{ode_1_gen}) and (\ref{ode_2}), respectively, are inversely proportional to the negatives of the $r$ and $z$ components of an acceleration (or pressure) field at $t = 0$, respectively.

Equations~(\ref{ode_1_gen}) and (\ref{ode_2}), subject to the initial conditions given by Eqs.~(\ref{Rr_zero})-(\ref{Rz_dot_zero}), have no known closed-form solution for arbitrary $\gamma$, $R_{r,0}$, $R_{z,0}$, $\dot{R}_{r,0}$, $\dot{R}_{z,0}$, $\ddot{R}_{r,0}$, and $\ddot{R}_{z,0}$. However, various numerical solutions of this same equation set may be broadly categorized according to their qualitative behavior in $R_r$ and $R_z$, in turn resulting from different combinations of the parameters $\dot{R}_{r,0}$, $\dot{R}_{z,0}$, $\ddot{R}_{r,0}$, and $\ddot{R}_{z,0}$. 

The salient initial parameter set including $\dot{R}_{r,0}$, $\dot{R}_{z,0}$, $\ddot{R}_{r,0}$, and $\ddot{R}_{z,0}$ features multiple possible generic combinations, hereafter referred to as ``cases.'' These cases are generated by the following possible parameterizations, considered combinatorically:
\begin{itemize}
\item $\dot{R}_{r,0}$: positive, negative, or zero,
\item $\dot{R}_{z,0}$: positive, negative, or zero,
\item $\ddot{R}_{r,0}$: positive or negative,
\item $\ddot{R}_{z,0}$: positive or negative.
\end{itemize}
These cases allow for two distinct solution behaviors for each $R_q$ $\left(q \in r,z\right)$:
\begin{enumerate}
\item $R_q \to \infty$ as $t \to -\infty$ and as $t \to \infty$. Globally concave-up solutions of this type are referred to as ``double-regular (DR),'' and manifest whenever the initial acceleration field is outward-directed or positive; that is $\ddot{R}_{q,0} > 0$. 
\item $R_q \to 0$ as $t \to -t^*$ and as $t \to t^*$, for some $t^* > 0$. Globally concave-down solutions of this type are referred to as ``double-singluar (DS),'' and manifest whenever the initial acceleration field is inward-directed or negative; that is $\ddot{R}_{q,0} < 0$. 
\end{enumerate}
The sign or value of each $\dot{R}_{q,0}$ affects only the slope of the associated $R_q$ curve at $t=0$; that is, these parameters only shift the $R_q$ curves to the left or right along the $t$-axis, and do not otherwise materially influence the qualitative global solution behavior.

In any event, as both double regular and double singlar behaviors are available for both $R_r$ and $R_z$, the solutions of Eqs.~(\ref{ode_1_gen}) and (\ref{ode_2}) manifest a total of three distinct behavioral archetypes (ignoring distinctions between DR-DS and DS-DR, for example). Examples of scenarios giving rise to these three behaviors are summarized in Table~\ref{initialconditionstable}), and are discussed further in Secs.~\ref{sec:DR-DR}-\ref{sec:DR-DS}.
\begin{table}[H]
\centering
\begin{tabular}{c c c c c c c}
Solution Type & \multicolumn{6}{c}{Initial Conditions} \\
\hline
&$R_{r,0}$ & $R_{z,0}$ & $\dot{R}_{r,0}$ & $\dot{R}_{z,0}$ &$\ddot{R}_{r,0}$ & $\ddot{R}_{z,0}$ \\
DR-DR & $1$ & $1$ & $1$ & $-1/4$ & $1$ & $1$ \\
DS-DS & $1$ & $1$ & $0$ & $1$ & $-1$ & $-1$ \\
DR-DS & $1$ & $1$ & $-2$ & $0$ & $1$ & $-1$ \\
\hline
\end{tabular}
\caption{Example scenarios that give rise to the three solution archetypes arising from numerical solution of Eqs.~(\ref{ode_1_gen}) and (\ref{ode_2}), subject to the initial conditions given by Eqs.~(\ref{Rr_zero})-(\ref{Rz_dot_zero}).}
\label{initialconditionstable}
\end{table}
%
%
%
%
%%%%%%%%%%%%%%%%%%%%%%%%%%%%%%%%%%%%%%%%%%%%%%%%%%
%%%%%%%%%%%%%%%%%%%%%%%%%%%%%%%%%%%%%%%%%%%%%%%%%%
\subsubsection{DR-DR Solutions}
\label{sec:DR-DR}
%%%%%%%%%%%%%%%%%%%%%%%%%%%%%%%%%%%%%%%%%%%%%%%%%%
%%%%%%%%%%%%%%%%%%%%%%%%%%%%%%%%%%%%%%%%%%%%%%%%%%
%
%
%
\begin{figure}[h]
        \includegraphics[scale=0.5]{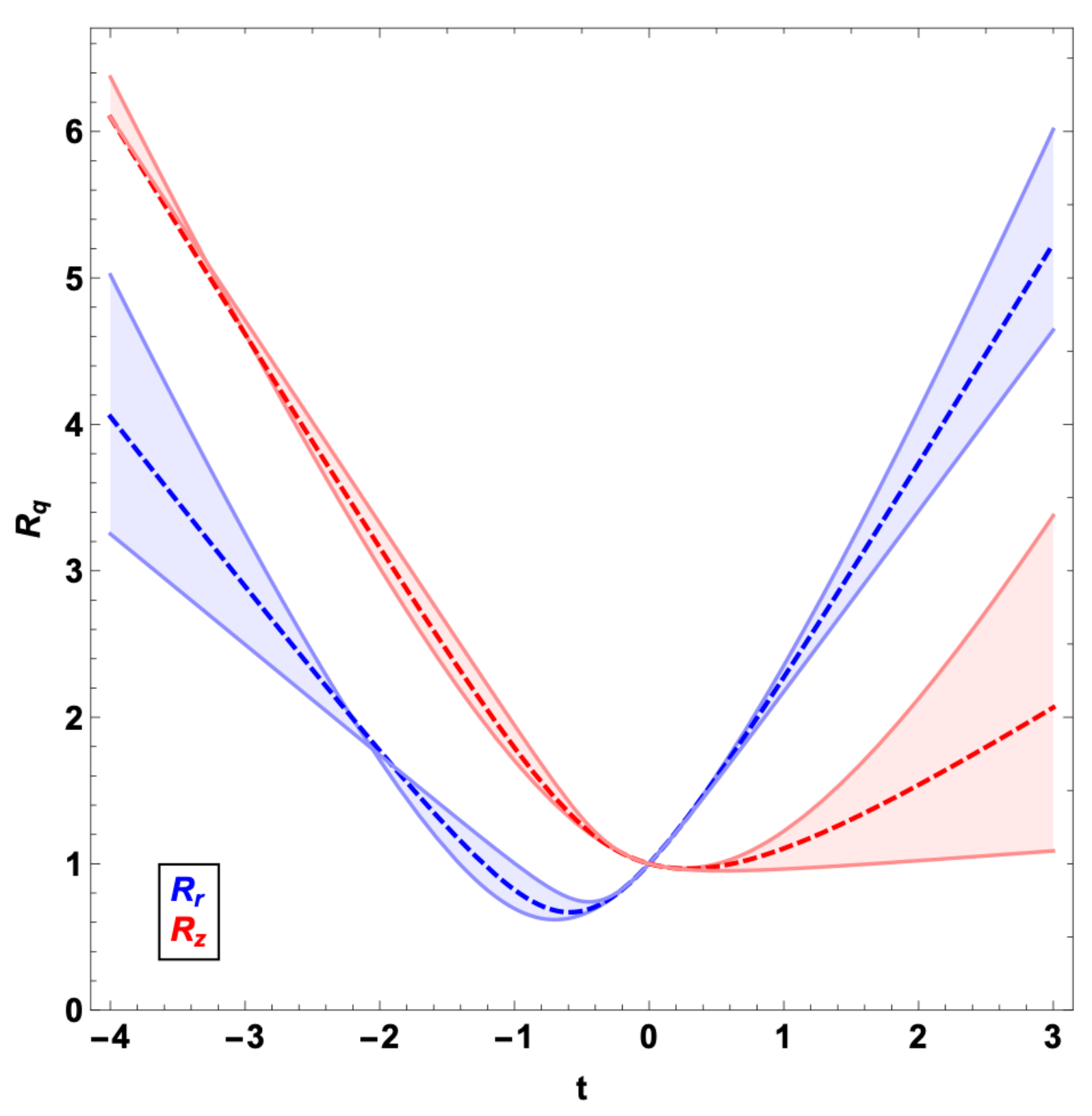}  
        \caption{ DR-DR solution of Eqs.~(\ref{ode_1_gen}) and (\ref{ode_2}), with Eqs.~(\ref{Rr_zero})-(\ref{Rz_dot_zero}) set to the values appearing in the first row of Table~\ref{initialconditionstable}. Shaded regions indicate a range of $\gamma$ parameterizations including $\gamma \in \left[1.1,3.0\right]$; in each case the dashed line correpsonds to $\gamma = 5/3$.}
\label{DR_DR_plot}
\end{figure}
An example of an initial data parameterization that gives rise to a DR-DR type solution (i.e., both $R_r$ and $R_z$ are double-regular, or globally concave-up) of Eqs.~(\ref{ode_1_gen}) and (\ref{ode_2}) with Eqs.~(\ref{Rr_zero})-(\ref{Rz_dot_zero}) is given in the first row of Table~\ref{initialconditionstable}, so that with Eqs.~(\ref{kappa_1_def}), (\ref{kappa_2_def}), and (\ref{kappas_relation}), $\kappa_1 = -1$, $\kappa_2 = -1$, and $\kappa_3 = 1$. The numerical solution of Eqs.~(\ref{ode_1_gen}) and (\ref{ode_2}) with Eqs.~(\ref{Rr_zero})-(\ref{Rz_dot_zero}) under the aforementioned parameterization is depicted in Fig.~\ref{DR_DR_plot}, for several choices of the adiabatic index $\gamma$. 

From the physical standpoint, the DR-DR behavior exemplified in Fig.~\ref{DR_DR_plot} manifests whenever $\ddot{R}_{r,0} > 0$ and $\ddot{R}_{z,0} > 0$, indicating that the global acceleration field is entirely positive at $t = 0$, and remains so for all $t$. Consequently, both $R_r$ and $R_z$ are observed to diverge as $|t| \to \infty$.

Otherwise, Fig.~\ref{DR_DR_plot} also features the trend that for sufficiently large $|t|$, both $R_r$ and $R_z$ increase with decreasing adiabatic index $\gamma$. In turn, for given initial conditions, this trend demonstrates that $R_r$ and $R_z$ evolve more rapidly in early or late time as $\gamma \to 1$. This trend is physically plausible in that the ideal gas compressibility increases with decreasing $\gamma$; in this sense, the ``small $\gamma$'' systems are expected to be more dynamically responsive (i.e., less rigid).
%
%%%%%%%%%%%%%%%%%%%%%%%%%%%%%%%%%%%%%%%%%%%%%%%%%%
%%%%%%%%%%%%%%%%%%%%%%%%%%%%%%%%%%%%%%%%%%%%%%%%%%
\subsubsection{DS-DS Solutions}
\label{sec:DS-DS}
%%%%%%%%%%%%%%%%%%%%%%%%%%%%%%%%%%%%%%%%%%%%%%%%%%
%%%%%%%%%%%%%%%%%%%%%%%%%%%%%%%%%%%%%%%%%%%%%%%%%%
%
%
%
\begin{figure}[h]
        \includegraphics[scale=0.515]{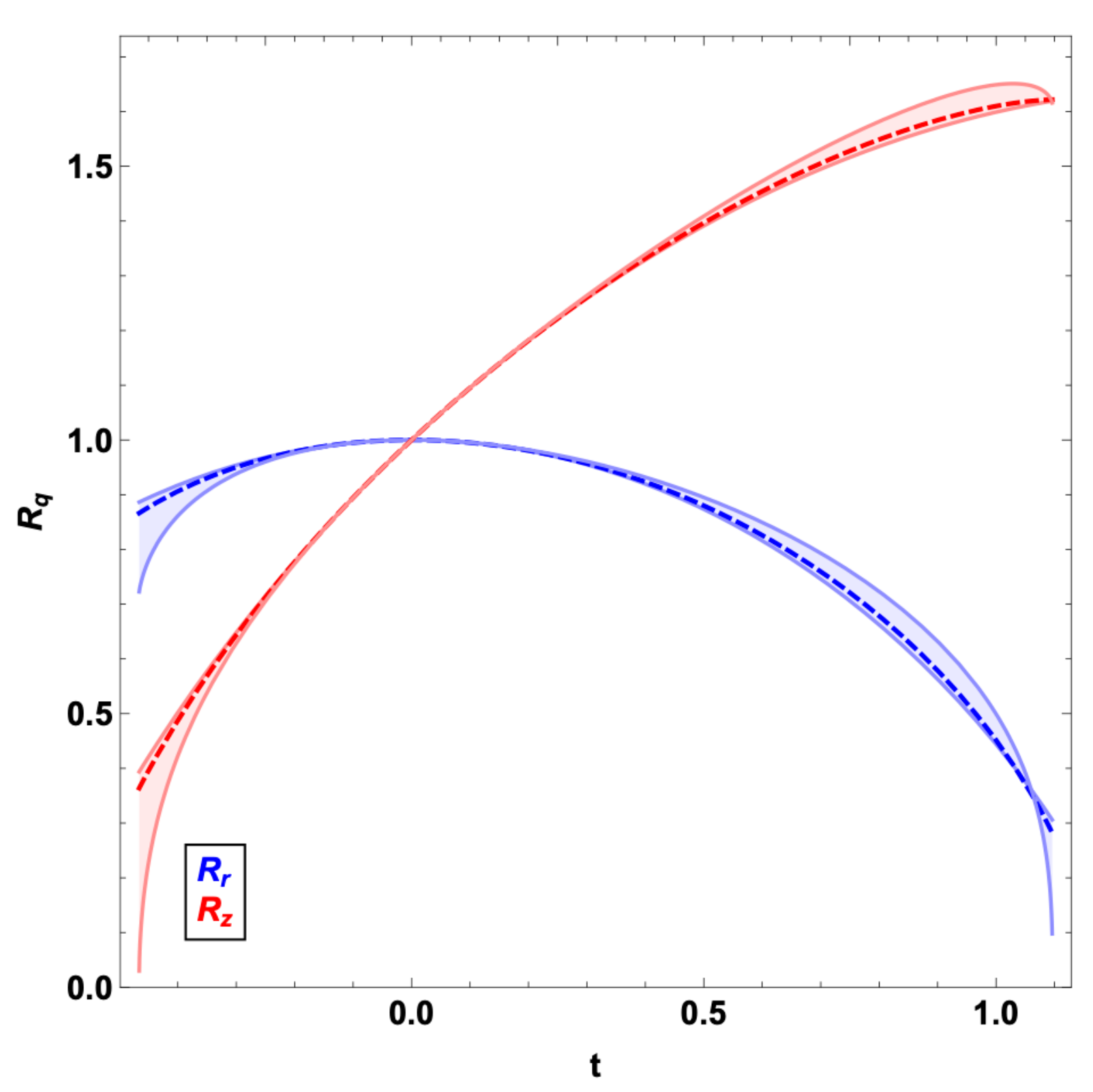}  
        \caption{DS-DS solution of Eqs.~(\ref{ode_1_gen}) and (\ref{ode_2}), with Eqs.~(\ref{Rr_zero})-(\ref{Rz_dot_zero}) set to the values appearing in the middle row of Table~\ref{initialconditionstable}. Shaded regions indicate a range of $\gamma$ parameterizations including $\gamma \in \left[1.1,3.0\right]$; in each case the dashed line correpsonds to $\gamma = 5/3$.}
\label{DS_DS_plot}
\end{figure}
An example of an initial data parameterization that gives rise to a DS-DS type solution (i.e., both $R_r$ and $R_z$ are double-singular, or globally concave-down) of Eqs.~(\ref{ode_1_gen}) and (\ref{ode_2}) with Eqs.~(\ref{Rr_zero})-(\ref{Rz_dot_zero}) is given by the second row of Table~\ref{initialconditionstable}, so that with Eqs.~(\ref{kappa_1_def}), (\ref{kappa_2_def}), and (\ref{kappas_relation}), $\kappa_1 = 1$, $\kappa_2 = 1$, and $\kappa_3 = 1$. The numerical solution of Eqs.~(\ref{ode_1_gen}) and (\ref{ode_2}) with Eqs.~(\ref{Rr_zero})-(\ref{Rz_dot_zero}) under the aforementioned parameterization is depicted in Fig.~\ref{DS_DS_plot}, for several choices of the adiabatic index $\gamma$. 

From the physical standpoint, the DS-DS behavior exemplified in Fig.~\ref{DS_DS_plot} manifests whenever $\ddot{R}_{r,0} < 0$ and $\ddot{R}_{z,0} < 0$, indicating that the global acceleration field is entirely negative at $t = 0$, and remains so for all $t$. Consequently, both $R_r$ and $R_z$ are observed to converge as $|t| > 0$. For the specific examples depicted in Fig.~\ref{DS_DS_plot}, in each featured case one of $R_r$ or $R_z$ reaches zero ``first'' (i.e., at some $t = |t*|$ smaller than the corresponding $t = |t*|$ associated with the other $R_q$), after which the overall solution ceases to have physical meaning. 

Otherwise, Fig.~\ref{DS_DS_plot} also features the same trends with respect to the adiabatic index $\gamma$ as observed and explained in Sec.~\ref{sec:DR-DR}. The value of $t = |t*|$ at which the solution terminates depends strongly on $\gamma$. 
%
%%%%%%%%%%%%%%%%%%%%%%%%%%%%%%%%%%%%%%%%%%%%%%%%%%
%%%%%%%%%%%%%%%%%%%%%%%%%%%%%%%%%%%%%%%%%%%%%%%%%%
\subsubsection{DR-DS Solutions}
\label{sec:DR-DS}
%%%%%%%%%%%%%%%%%%%%%%%%%%%%%%%%%%%%%%%%%%%%%%%%%%
%%%%%%%%%%%%%%%%%%%%%%%%%%%%%%%%%%%%%%%%%%%%%%%%%%
%
%
%
\begin{figure}[h]
        \includegraphics[scale=0.5]{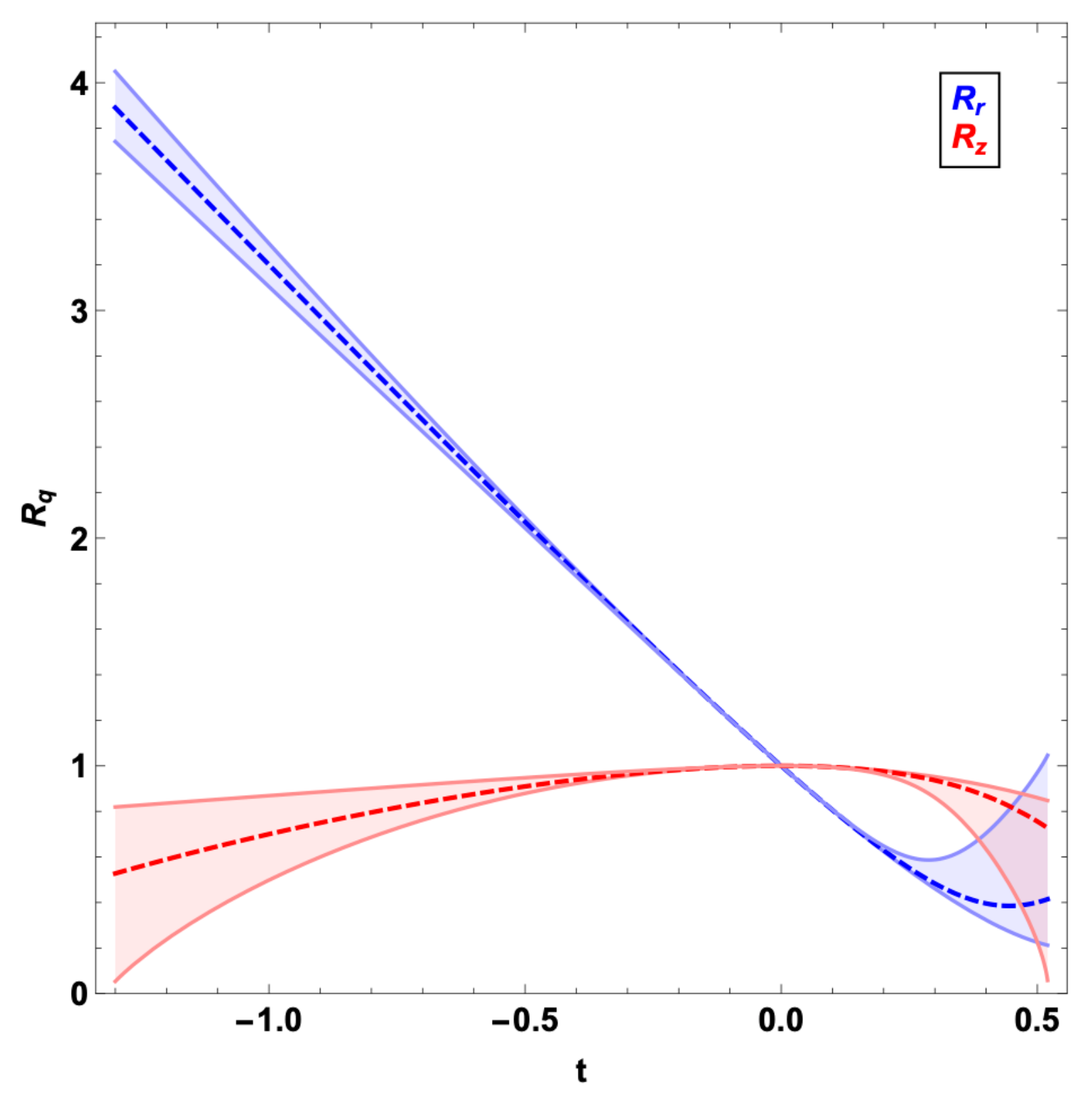}  
        \caption{DR-DS solution of Eqs.~(\ref{ode_1_gen}) and (\ref{ode_2}), with Eqs.~(\ref{Rr_zero})-(\ref{Rz_dot_zero}) set to the values appearing in the final row of Table~\ref{initialconditionstable}. Shaded regions indicate a range of $\gamma$ parameterizations including $\gamma \in \left[1.1,3.0\right]$; in each case the dashed line correpsonds to $\gamma = 5/3$. }
\label{DR_DS_plot}
\end{figure}

An example of an initial data parameterization that gives rise to a DR-DS type solution (i.e., one each of $R_r$ and $R_z$ is double-regular or globally concave-up and double-singluar or globally concave-down) of Eqs.~(\ref{ode_1_gen}) and (\ref{ode_2}) with Eqs.~(\ref{Rr_zero})-(\ref{Rz_dot_zero}) is given by the final row of Table~\ref{initialconditionstable} so that with Eqs.~(\ref{kappa_1_def}), (\ref{kappa_2_def}), and (\ref{kappas_relation}), $\kappa_1 = -1$, $\kappa_2 = 1$, and $\kappa_3 = -1$. The numerical solution of Eqs.~(\ref{ode_1_gen}) and (\ref{ode_2}) with Eqs.~(\ref{Rr_zero})-(\ref{Rz_dot_zero}) under the aforementioned parameterization is depicted in Fig.~\ref{DR_DS_plot}, for several choices of the adiabatic index $\gamma$. 

From the physical standpoint, the DR-DS behavior exemplified in Fig.~\ref{DR_DS_plot} manifests whenever $\ddot{R}_{r,0} > 0$ and $\ddot{R}_{z,0} < 0$ (or vice versa), indicating that the global acceleration field is positive in one direction and negative in the other at $t = 0$, and remains so for all $t$. Consequently, one of $R_r$ and $R_z$ is observed to diverge as $|t| > 0$, while the other is observed to converge. For the specific examples depicted in Fig.~\ref{DR_DS_plot}, in each featured case only $R_z$ reaches zero at two times, after which the overall solution ceases to have physical meaning. 

Otherwise, Fig.~\ref{DR_DS_plot} also features the same trends with respect to the adiabatic index $\gamma$ as observed and explained in Secs.~\ref{sec:DR-DR} and \ref{sec:DS-DS}.
%
%%%%%%%%%%%%%%%%%%%%%%%%%%%%%%%%%%%%%%%%%%%%%%%%%%
%%%%%%%%%%%%%%%%%%%%%%%%%%%%%%%%%%%%%%%%%%%%%%%%%%
\subsubsection{The Asymptotic Scale Radius Ratio}
\label{sec:asymptotic}
%%%%%%%%%%%%%%%%%%%%%%%%%%%%%%%%%%%%%%%%%%%%%%%%%%
%%%%%%%%%%%%%%%%%%%%%%%%%%%%%%%%%%%%%%%%%%%%%%%%%%

%
%
%
\begin{figure*}[]
        \includegraphics[scale=0.5]{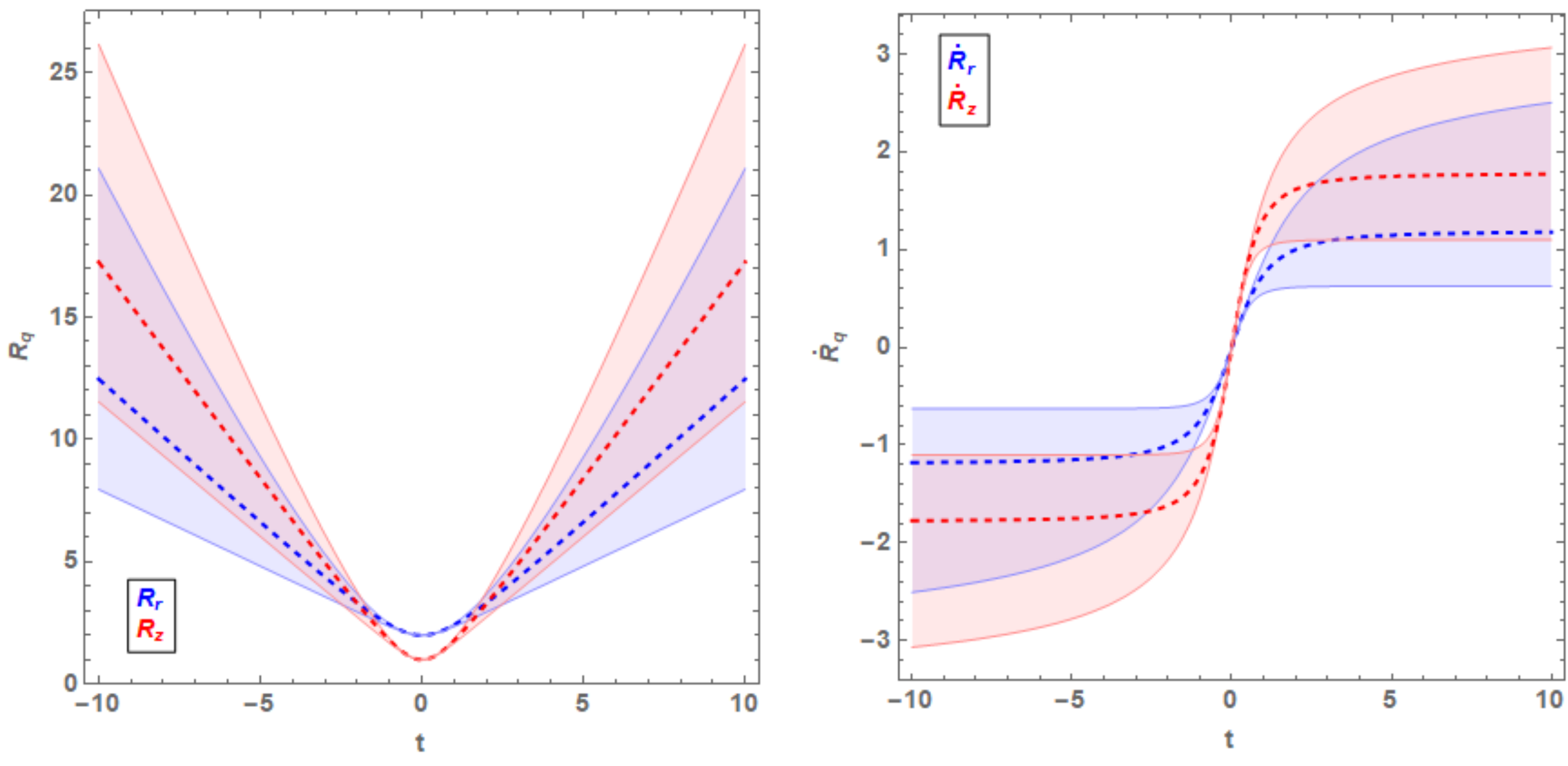}  
        \caption{DR-DR solution of Eqs.~(\ref{ode_1_gen}) and (\ref{ode_2}) for $R_r$ and $R_z$ (left) and $\dot{R}_r$ and $\dot{R}_z$ (right), with Eqs.~(\ref{Rr_zero})-(\ref{Rz_dot_zero}) set to $R_{r,0} = 2$, $R_{z,0} = 1$, $\dot{R}_{r,0} = \dot{R}_{z,0} = 0$; also $\kappa_1 = \kappa_2$ given by Eq.~(\ref{kappas_asymp}), so that $\kappa_3 = 1$. Shaded regions indicate a range of $\gamma$ parameterizations including $\gamma \in \left[1.1,3.0\right]$; in each case the dashed line correpsonds to $\gamma = 5/3$. }
\label{asymptotic_plot}
\end{figure*}
\begin{figure*}[]
        \includegraphics[scale=0.5]{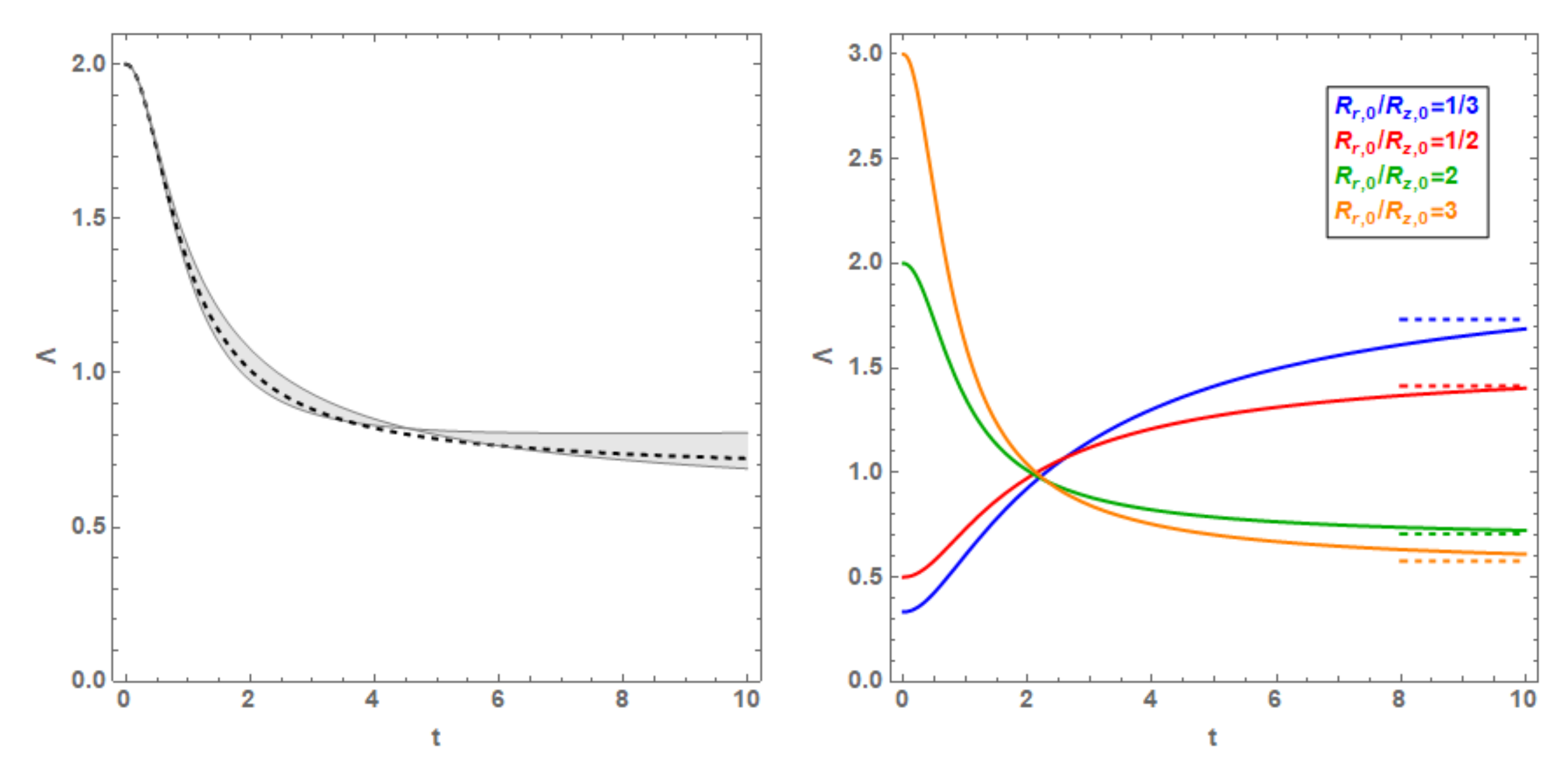}  
        \caption{The scale radius ratio $\Lambda\left(t\right)$ defined by Eq.~(\ref{Lambda_def}), and corresponding to the example solution provided in Fig.~\ref{asymptotic_plot}. Left: Solutions for $R_{r,0} = 2$ and $R_{z,0} = 1$; shaded regions indicate a range of $\gamma$ parameterizations including $\gamma \in \left[1.1,3.0\right]$, where the dashed line correpsonds to $\gamma = 5/3$. Right: Solutions for $\gamma = 5/3$ and indicated choices of $\frac{R_{r,0}}{R_{z,0}}$; dashed lines indicate the asymptotic values as given by Eq.~(\ref{Lambda_inf_HL}). }
\label{Lambda_plot}
\end{figure*}

A matter that has received considerable attention in the established literature on DR-DR type Nemchinov-Dyson solutions is the asymptotic (i.e, late-time) ratio of the scale radii $R_r$ and $R_z$ under a certain set of initial conditions, namely Eqs.~(\ref{Rr_zero})-(\ref{Rz_dot_zero}) with $\dot{R}_{r,0} = \dot{R}_{z,0} = 0$ and $R_{r,0} \neq R_{z,0}$. To explore this notion, an example of an initial data parameterization that gives rise to the relevant DR-DR type solution (i.e., both $R_r$ and $R_z$ are double-regular, or globally concave-up) of Eqs.~(\ref{ode_1_gen}) and (\ref{ode_2}) is given by $\kappa_3 = 1$, whence
\begin{equation}
\frac{R_r}{R_z} = \frac{\ddot{R}_z}{\ddot{R}_r} ,
\label{kappa_is_one}
\end{equation}
one possible consequence of which is
\begin{equation}
\kappa_1 = \kappa_2 = -R_{r,0}^{1-2\gamma} R_{z,0}^{-\gamma} .
\label{kappas_asymp}
\end{equation}

The numerical solution of Eqs.~(\ref{ode_1_gen}) and (\ref{ode_2}) with Eqs.~(\ref{Rr_zero})-(\ref{Rz_dot_zero}) and (\ref{kappas_asymp}), and under the aforementioned parameterization is depicted in Fig.~\ref{asymptotic_plot}, for $R_{r,0} = 2$, $R_{z,0} = 1$, and several choices of the adiabatic index $\gamma$. Two trends are immediately evident from Fig.~\ref{asymptotic_plot}, in addition to those discussed in Sec.~\ref{sec:DR-DR}:
\begin{enumerate}
\item The $R_r$ and $R_z$ curves cross at some finite $t > 0$,
\item For some late $t > 0$, both $R_r$ and $R_z$ feature constant slopes, or $\dot{R}_r$ and $\dot{R}_z$ approach constant values. This phenomenon is characterized by Nemchinov~\cite{nemchinov_1965} as corresponding to ``when the expansion becomes intertial.'' 
\end{enumerate}
These essential phenonema may also be encoded in the time-evolution of the ratio $\Lambda\left(t\right)$, for example defined by
\begin{equation}
\Lambda \equiv \frac{R_r}{R_z} ,
\label{Lambda_def}
\end{equation}
and depcited in the left member of Fig.~\ref{Lambda_plot} for the same example scenario shown in Fig.~\ref{asymptotic_plot}. 

The left member of Fig.~\ref{Lambda_plot} indicates that in the illustrated example scenario the scale radius ratio features $\Lambda > 1$ at $t = 0$, reaches $\Lambda = 1$ at the same time where the $R_r$ and $R_z$ curves cross in Fig.~\ref{asymptotic_plot}, and features $\Lambda < 1$ thereafter. As otherwise suggested by Fig.~\ref{asymptotic_plot}, the left member of Fig.~\ref{Lambda_plot} also indicates $\Lambda$ approaches a constant value $\Lambda_{\infty}$ for late times; clearly $\Lambda_{\infty}$ depends on the value of $\gamma$. 

These trends have been thoroughly investigated by authors such as Dyson~\cite{dyson_original}, Nemchinov~\cite{nemchinov_1965}, Anisimov and Lysikov~\cite{anisimov_1970}, and Hunter and London~\cite{hunter1988multidimensional} to name a few. In particular, both Dyson~\cite{dyson_original} and Nemchinov~\cite{nemchinov_1965} provide tables of $\Lambda_{\infty}$ not only for various choices of $\gamma$, but also the initial scale radius ratio $\Lambda_0$ defined by
\begin{equation}
\Lambda_0 \equiv \frac{R_{r,0}}{R_{z,0}} .
\label{Lambda0_def}
\end{equation}
Moreover, for the special case of $\gamma = 5/3$, Anisimov and Lysikov~\cite{anisimov_1970} provide an approximate analytical expression for $\Lambda_{\infty}$, valid under the circumstances where $R_{r,0} \approx R_{z,0}$. In turn, this result has since been further rigorized by Hunter and London~\cite{hunter1988multidimensional} (see also references to H. Hietarinta therein), who based on both rigorous analytical calculations and substantiating numerical evidence arrived at
\begin{equation}
\Lambda_{\infty}\left(\gamma = 5/3 \right) = \sqrt{\frac{R_{z,0}}{R_{r,0}}} ,
\label{Lambda_inf_HL}
\end{equation}
but is notably not in agreement with with Nemchinov's~\cite{nemchinov_1965} counterpart result. However, the right member of Fig.~\ref{Lambda_plot} is consistent with Eq.~(\ref{Lambda_inf_HL}) as indicated, potentially casting doubt on the veracity of some of Nemchinov's~\cite{nemchinov_1965} numerical results.

Beyond even this independent confirmation of Hunter and London's~\cite{hunter1988multidimensional} conclusions, there remain a variety of outstanding matters pertaining to the asymptotic scale radius ratio. For example, reconciliation of Hunter and London's~\cite{hunter1988multidimensional} and the current results with Anisimov and Lysikov's~\cite{anisimov_1970} analytical formula remains to be investigated, as does the construction of analytical results like Eq.~(\ref{Lambda_inf_HL}) for $\gamma \neq 5/3$. Given the likely non-trivial effort necessary to embark upon some of these and related endeavors, their discussion as part of future programs of study is relegated to Sec.~\ref{sec:recommendations}.  
%
%
%
%%%%%%%%%%%%%%%%%%%%%%%%%%%%%%%%%%%%%%%%%%%%%%%%%%
%%%%%%%%%%%%%%%%%%%%%%%%%%%%%%%%%%%%%%%%%%%%%%%%%%
\subsection{Solution Sets for $\Pi$}
\label{sec:pi_sols}
%%%%%%%%%%%%%%%%%%%%%%%%%%%%%%%%%%%%%%%%%%%%%%%%%%
%%%%%%%%%%%%%%%%%%%%%%%%%%%%%%%%%%%%%%%%%%%%%%%%%%
%
As noted by Sedov~\cite{sedov} in the context of the 1D linear velocity solutions, in the axisymmetric Nemchinov-Dyson soltution given by Eqs.~(\ref{ur_assump}), (\ref{uz_assump}), (\ref{rho_dyson_solution_zeta}), (\ref{p_dyson_solution_zeta}), (\ref{sie_dyson_solution_zeta}), and (\ref{entropy_dyson_solution_zeta}), the arbitrary functions $\Pi$, $\Gamma$, $\Upsilon$, and $\Sigma$ are
\begin{quote}
``...directly related to the entropy distribution through the gas.''
\end{quote}
This notion is of course explicitly true by definition [i.e., in light of Eq.~(\ref{entropy_dyson_solution_zeta})] for the function $\Sigma$, and in due course $\Pi$, $\Gamma$, and $\Upsilon$ according to Eqs.~(\ref{big_gamma_def}), (\ref{upsilon_def}), and (\ref{sigma_def}). However, aside from Eq.~(\ref{pi_solution}), no additional constraints are available in the underlying formulation of the ideal gas inviscid Euler equations for the resolution of the otherwise arbitrary functional forms in $\zeta$ of $\Pi$, $\Gamma$, $\Upsilon$, and $\Sigma$. 

Indeed, this degree of arbitrariness appearing in the axisymmetric Nemchinov-Dyson solution (or, more broadly, any solution of the ideal gas inviscid Euler equations featuring linear velocity assumptions) is a direct result of the lack of ancillary mechanisms (e.g., gravity, viscosity, or heat conduction) appearing in Eqs.~(\ref{cons_mass})-(\ref{cons_enegy_E}). On the other hand, the inclusion of any such mechanism in Eqs.~(\ref{cons_mass})-(\ref{cons_enegy_E}) provides an additional constraint that must be satisfied in addition to Eq.~(\ref{pi_solution}), and so selects unique (but self-consistent) forms of $\Pi$, $\Gamma$, $\Upsilon$, and $\Sigma$. An example of this phenomenology is provided by Hendon and Ramsey~\cite{hendon2012radiation}, in the context of 1D linear velocity solutions featuring a thermal radiation diffusion process in an ideal gas. Other examples are provided in the voluminous body of literature featuring linear velocity ellipsoidal gas cloud motions coupled to gravitational processes (see, for example, Bogoyavlensky~\cite{bogoyavlenskymethods} and numerous references found therein).

When featuring a dissipative process, this outcome is expected due to the classically established thermodynamical connections between dissipation and entropy generation. In particular, according to the Second Law of Thermodynamics, a dissipative or irreversible process not only transforms energy from one form to another, but also produces entropy at a specified rate. In the context of fluid flow scenarios, the prescribed functional form of a dissipative process sets this rate, and thus constrains the entropy $S$ (or the function $\Sigma$, and so $\Pi$, $\Gamma$, and $\Upsilon$). In the absence of such a mechanism - such as in Eqs.~(\ref{cons_mass})-(\ref{cons_enegy_E}) - no dissipation rate is available to be calculated and thus further constrain the underlying thermodynamics, and $\Pi$, $\Gamma$, $\Upsilon$, and $\Sigma$ therefore remain arbitrary unless another ancillary constraint is provided.  

\begin{figure*}[]
    \begin{subfigure}[t]{0.4\textwidth}
 %       \centering
        \includegraphics[scale=0.30]{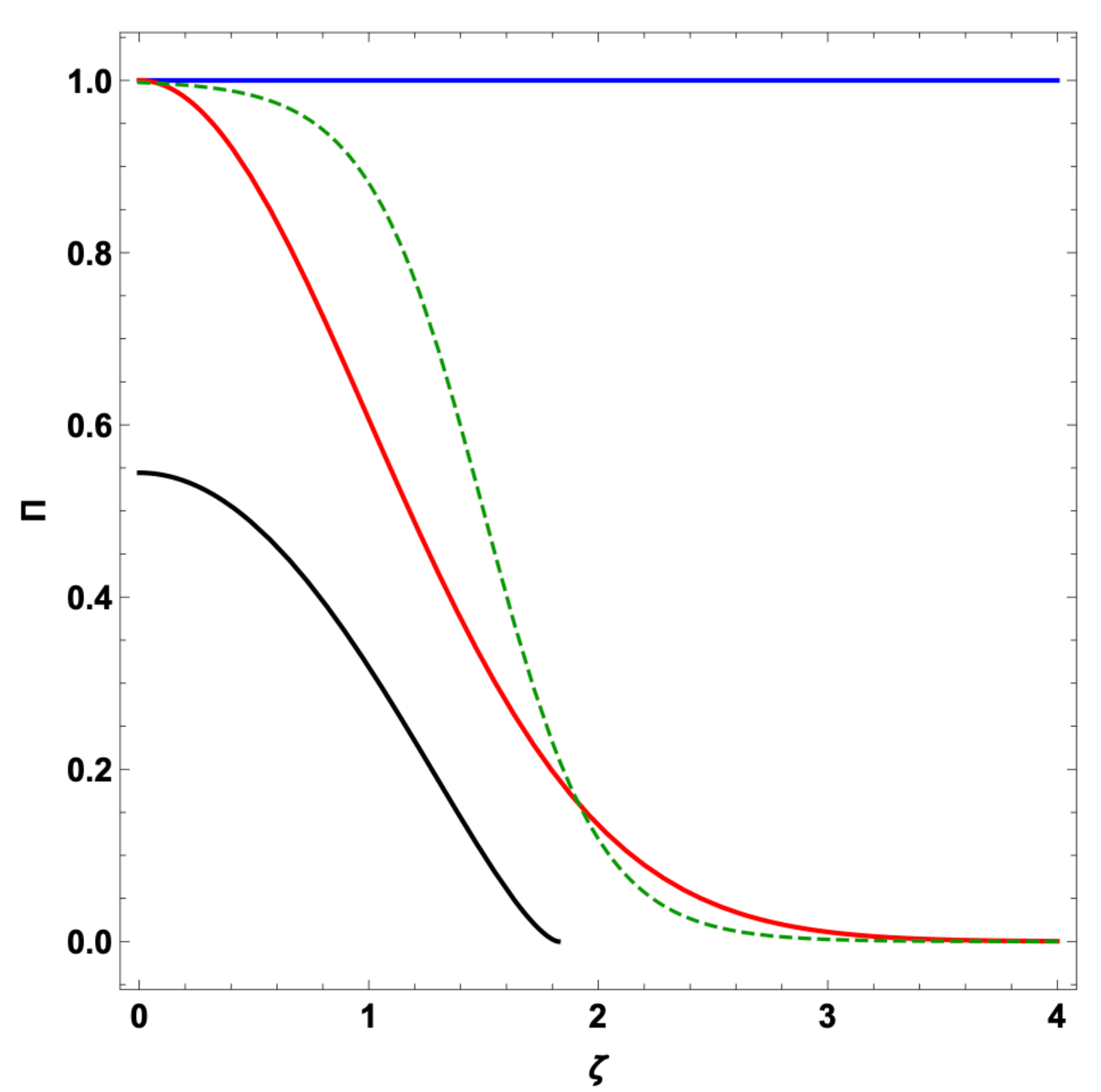}\caption{}\label{piplot}
    \end{subfigure} 
    \begin{subfigure}[t]{0.4\textwidth}
 %       \centering
        \includegraphics[scale=0.30]{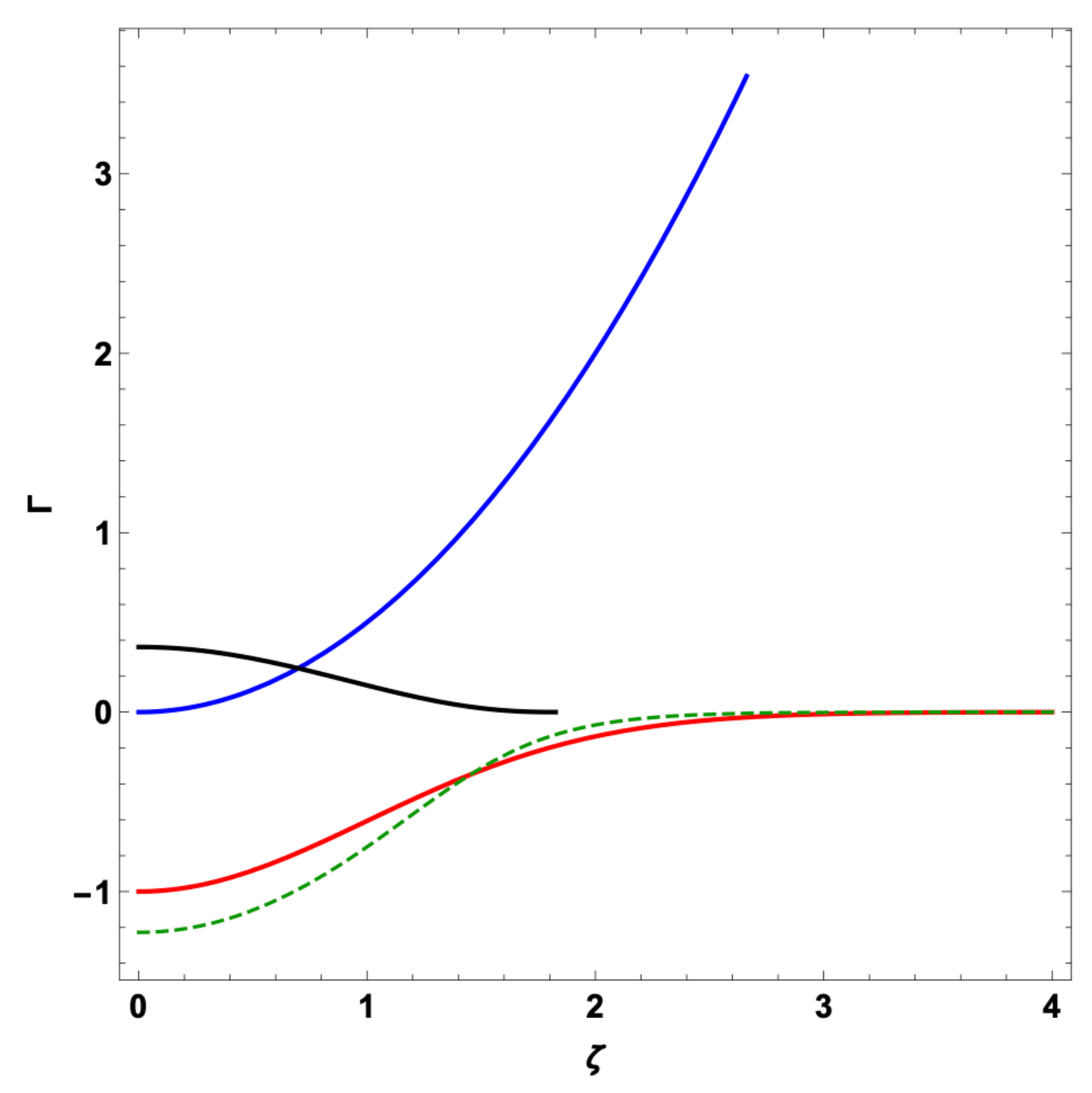}\caption{}\label{gammaplot}
    \end{subfigure}   
    \begin{subfigure}[t]{0.4\textwidth}
 %       \centering
        \includegraphics[scale=0.28]{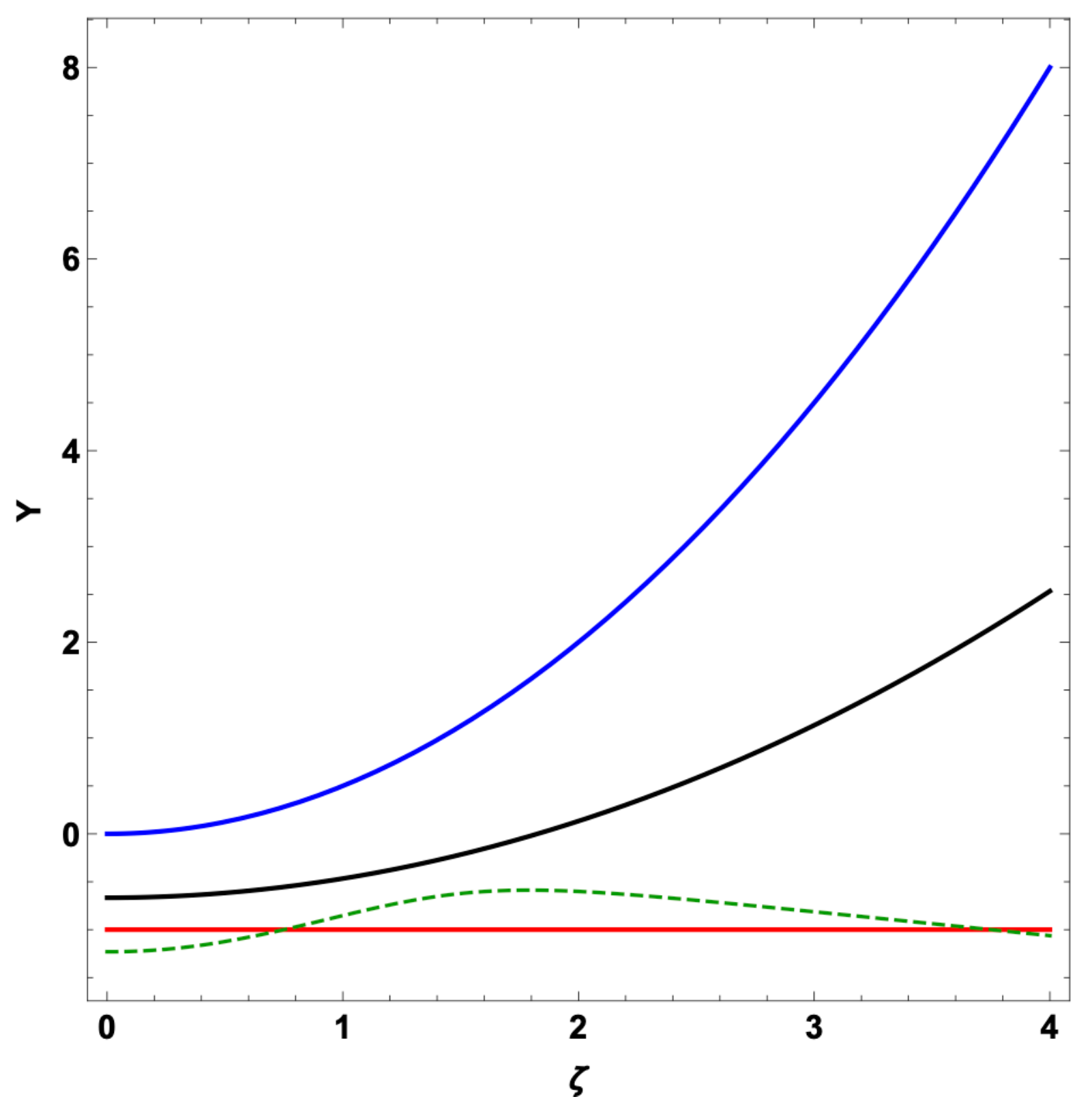}\caption{}\label{upsilonplot}
    \end{subfigure}   
    \begin{subfigure}[t]{0.4\textwidth}
 %       \centering
        \includegraphics[scale=0.30]{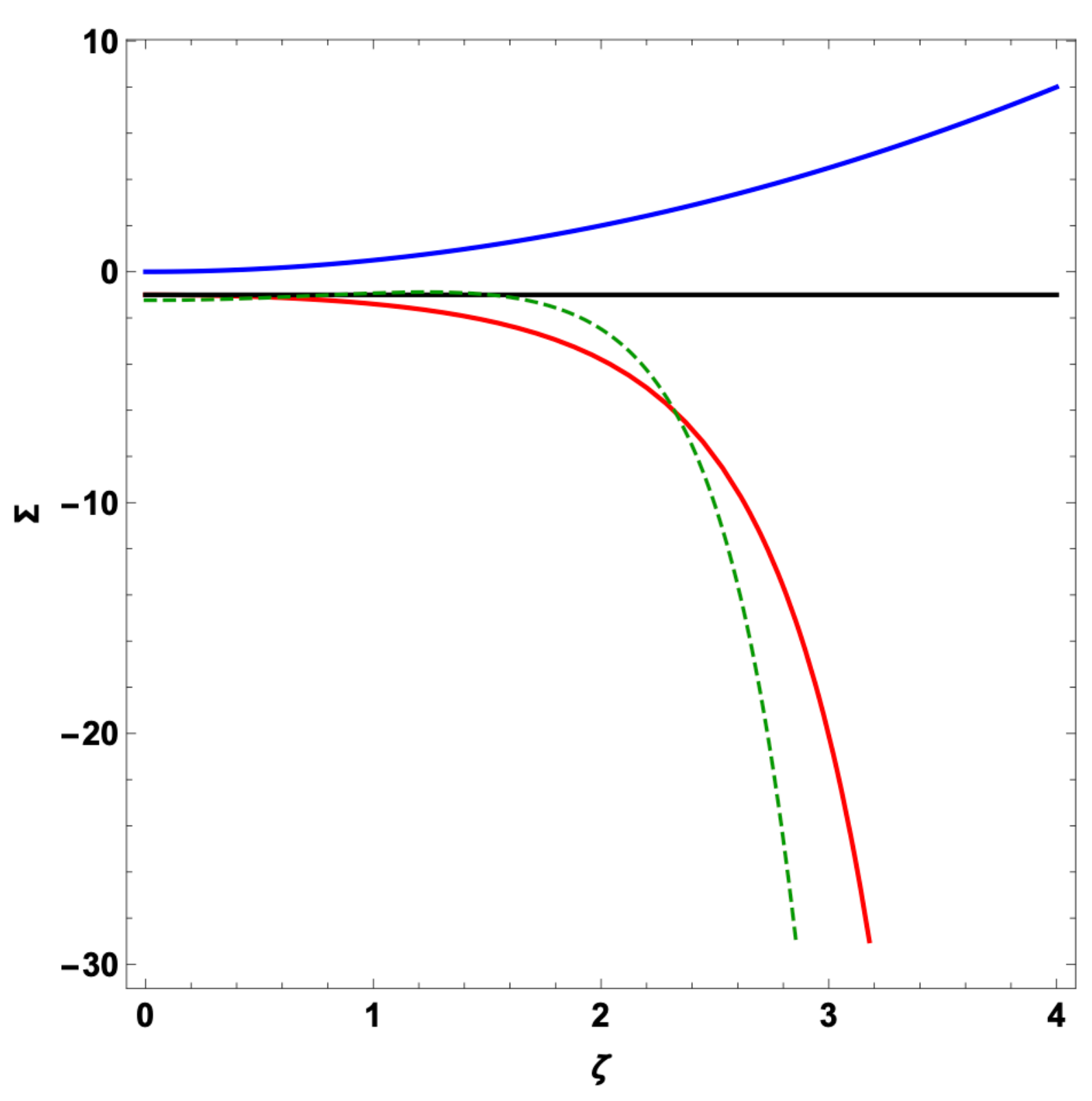}\caption{}\label{sigmaplot}
    \end{subfigure}   
        \caption{The functions $\Pi\left(\zeta\right)$ (a), $\Gamma\left(\zeta\right)$ (b), $\Upsilon\left(\zeta\right)$ (c), and $\Sigma\left(\zeta\right)$ (d) for the uniform density (solid blue), uniform SIE (solid red), uniform entropy (solid black), and diffuse surface (dashed green) solutions.}
\label{spatial_solution_set_plots}
\end{figure*}

In this case, the functional form of $\Pi$ in $\zeta$ [and through their definitions given by Eqs.~(\ref{big_gamma_def}), (\ref{upsilon_def}), and (\ref{sigma_def}), $\Gamma$, $\Upsilon$, and $\Sigma$] appearing in the axisymmetric Nemchinov-Dyson solution therefore remains arbitrary, and thus may be prescribed according to target flow patterns of interest. Four such examples are provided in Secs.~\ref{subsubsec:uniform_density}-\ref{subsubsec:woodsaxon}, and are depicted in Fig.~\ref{spatial_solution_set_plots}.
%
%%%%%%%%%%%%%%%%%%%%%%%%%%%%%%%%%%%%%%%%%%%%%%%%%%
%%%%%%%%%%%%%%%%%%%%%%%%%%%%%%%%%%%%%%%%%%%%%%%%%%
\subsubsection{Uniform Density Solutions}
\label{subsubsec:uniform_density}
%%%%%%%%%%%%%%%%%%%%%%%%%%%%%%%%%%%%%%%%%%%%%%%%%%
%%%%%%%%%%%%%%%%%%%%%%%%%%%%%%%%%%%%%%%%%%%%%%%%%%
%
The form of the function $\Pi$ associated with a uniform (i.e., constant in spatial coordinates, but not necessarily in time) density distribution is given by
\begin{equation}
\Pi\left(\zeta\right) = \Pi_0 ,
\label{pi_uniform_density}
\end{equation}
where the otherwise arbitrary constant $\Pi_0 > 0$ so that the density given by Eq.~(\ref{rho_dyson_solution_zeta}) is positive definite and thus physically realistic. With Eq.~(\ref{pi_uniform_density}), the related functions $\Gamma$, $\Upsilon$, and $\Sigma$ defined by Eqs.~(\ref{big_gamma_def}), (\ref{upsilon_def}), and (\ref{sigma_def}), respectively, realize as
\begin{eqnarray}
\Gamma\left(\zeta\right) &=& \frac{\Pi_0 \zeta^2}{2} + \Gamma_0, \label{gamma_uniform_density} \\
\Upsilon\left(\zeta\right) &=& \frac{\zeta^2}{2} + \frac{\Gamma_0}{\Pi_0},\label{upsilon_uniform_density} \\
\Sigma\left(\zeta\right) &=& \frac{\Pi_0^{1-\gamma}\zeta}{2} + \Pi_0^{-\gamma}\Gamma_0 ,\label{sigma_uniform_density}
\end{eqnarray}
where $\Gamma_0$ is an arbitrary integration constant, the sign of which must be selected so that the pressure, SIE, and entropy given by Eqs.~(\ref{p_dyson_solution_zeta}), (\ref{sie_dyson_solution_zeta}), and (\ref{entropy_dyson_solution_zeta}) are positive definite and thus physically realistic. The pressure and SIE distributions associated with a uniform density distribution are thus revealed to be parabolic in $\zeta$, while the entropy distribution is revealed to be linear in $\zeta$. 

Equations~(\ref{pi_uniform_density})-(\ref{sigma_uniform_density}) are depicted in Fig.~\ref{spatial_solution_set_plots} for the example parameterization $\gamma = 5/3$, $\Pi_0 = 1$, and $\Gamma_0 = 0$. 
%
%%%%%%%%%%%%%%%%%%%%%%%%%%%%%%%%%%%%%%%%%%%%%%%%%%
%%%%%%%%%%%%%%%%%%%%%%%%%%%%%%%%%%%%%%%%%%%%%%%%%%
\subsubsection{Uniform SIE Solutions}
\label{subsubsec:uniform_sie}
%%%%%%%%%%%%%%%%%%%%%%%%%%%%%%%%%%%%%%%%%%%%%%%%%%
%%%%%%%%%%%%%%%%%%%%%%%%%%%%%%%%%%%%%%%%%%%%%%%%%%
%
The form of the function $\Upsilon$ associated with a uniform (i.e., constant in $r$ and $z$, but not necessarily in $t$) SIE (or temperature) distribution is given by
\begin{equation}
\Upsilon\left(\zeta\right) = \Upsilon_0 ,
\label{upsilon_uniform_sie}
\end{equation}
where $\Upsilon_0$ is an arbitrary constant, the sign of which must be selected so that the SIE given by Eq.~(\ref{sie_dyson_solution_zeta}) is positive definite and thus physically realistic. With Eqs.~(\ref{upsilon_def}) and (\ref{upsilon_uniform_density}), the related function $\Pi$ then satisfies
\begin{equation}
\Upsilon_0 = \frac{\int \zeta \Pi d\zeta}{\Pi} ,
\label{upsilon_def_uniform_sie}
\end{equation}
or, equivalently, after differentiating Eq.~(\ref{upsilon_def_uniform_sie}) with respect to $\zeta$,
\begin{equation}
\Upsilon_0 \frac{d\Pi}{d\zeta} - \zeta \Pi = 0 ,
\label{pi_ode_uniform_sie}
\end{equation}
the solution of which is given by
\begin{equation}
\Pi\left(\zeta\right) = \Pi_0 \rm{exp}\left(\frac{\zeta^2}{2\Upsilon_0}\right) ,
\label{pi_uniform_sie}
\end{equation}
where the otherwise arbitrary integration constant $\Pi_0 > 0$ so that the density given by Eq.~(\ref{rho_dyson_solution_zeta}) is positive definite and thus physically realistic. With Eq.~(\ref{pi_uniform_sie}), the related functions $\Gamma$ and $\Sigma$ defined by Eqs.~(\ref{big_gamma_def}) and (\ref{sigma_def}), respectively, realize as
\begin{eqnarray}
\Gamma\left(\zeta\right) &=& \Pi_0 \Upsilon_0 \rm{exp}\left(\frac{\zeta^2}{2\Upsilon_0}\right),
\label{gamma_uniform_sie} \\
\Sigma\left(\zeta\right) &=&
\Pi_0^{1-\gamma} \Upsilon_0 \rm{exp}\left[\frac{\left(1-\gamma\right) \zeta^2}{2\Upsilon_0}\right] .
\label{sigma_uniform_sie}
\end{eqnarray}
The density, pressure, and entropy distributions associated with a uniform SIE distribution are thus revealed to be Gaussian in $\zeta$. 

Equations~(\ref{upsilon_uniform_sie}), (\ref{pi_uniform_sie}), (\ref{gamma_uniform_sie}), and (\ref{sigma_uniform_sie}) are depicted in Fig.~\ref{spatial_solution_set_plots} for the example parameterization $\gamma = 5/3$, $\Pi_0 = 1$, and $\Upsilon_0 = -1$. 
%
%%%%%%%%%%%%%%%%%%%%%%%%%%%%%%%%%%%%%%%%%%%%%%%%%%
%%%%%%%%%%%%%%%%%%%%%%%%%%%%%%%%%%%%%%%%%%%%%%%%%%
\subsubsection{Uniform Entropy Solutions}
\label{subsubsec:uniform_entropy}
%%%%%%%%%%%%%%%%%%%%%%%%%%%%%%%%%%%%%%%%%%%%%%%%%%
%%%%%%%%%%%%%%%%%%%%%%%%%%%%%%%%%%%%%%%%%%%%%%%%%%
%
The form of the function $\Sigma$ associated with a uniform (i.e., constant in $r$ and $z$, but not necessarily in $t$) entropy distribution is given by
\begin{equation}
\Sigma\left(\zeta\right) = \Sigma_0 ,
\label{sigma_uniform_entropy}
\end{equation}
where $\Sigma_0$ is an arbitrary constant, the sign of which must be selected so that the entropy given by Eq.~(\ref{entropy_dyson_solution_zeta}) is positive definite and thus physically realistic. With Eqs.~(\ref{sigma_def}) and (\ref{sigma_uniform_entropy}), the related function $\Pi$ then satisfies
\begin{equation}
\Sigma_0 = -\Pi^{-\gamma} \int \zeta \Pi d\zeta ,
\label{sigma_def_uniform_entropy}
\end{equation}
or, equivalently, after differentiating Eq.~(\ref{sigma_def_uniform_entropy}) with respect to $\zeta$,
\begin{equation}
\gamma \Sigma_0 \Pi^{\gamma-1} \frac{d\Pi}{d\zeta} + \zeta \Pi = 0 ,
\label{pi_ode_uniform_entropy}
\end{equation}
the solution of which is given by
\begin{equation}
\Pi\left(\zeta\right) = \left[ \left(\gamma-1\right)\Pi_0 + \frac{\left(\gamma-1\right) \zeta^2}{2\gamma\Sigma_0}\right]^{\frac{1}{\gamma-1}} ,
\label{pi_uniform_entropy}
\end{equation}
where $\Pi_0$ is an arbitrary integration constant. With Eq.~(\ref{pi_uniform_sie}), the related functions $\Gamma$ and $\Upsilon$ defined by Eqs.~(\ref{big_gamma_def}) and (\ref{upsilon_def}), respectively, realize as
\begin{eqnarray}
\Gamma\left(\zeta\right) &=&
-\Pi_0 S_0
\left[ \left(\gamma-1\right)\Pi_0 + \frac{\left(\gamma-1\right) \zeta^2}{2\gamma\Sigma_0}\right]^{\frac{\gamma}{\gamma-1}}
\label{gamma_uniform_entropy}, \\
\Upsilon\left(\zeta\right) &=&
\frac{\left(\gamma-1\right)\left(2\gamma\Pi_0\Sigma_0 + \zeta^2\right)}{2\gamma} ,
\label{upsilon_uniform_entropy}
\end{eqnarray}
indicating the sign of the otherwise arbitrary constant $\Pi_0$ must be selected in conjunction with that of $\Sigma_0$ so that the pressure given by Eq.~(\ref{p_dyson_solution_zeta}) is positive definite and thus physically realistic. The density and pressure distributions associated with a uniform entropy distribution are thus revealed to follow power laws in $\zeta$, while the SIE distribution is revealed to be quadratic in $\zeta$. 

Equations~(\ref{sigma_uniform_entropy}), (\ref{pi_uniform_entropy}), (\ref{gamma_uniform_entropy}), and (\ref{upsilon_uniform_entropy}) are depicted in Fig.~\ref{spatial_solution_set_plots} for the example parameterization $\gamma = 5/3$, $\Pi_0 = 1$, and $\Sigma_0 = -1$. 
%
%%%%%%%%%%%%%%%%%%%%%%%%%%%%%%%%%%%%%%%%%%%%%%%%%%
%%%%%%%%%%%%%%%%%%%%%%%%%%%%%%%%%%%%%%%%%%%%%%%%%%
\subsubsection{Diffuse Surface Solutions}
\label{subsubsec:woodsaxon}
%%%%%%%%%%%%%%%%%%%%%%%%%%%%%%%%%%%%%%%%%%%%%%%%%%
%%%%%%%%%%%%%%%%%%%%%%%%%%%%%%%%%%%%%%%%%%%%%%%%%%
%
For an otherwise arbitrary constant $\Pi_0 > 0$, so that the density given by Eq.~(\ref{rho_dyson_solution_zeta}) is positive definite and thus physically realistic, a form of the function $\Pi$ given by
\begin{equation}
\Pi\left(\zeta\right)=\frac{\Pi_0}{1+\rm{exp}\left(\frac{\zeta}{\zeta_0}-\zeta_1\right)} ,
\label{pi_woodsaxon}
\end{equation}
corresponds to a distribution approximately satisfying
\begin{eqnarray}
\Pi\left(\zeta\right) \approx
\begin{cases}
\Pi_0 & \zeta \leq \zeta^{*} \\
0 & \zeta \geq \zeta^{*}
\end{cases}
,
\label{pi_woodsaxon_approx}
\end{eqnarray}
where the parameter $\zeta^{*}$ is defined in terms of the constants $\zeta_0 > 0$ and $\zeta_1 > 0$ by
\begin{equation}
\zeta^{*} \equiv \zeta_0 \zeta_1 ,
\label{zeta_star_def}
\end{equation}
that is, $\zeta^{*}$ represents the point where $\Pi = \frac{\Pi_0}{2}$. This ``boundary'' between the ``non-zero'' and ``zero'' portions Eq.~(\ref{pi_woodsaxon}) is in fact continuous, but becomes increasingly sharp or less ``diffuse'' as $\zeta_1 \to \infty$. As such, axisymmetric Nemchinov-Dyson solutions featuring Eq.~(\ref{pi_woodsaxon}) are referred to as ``diffuse surface'' solutions.

With Eq.~(\ref{pi_woodsaxon}), the related functions $\Gamma$, $\Upsilon$, and $\Sigma$ defined by Eqs.~(\ref{big_gamma_def}), (\ref{upsilon_def}), and (\ref{sigma_def}), respectively, realize as
\begin{eqnarray}
\Gamma\left(\zeta\right) &=& \zeta_0\Pi_0\bigg[\zeta \ln\left(\frac{1}{1+\mathrm{exp}\left(\zeta_1 -\frac{\zeta}{\zeta_0}\right)}\right)\nonumber\\
&&+\zeta_0 \mathrm{Li}_2\left(-\mathrm{exp}\left(\zeta_1-\frac{\zeta}{\zeta_0}\right)\right)\bigg] + \Gamma_0
\label{gamma_woodsaxon} , \\
\Upsilon\left(\zeta\right) &=&\zeta_0\left[1+\mathrm{exp}\left(\frac{\zeta}{\zeta_0}-\zeta_1\right)\right]\nonumber\\
&&\times\bigg[\zeta\ln\left(\frac{1}{1+\mathrm{exp}\left(\zeta_1 -\frac{\zeta}{\zeta_0}\right)}\right)\nonumber\\
&&+\zeta_0 \mathrm{Li}_2\left(-\mathrm{exp}\left(\zeta_1-\frac{\zeta}{\zeta_0}\right)\right)\bigg]\nonumber\\
&&+ \frac{1+\mathrm{exp}\left(\frac{\zeta}{\zeta_0}-\zeta_1\right)}{\Pi_0}\Gamma_0
\label{upsilon_woodsaxon} , \\
\Sigma\left(\zeta\right) &=& \zeta_0\Pi_0^{1-\gamma}\left[1+\mathrm{exp}\left(\frac{\zeta}{\zeta_0}-\zeta_1\right)\right]^\gamma\nonumber\\
&&\times\bigg[\zeta \ln\left(\frac{1}{1+\mathrm{exp}\left(\zeta_1 -\frac{\zeta}{\zeta_0}\right)}\right)\nonumber\\
&&+\zeta_0 \mathrm{Li}_2\left(-\mathrm{exp}\left(\zeta_1-\frac{\zeta}{\zeta_0}\right)\right)\bigg]\nonumber\\
&&+\frac{\left[1+\mathrm{exp}\left(\frac{\zeta}{\zeta_0}-\zeta_1\right)\right]^\gamma}{\Pi^{\gamma}}\Gamma_0
\label{sigma_woodsaxon} ,
\end{eqnarray}
where $\Gamma_0$ is an arbitrary integration constant, the sign of which must be selected so that the pressure, SIE, and entropy given by Eqs.~(\ref{p_dyson_solution_zeta}), (\ref{sie_dyson_solution_zeta}), and (\ref{entropy_dyson_solution_zeta}) are positive definite and thus physically realistic, and $\mathrm{Li}_2$ is the second order Jonqui\`{e}re's (polylogarithm) function\footnote{An alternate form can be written as the complete Fermi-Dirac integral. Indeed, we find that $\mathrm{Li}_2\left(-e^x\right) = -\frac{1}{\Gamma(2)}\int^\infty_0\frac{t}{e^{t-x}+1}dt$ where the gamma function is $\Gamma(2)=1$ and $x=4\zeta-6$.}. The pressure, SIE, and entropy distributions associated with a diffuse surface density distribution are thus revealed to be highly non-trivial in $\zeta$.

Equations~(\ref{pi_woodsaxon})-(\ref{sigma_woodsaxon}) are depicted in Fig.~\ref{spatial_solution_set_plots} for the example parameterization including
\begin{equation}
\Gamma_0 = -\Pi_0 \zeta_0^2 \rm{Li}_2\left[-\rm{exp}\left(\zeta_1\right)\right] ,
\label{gamma_0_woodsaxon}
\end{equation}
so that $\Gamma\left(\zeta = 0\right) = 0$, and $\gamma = 5/3$, $\Pi_0 = 1$, $\zeta_0 = 1/4$, and $\zeta_1 = 6$.
%
%%%%%%%%%%%%%%%%%%%%%%%%%%%%%%%%%%%%%%%%%%%%%%%%%%
%%%%%%%%%%%%%%%%%%%%%%%%%%%%%%%%%%%%%%%%%%%%%%%%%%
\section{Example Axisymmetric Nemchinov-Dyson Solutions}
\label{subsec:examples}
%%%%%%%%%%%%%%%%%%%%%%%%%%%%%%%%%%%%%%%%%%%%%%%%%%
%%%%%%%%%%%%%%%%%%%%%%%%%%%%%%%%%%%%%%%%%%%%%%%%%%
%
The various elements appearing in Secs.~\ref{sec:scale_radius_sols} and \ref{sec:pi_sols} (and their many alternate parameterizations) may be combined into Eqs.~(\ref{ur_assump}), (\ref{uz_assump}), (\ref{rho_dyson_solution_zeta}), (\ref{p_dyson_solution_zeta}), (\ref{sie_dyson_solution_zeta}), and (\ref{entropy_dyson_solution_zeta}) to yield a limitless number of axisymmetric Nemchinov-Dyson solutions of Eqs.~(\ref{cons_mass_cyl})-(\ref{cons_mom_uz_cyl}) and (\ref{cons_energy_P_cyl_ideal}). A common feature among these possible solutions is their manifestation as a collection of axisymmetric conic sections of revolution whose eccentricities vary with time. 

More precisely, and inclusive of figures of infinite spatial extent, in light of Eq.~(\ref{zeta_def}) axisymmetric Nemchinov-Dyson solutions include density, pressure, SIE, and entropy solution fields featuring ellipsoidal, circular, or hyperbolic constant-value contours; the eccentricity of each of these ``level surfaces'' varies with time. In particular, if each of Eqs.~(\ref{rho_dyson_solution_zeta}), (\ref{p_dyson_solution_zeta}), (\ref{sie_dyson_solution_zeta}), and (\ref{entropy_dyson_solution_zeta}) is generically written as
\begin{equation}
\varphi\left(r,z,t\right) = \tau\left(t\right) \Psi\left(\zeta^2\right) ,
\label{separable}
\end{equation}
for a non-constant but otherwise invertible function $\Psi$ (i.e., with inverse function $\Psi^{-1}$), the time-dependent level surface associated with the constant state variable $\varphi = \varphi_0$ is defined by
\begin{equation}
\zeta^2 = \Psi^{-1}\left[\frac{\varphi_0}{\tau\left(t\right)}\right] ,
\label{separable_invert}
\end{equation}
or, with Eq.~\ref{zeta_def},
\begin{equation}
\frac{\kappa_3 r^2}{a_{r}^2} + \frac{z^2}{a_{z}^2} = 1 ,
\label{separable_invert_rz}
\end{equation}
where $a_{r}^2\left(t\right)$ and $a_{z}^2\left(t\right)$ are defined by
\begin{eqnarray}
a_{r}^2 && \equiv R_r^2 \Psi^{-1}\left[\frac{\varphi_0}{\tau\left(t\right)}\right], \label{semi-major} \\
a_{z}^2 && \equiv R_z^2 \Psi^{-1}\left[\frac{\varphi_0}{\tau\left(t\right)}\right]. \label{semi-minor}
\end{eqnarray}

Depending on the sign of the constant $\kappa_3$, Eq.~(\ref{separable_invert_rz}) defines either an ellipse ($\kappa_3 > 0$) or a hyperbola ($\kappa_3 < 0$) in $\left(r,z\right)$-space, the eccentricity of which varies according to the behavior in time of both $a_r$ and $a_z$. In particular, for $\kappa_3 > 0$, the eccentricity of any ellipsoidal level surfaces is given by
\begin{equation}\label{eccentricity}
e = \sqrt{1 - \rm{min} \left( \frac{a_{r}^2}{a_{z}^2},\frac{a_{z}^2}{a_{r}^2} \right) },
\end{equation}
or, with Eqs.~(\ref{semi-major}) and (\ref{semi-minor}), 
\begin{equation}\label{eccentricity_R}
e = \sqrt{1 - \rm{min} \left( \frac{R_{r}^2}{R_{z}^2},\frac{R_{z}^2}{R_{r}^2} \right) },
\end{equation}
which is independent of both the level surface value $\varphi_0$ and the functional form $\Psi$ of the solution field; as such, the eccentricity of all ellipsoidal level surfaces in all state variables for a given Nemchinov-Dyson solution depends only on the scale radii $R_r\left(t\right)$ and $R_z\left(t\right)$. 

Similarly, for $\kappa_3 < 0$, the eccentricity of any hyperbolic level surfaces is given by
\begin{equation}\label{eccentricity_hyper}
e = \sqrt{1 + \frac{a_{r}^2}{a_{z}^2} },
\end{equation}
or, with Eqs.~(\ref{semi-major}) and (\ref{semi-minor}), 
\begin{equation}\label{eccentricity_hyper_R}
e = \sqrt{1 + \frac{R_{r}^2}{R_{z}^2} },
\end{equation}
which is again independent of both the level surface value $\phi_0$ and the functional form $\Psi$ of the solution field; as such, the eccentricity of all hyperbolic level surfaces in all state variables for a given Nemchinov-Dyson solution again depends only on the scale radii $R_r\left(t\right)$ and $R_z\left(t\right)$. 

Under this construction, three example axisymmetric Nemchinov-Dyson solutions are given in Secs.~\ref{subsubsec:isothermal_DR_DR}-\ref{subsubsec:woodsaxon_DR_DS}.  
%
%%%%%%%%%%%%%%%%%%%%%%%%%%%%%%%%%%%%%%%%%%%%%%%%%%
%%%%%%%%%%%%%%%%%%%%%%%%%%%%%%%%%%%%%%%%%%%%%%%%%%
\subsection{Uniform SIE DR-DR Solution}
\label{subsubsec:isothermal_DR_DR}
%%%%%%%%%%%%%%%%%%%%%%%%%%%%%%%%%%%%%%%%%%%%%%%%%%
%%%%%%%%%%%%%%%%%%%%%%%%%%%%%%%%%%%%%%%%%%%%%%%%%%
%
%
%
\begin{figure*}[]
        \includegraphics[scale=0.6]{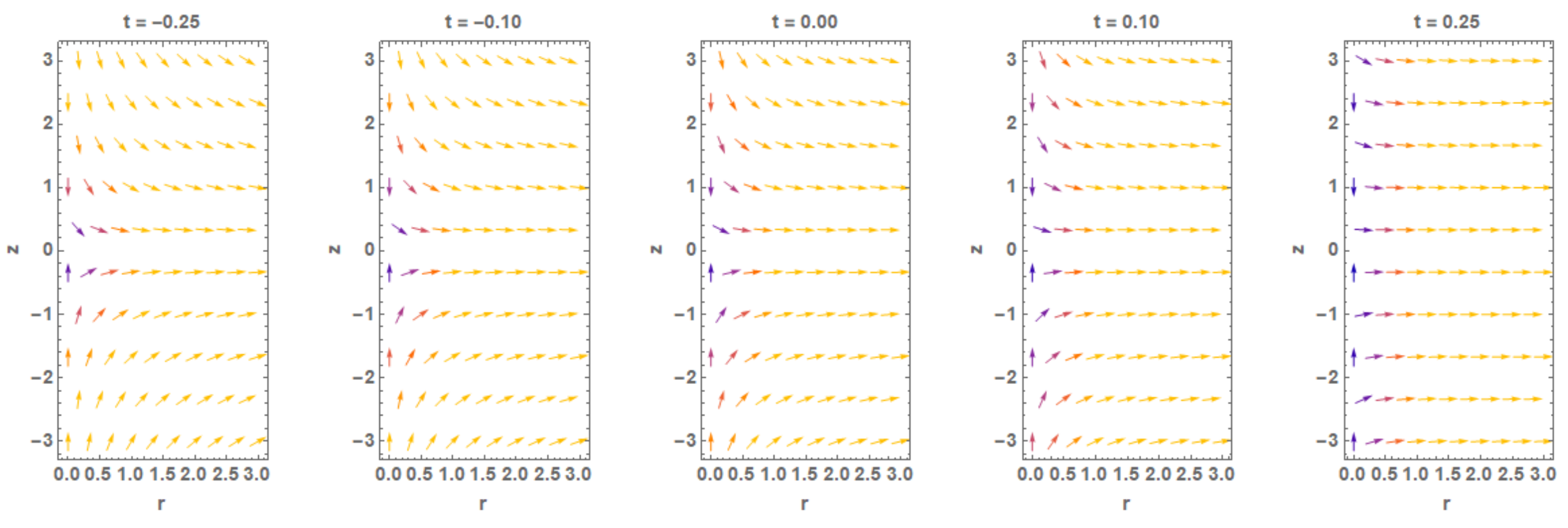}
        \caption{Equations~(\ref{ur_assump}) and (\ref{uz_assump}) evaluated at various times, featuring the $\gamma = 5/3$ DR-DR numerical solution depicted in Fig.~\ref{DR_DR_plot}.}
\label{isothermal_DR_DR_velocity_plot}
\end{figure*}
\begin{figure*}[]
        \begin{subfigure}[t]{1.1\textwidth}
      %  \centering
        \includegraphics[scale=0.5]{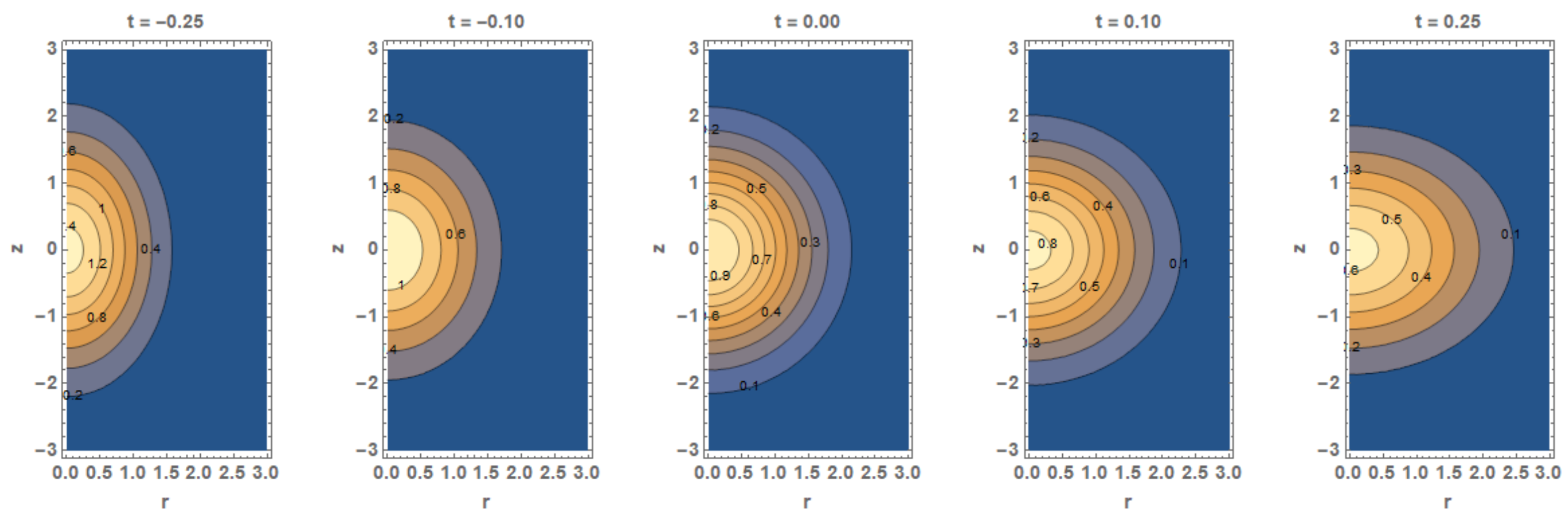}
		\caption{}\label{isothermal_dr_dr_density_plot}
    \end{subfigure} 
                \begin{subfigure}[t]{1.1\textwidth}
       % \centering
        \includegraphics[scale=0.5]{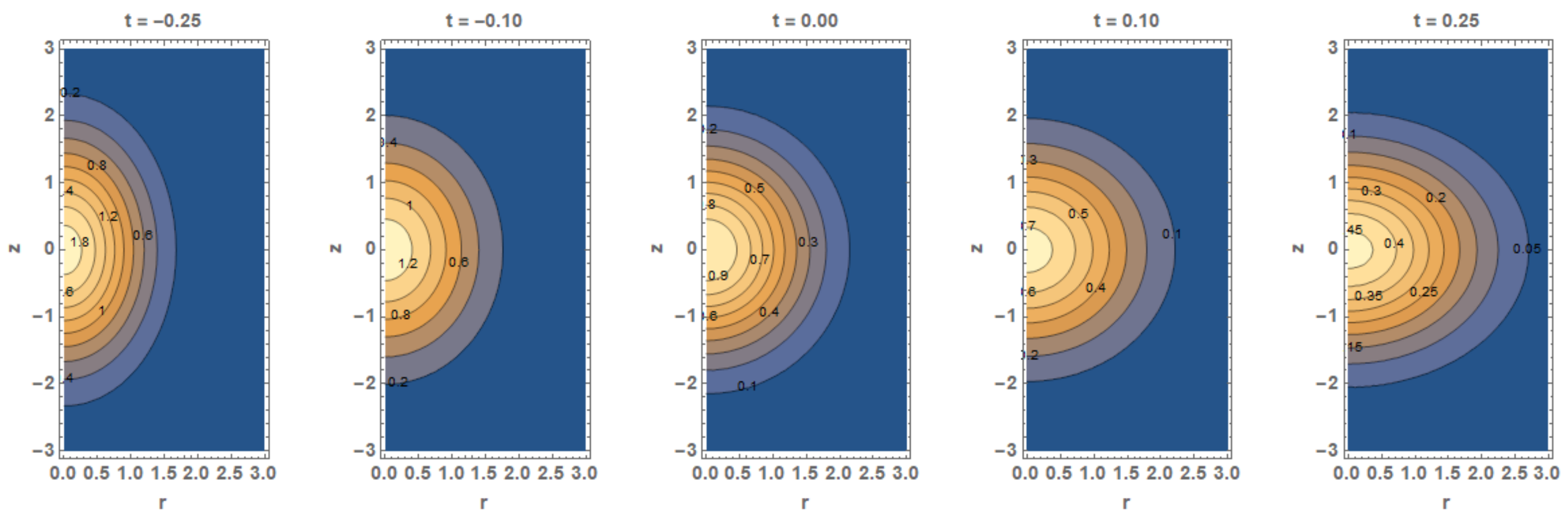}
        \caption{}\label{isothermal_dr_dr_pressure_plot}
    \end{subfigure} 
                   \begin{subfigure}[t]{1.1\textwidth}
       % \centering
        \includegraphics[scale=0.5]{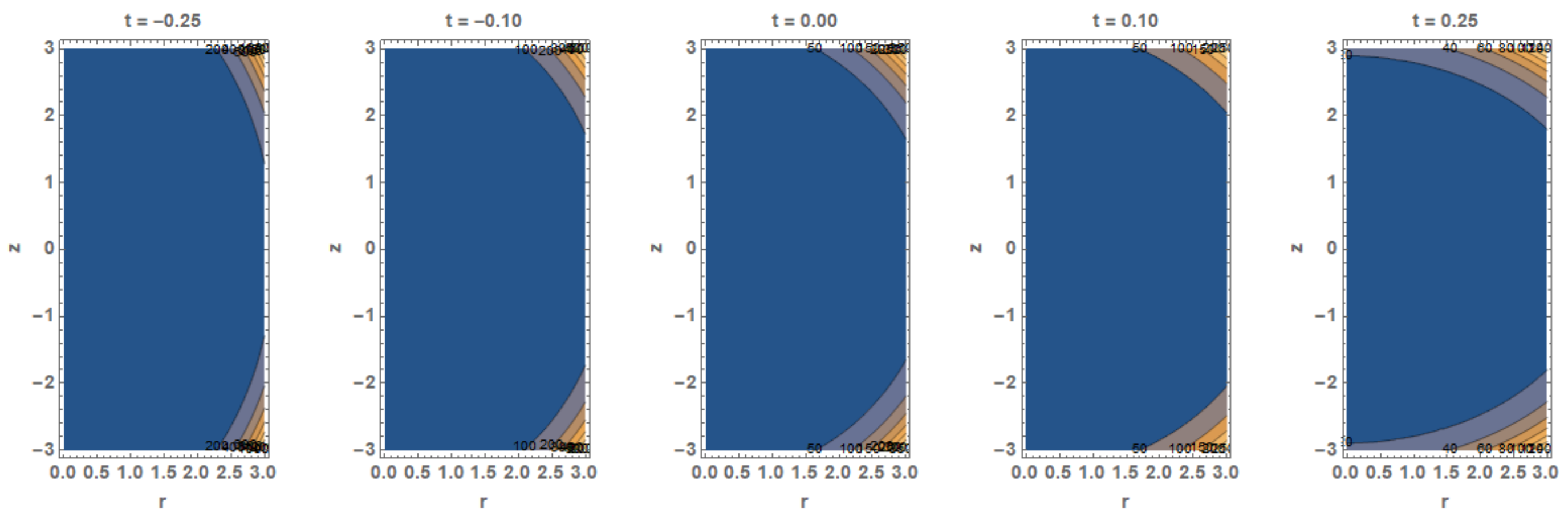}
        \caption{}\label{isothermal_dr_dr_entropy_plot}
    \end{subfigure} 
        \caption{Equations~(\ref{density_uniform_sie_dr_dr}), (\ref{pressure_uniform_sie_dr_dr}), and (\ref{entropy_uniform_sie_dr_dr}) ((a), (b), and (c), respectively) evaluated at various times, under the example parameterization $\Pi_0 = 1$, $\Upsilon_0 = -1$, and $\gamma = 5/3$, and featuring the $\gamma = 5/3$, $\kappa_3 = 1$ DR-DR numerical solution depicted in Fig.~\ref{DR_DR_plot}.}
\label{isothermal_DR_DR_plot}
\end{figure*}
\begin{figure}[t]
        \includegraphics[scale=0.4]{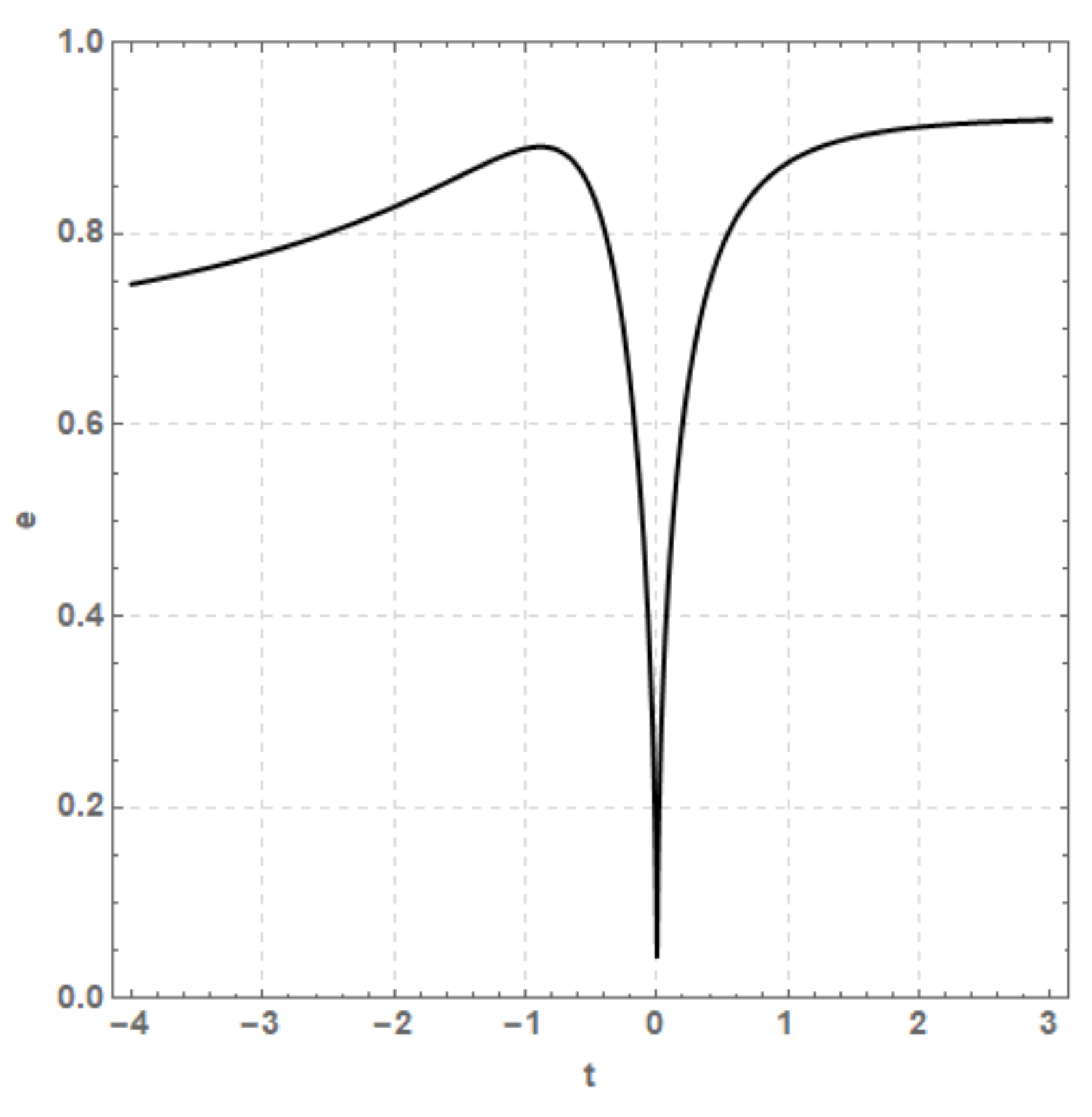}
        \caption{Equation~(\ref{eccentricity_R}) evaluated at various times, featuring the $\gamma = 5/3$, $\kappa_3 =1$ DR-DR numerical solution depicted in Fig.~\ref{DR_DR_plot}.}
\label{isothermal_DR_DR_essentricity_plot}
\end{figure}
\begin{figure}[t]
        \includegraphics[scale=0.4]{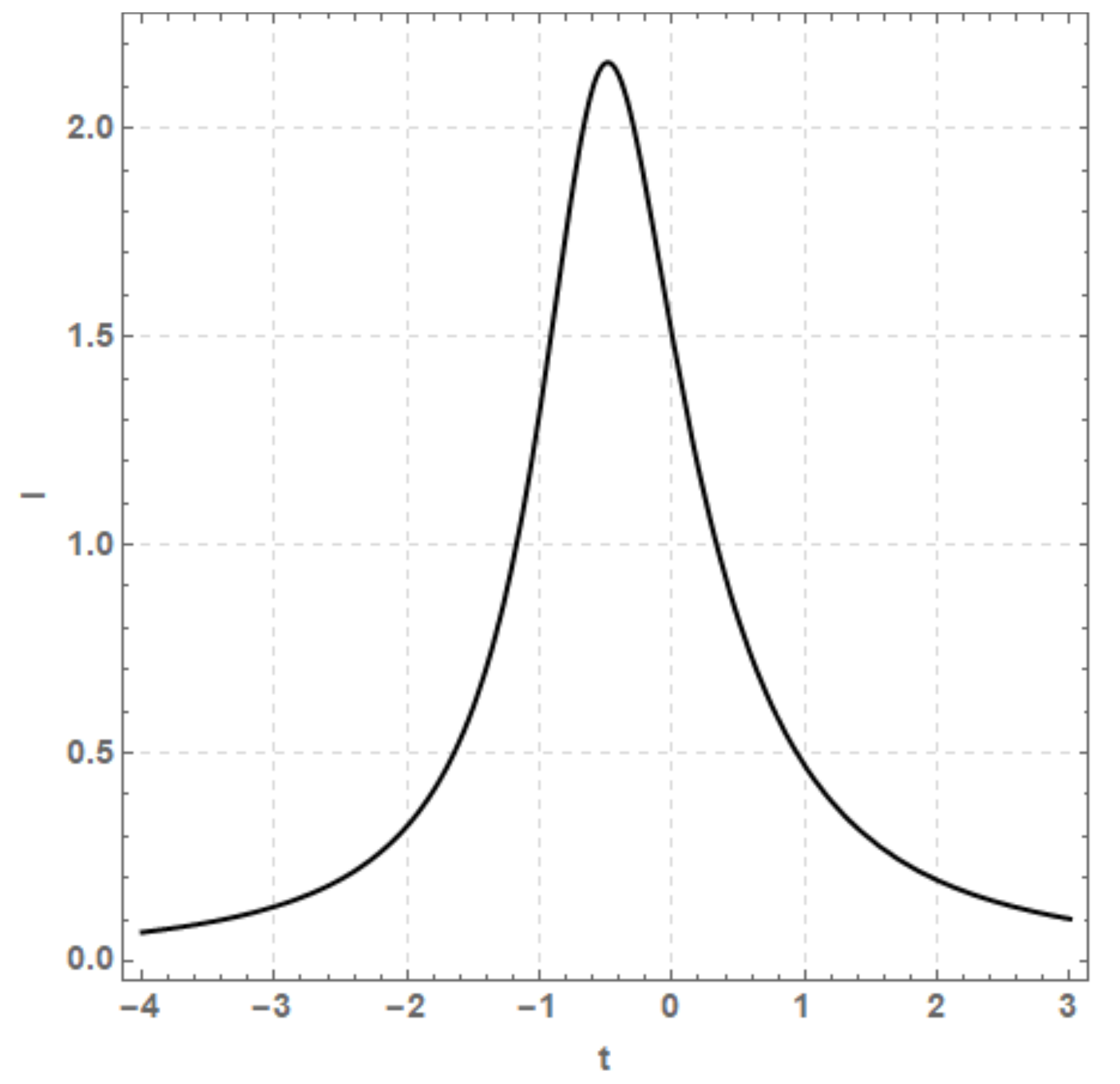}
        \caption{Equation~(\ref{sie_uniform_sie_dr_dr}) evaluated at various times, under the example parameterization $\Pi_0 = 1$, $\Upsilon_0 = -1$, and $\gamma = 5/3$, and featuring the $\gamma = 5/3$, $\kappa_3 = 1$ DR-DR numerical solution depicted in Fig.~\ref{DR_DR_plot}.}
\label{isothermal_DR_DR_sie_plot}
\end{figure}

As a first example, a ``uniform SIE DR-DR'' axisymmetric Nemchinov-Dyson solution follows from combining the results appearing in Secs.~\ref{sec:DR-DR} and \ref{subsubsec:uniform_sie}. This solution is given by Eqs.~(\ref{ur_assump}) and (\ref{uz_assump}) for $u_r\left(r,z,t\right)$ and $u_z\left(r,z,t\right)$, respectively, and, with Eqs.~(\ref{xi_sim}), (\ref{eta_sim}), (\ref{zeta_def}), (\ref{rho_dyson_solution_zeta}), (\ref{p_dyson_solution_zeta}), (\ref{sie_dyson_solution_zeta}), (\ref{entropy_dyson_solution_zeta}), (\ref{upsilon_uniform_sie}), (\ref{pi_uniform_sie}), (\ref{gamma_uniform_sie}), and (\ref{sigma_uniform_sie}),
\begin{eqnarray}
\rho\left(r,z,t\right) &=& 
\frac{\Pi_0}{R_r^2 R_z} 
\left[ \left(\gamma-1\right)\Pi_0 + \frac{\left(\gamma-1\right) \zeta^2}{2\gamma\Sigma_0}\right]^{\frac{1}{\gamma-1}}\nonumber\\
&&\times\;\rm{exp}\left(\frac{\kappa_3 r^2}{2\Upsilon_0 R_r^2} + \frac{z^2}{2\Upsilon_0 R_z^2} \right)
\label{density_uniform_sie_dr_dr}, \\ 
P\left(r,z,t\right) &=& 
-\Pi_0 \Upsilon_0 
\frac{R_r \ddot{R}_r}{\kappa_3 R_r^2 R_z} 
\rm{exp}\left(\frac{\kappa_3 r^2}{2\Upsilon_0 R_r^2} + \frac{z^2}{2\Upsilon_0 R_z^2} \right)\nonumber\\
\label{pressure_uniform_sie_dr_dr}, \\ 
I\left(r,z,t\right) &=& 
-\frac{R_r \ddot{R}_r}{\kappa_3 \left(\gamma-1\right)} 
\Upsilon_0 
\label{sie_uniform_sie_dr_dr}, \\ 
S\left(r,z,t\right) &=& 
-\Pi_0^{1-\gamma} \Upsilon_0 
\frac{\ddot{R}_r R_r^{2\gamma-1} R_z^{\gamma-1}}{\kappa_3} 
\rm{exp}\bigg[\frac{\left(1-\gamma\right)\kappa_3 r^2}{2\Upsilon_0 R_r^2}\nonumber\\
&&+ \frac{\left(1-\gamma\right)z^2}{2\Upsilon_0 R_z^2} \bigg]
\label{entropy_uniform_sie_dr_dr},
\end{eqnarray}
where $\Pi_0 > 0$ and $\Upsilon_0 < 0$ are otherwise arbitrary constants. For the DR-DR type solution provided in Sec.~\ref{sec:DR-DR}, $\kappa_3 = 1$ as appearing in Eqs.~(\ref{density_uniform_sie_dr_dr})-(\ref{entropy_uniform_sie_dr_dr}), and the numerical representations of $R_r$ and $R_z$ are depicted in Fig.~\ref{DR_DR_plot}. 

Equations~(\ref{ur_assump}), (\ref{uz_assump}), (\ref{density_uniform_sie_dr_dr}), (\ref{pressure_uniform_sie_dr_dr}), and (\ref{entropy_uniform_sie_dr_dr}) are depicted in Figs.~\ref{isothermal_DR_DR_velocity_plot} and \ref{isothermal_DR_DR_plot}, for the example parameterization $\Pi_0 = 1$, $\Upsilon_0 = -1$, and $\gamma = 5/3$, and featuring the $\gamma = 5/3$ DR-DR numerical solution depicted in Fig.~\ref{DR_DR_plot}. The associated time-dependent eccentricity of all ellipsoidal level surfaces in the density, pressure, and entropy state variables is given by Eq.~(\ref{eccentricity_R}), and is provided in Fig.~\ref{spatial_solution_set_plots}. Finally, the associated Eq.~(\ref{sie_uniform_sie_dr_dr}) is independent of both $r$ and $z$ by construction (i.e., it assumes the same time-dependent value at every spatial point within the solution field); therefore only its time-dependence is depicted in Fig.~\ref{isothermal_DR_DR_sie_plot}. 

Figure~\ref{isothermal_DR_DR_velocity_plot} depicts the total velocity vector field associated with the DR-DR type solution [as Eqs.~(\ref{ur_assump}) and (\ref{uz_assump}) hold whether the spatial portion of the associated Nemchinov-Dyson solution is of uniform SIE type or not] featured in Sec.~\ref{sec:DR-DR}, including the appropriate, conjoined linear behavior in both $r$ and $z$. The directions of the various velocity vectors appearing in Fig.~\ref{isothermal_DR_DR_velocity_plot} are directly proportional to the slopes of the $R_r$ and $R_z$ curves appearing in Fig.~\ref{DR_DR_plot}, as otherwise explicitly revealed by Eqs.~(\ref{ur_assump}) and (\ref{uz_assump}). That is, whenver $\dot{R}_r > 0$ or $\dot{R}_z > 0$ in Fig.~\ref{DS_DS_plot}, the associated velocity vector is pointed ``outward'' in the appropriate direction in Fig. ~\ref{isothermal_DR_DR_velocity_plot}, and whenever $\dot{R}_r < 0$ or $\dot{R}_z < 0$ in Fig.~\ref{DS_DS_plot}, the associated velocity vector is pointed ``inward'' in the appropriate direction in Fig.~\ref{isothermal_DR_DR_velocity_plot}. For the DR-DR example depicted in Figs.~\ref{DR_DR_plot} and \ref{isothermal_DR_DR_velocity_plot}, the global motion of the associated uniform SIE solution is therefore largely dominated by motion in $r$ for all times. 

Figure~\ref{isothermal_DR_DR_plot} indicates that all density, pressure, and entropy level surfaces are indeed ellipsoidal in shape, and (as depicted) continuously deform from prolate to oblate with increasing time, while also rarefying and depressurizing. Furthermore, Fig.~\ref{isothermal_DR_DR_essentricity_plot} shows that these level surfaces vary rapidly between high eccentricity $\left(e > 0.8\right)$ and perfect sphericity $\left( e = 0.0\right)$ in the neighborhood of $t=0$ [by design, in light of Eqs.~(\ref{Rr_zero}) and (\ref{Rz_zero})]. Otherwise, the $r$ and $z$ variation of the density, pressure, and entropy solutions proceeds according to variously sharp Gaussian distributions, as also indicated by Eqs.~(\ref{density_uniform_sie_dr_dr}), (\ref{pressure_uniform_sie_dr_dr}), and (\ref{entropy_uniform_sie_dr_dr}).

Finally, Fig.~\ref{isothermal_DR_DR_sie_plot} shows that the spatially constant SIE increases rapidly from small to peak values near $t = 0$, and afterward again decreasing rapidly. The maximum SIE occurs shortly before $t = 0$, thus corresponding to an event other than the solution field attaining perfect spherical symmetry. Rather, on the grounds of physical intuition, the SIE is maximized whenever the specific kinetic energy of the solution field is simultaneously minimized, which may be verified by inspection of Fig.~\ref{DR_DR_plot}. From Fig.~\ref{DR_DR_plot} and Eqs.~(\ref{ur_assump}) and (\ref{uz_assump}), the specific kinetic energy of the Nemchinov-Dyson solution is proportional to the square of the slopes of the $R_r$ and $R_z$ curves. Therefore, the time at which these slopes (i.e., $\dot{R}_r$ and $\dot{R}_z$) are jointly minimized is the same time at which the maximum SIE is observed to occur in Fig.~\ref{isothermal_DR_DR_sie_plot}.
%
%%%%%%%%%%%%%%%%%%%%%%%%%%%%%%%%%%%%%%%%%%%%%%%%%%
%%%%%%%%%%%%%%%%%%%%%%%%%%%%%%%%%%%%%%%%%%%%%%%%%%
\subsection{Uniform Entropy DS-DS Solution}
\label{subsubsec:isentropic_DS_DS}
%%%%%%%%%%%%%%%%%%%%%%%%%%%%%%%%%%%%%%%%%%%%%%%%%%
%%%%%%%%%%%%%%%%%%%%%%%%%%%%%%%%%%%%%%%%%%%%%%%%%%
%
%
%
%
\begin{figure*}[]
        \includegraphics[scale=0.6]{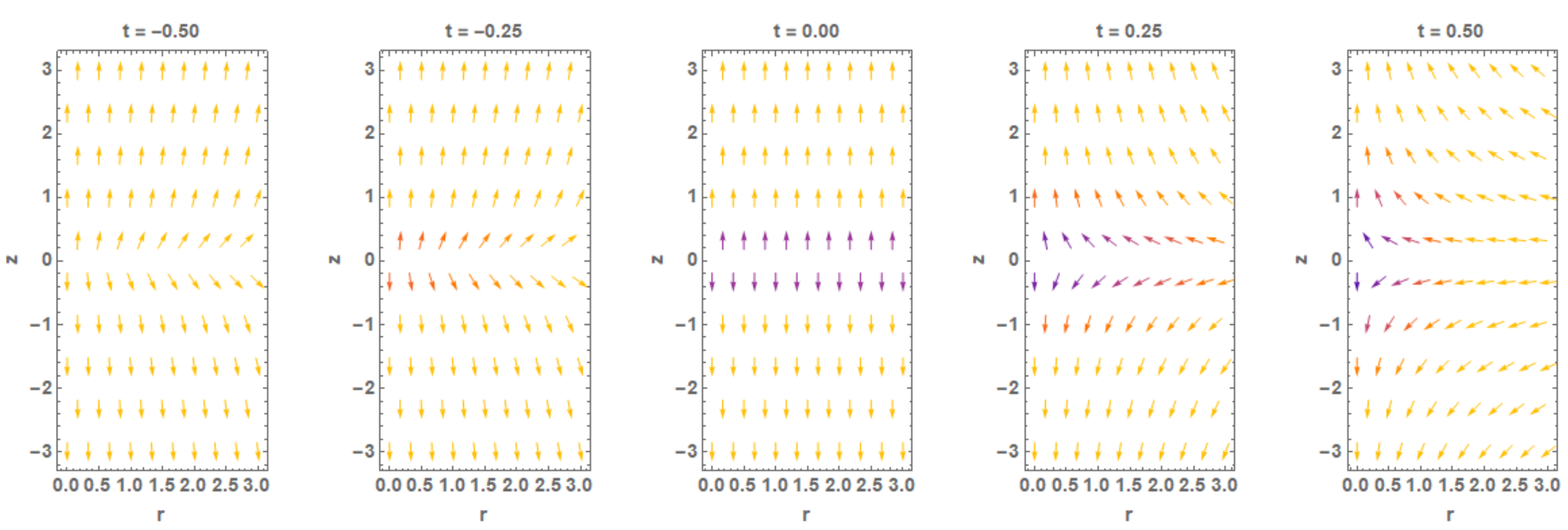}
        \caption{Equations~(\ref{ur_assump}) and (\ref{uz_assump}) evaluated at various times, featuring the $\gamma = 5/3$ DS-DS numerical solution depicted in Fig.~\ref{DS_DS_plot}.}
\label{isentropic_DS_DS_velocity_plot}
\end{figure*}
\begin{figure*}[]
        \begin{subfigure}[t]{1.1\textwidth}
      %  \centering
        \includegraphics[scale=0.5]{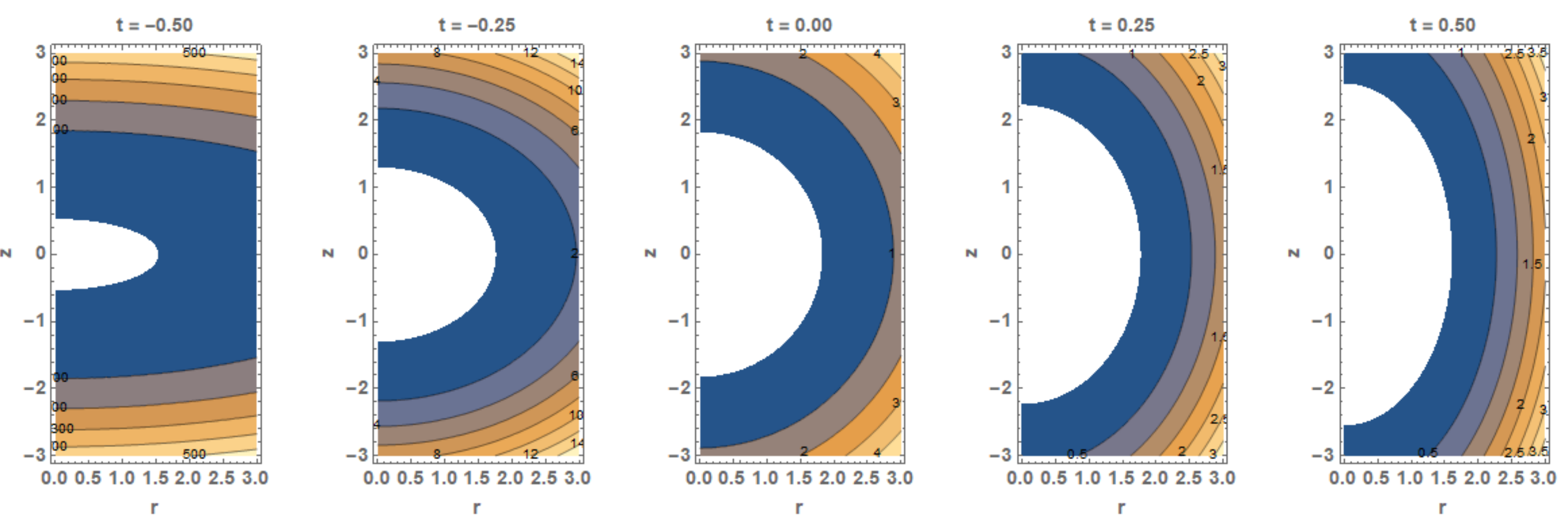}
        \caption{}\label{isentropic_ds_ds_density_plot}
    \end{subfigure} 
                \begin{subfigure}[t]{1.1\textwidth}
       % \centering
        \includegraphics[scale=0.5]{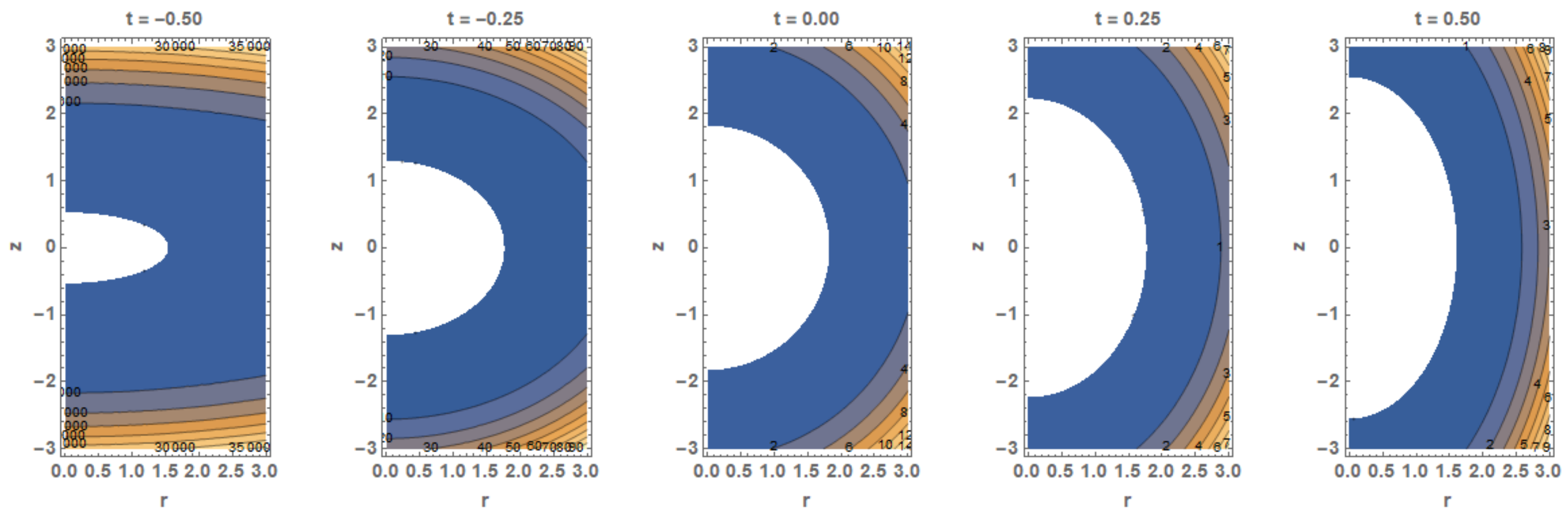}
        \caption{}\label{isentropic_ds_ds_pressure_plot}
    \end{subfigure} 
            \begin{subfigure}[t]{1.1\textwidth}
 %       \centering
        \includegraphics[scale=0.5]{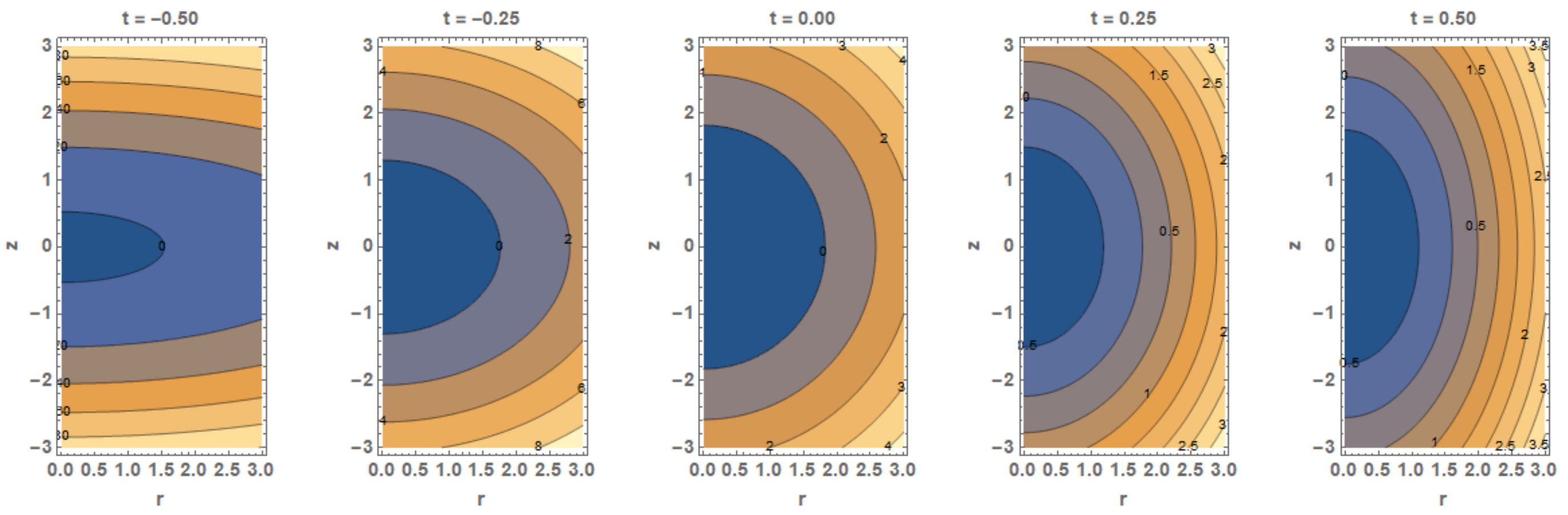}
        \caption{}\label{isentropic_ds_ds_sie_plot}
    \end{subfigure} 
        \caption{Equations~(\ref{density_uniform_entropy_ds_ds}), (\ref{pressure_uniform_entropy_ds_ds}), and (\ref{sie_uniform_entropy_ds_ds}) ((a), (b), and (c), respectively) evaluated at various times, under the example parameterization $\Pi_0 = -1$, $\Sigma_0 = -1$, and $\gamma = 5/3$, and featuring the $\gamma = 5/3$, $\kappa_3=1$ DS-DS numerical solution depicted in Fig.~\ref{DS_DS_plot}.}
\label{isentropic_DS_DS_plot}
\end{figure*}
\begin{figure}[t]
        \includegraphics[scale=0.4]{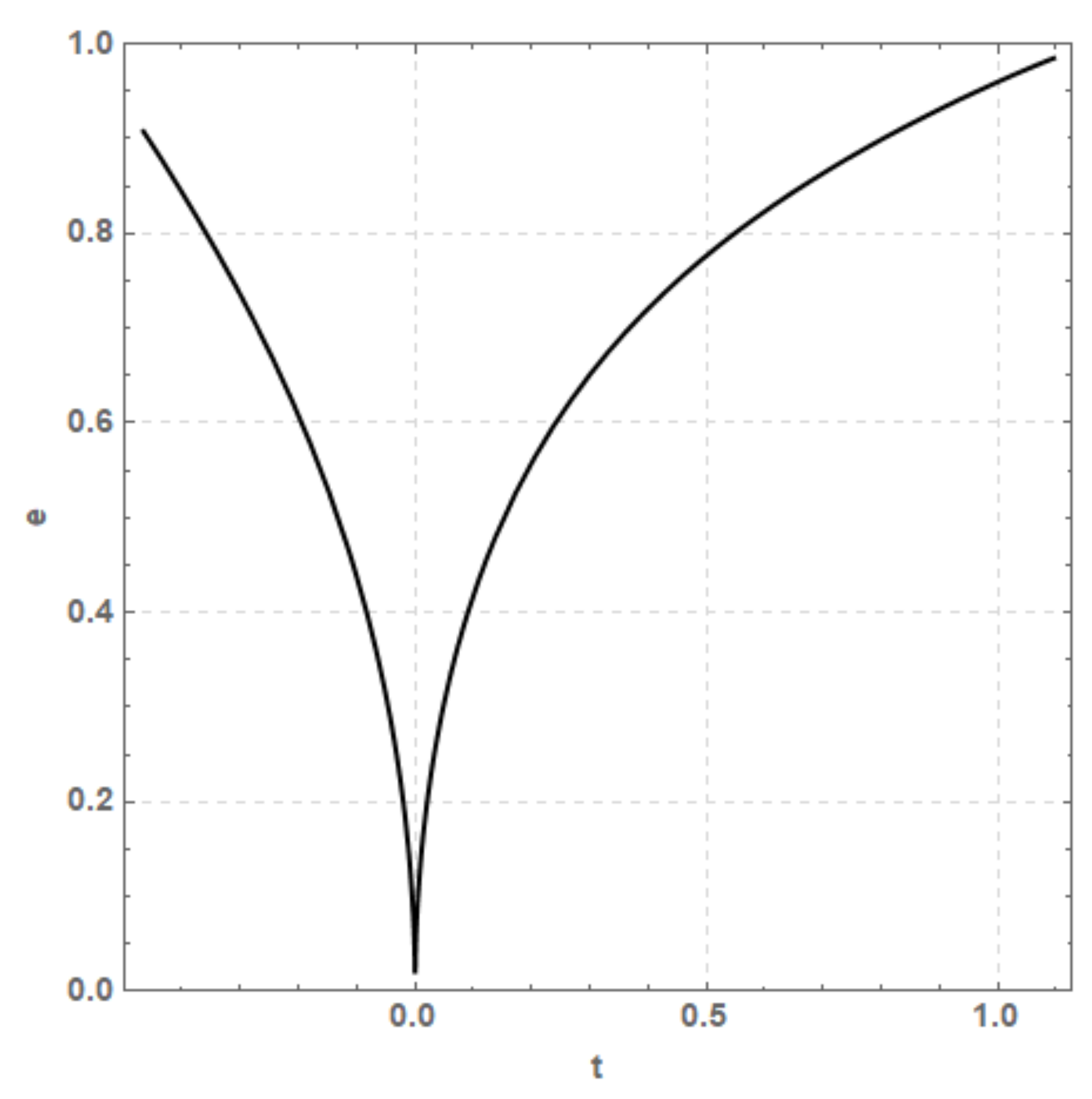}
        \caption{Equation~(\ref{eccentricity_R}) evaluated at various times, featuring the $\gamma = 5/3$, $\kappa_3 =1$ DS-DS numerical solution depicted in Fig.~\ref{DS_DS_plot}.}
\label{isentropic_DS_DS_eccentricity_plot}
\end{figure}
\begin{figure}[t]
        \includegraphics[scale=0.4]{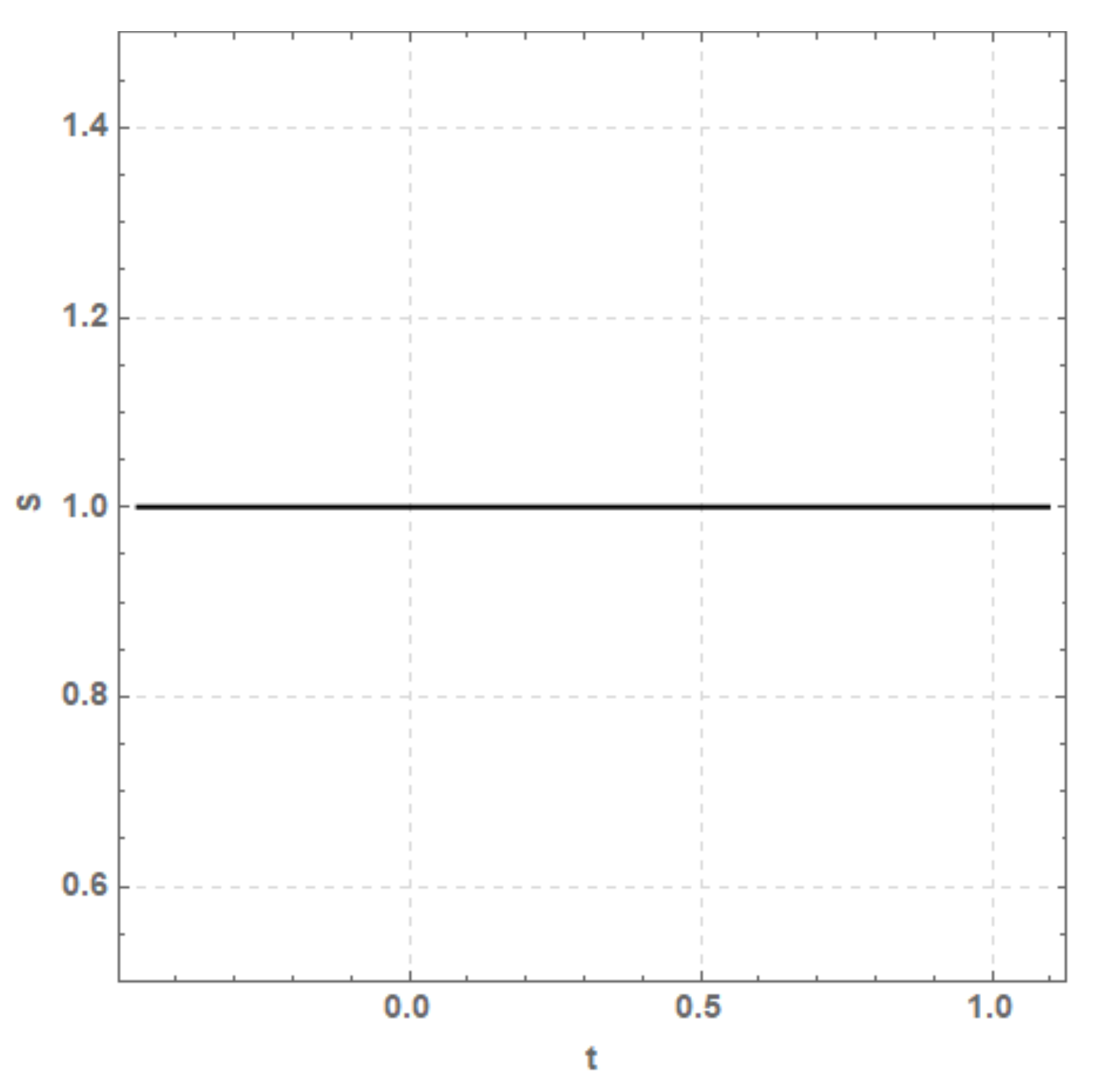}
        \caption{Equation~(\ref{entropy_uniform_entropy_ds_ds}) evaluated at various times, under the example parameterization $\Pi_0 = -1$, $\Sigma_0 = 1$, and $\gamma = 5/3$, and featuring the $\gamma = 5/3$, $\kappa_3 = 1$ DS-DS numerical solution depicted in Fig.~\ref{DS_DS_plot}.}
\label{isentropic_DS_DS_entropy_plot}
\end{figure}
As a second example, a ``uniform entropy DS-DS'' axisymmetric Nemchinov-Dyson solution follows from combining the results appearing in Secs.~\ref{sec:DS-DS} and \ref{subsubsec:uniform_entropy}. This solution is again given by Eqs.~(\ref{ur_assump}) and (\ref{uz_assump}) for $u_r\left(r,z,t\right)$ and $u_z\left(r,z,t\right)$, respectively, and, with Eqs.~(\ref{xi_sim}), (\ref{eta_sim}), (\ref{zeta_def}), (\ref{rho_dyson_solution_zeta}), (\ref{p_dyson_solution_zeta}), (\ref{sie_dyson_solution_zeta}), (\ref{entropy_dyson_solution_zeta}), (\ref{sigma_uniform_entropy}), (\ref{pi_uniform_entropy}), (\ref{gamma_uniform_entropy}), and (\ref{upsilon_uniform_entropy}),
\begin{eqnarray}
\rho\left(r,z,t\right) &=& 
\frac{1}{R_r^2 R_z}
\bigg[ \left(\gamma-1\right)\Pi_0 + \frac{\gamma-1}{2\gamma\Sigma_0}\left(\frac{\kappa_3 r^2}{R_r^2}\right.\nonumber\\
&&\left. + \frac{z^2}{R_z^2}\right) \bigg]^{\frac{1}{\gamma -1}} 
\label{density_uniform_entropy_ds_ds}, \\ 
P\left(r,z,t\right) &=& 
\frac{\Pi_0 \Sigma_0 R_r \ddot{R}_r}{\kappa_3 R_r^2 R_z}
\left[ \left(\gamma-1\right)\Pi_0 + \frac{\gamma-1}{2\gamma\Sigma_0}\left(\frac{\kappa_3 r^2}{R_r^2} \right.\right.\nonumber\\
&&\left.\left.+ \frac{z^2}{R_z^2} \right)\right]^{\frac{\gamma}{\gamma-1}} 
\label{pressure_uniform_entropy_ds_ds},  \\
I\left(r,z,t\right) &=& 
-\frac{R_r \ddot{R}_r}{2\gamma\kappa_3} 
\left[2\gamma\Pi_0\Sigma_0\left( \frac{\kappa_3 r^2}{R_r^2} + \frac{z^2}{R_z^2}\right) \right]
\label{sie_uniform_entropy_ds_ds}, \\ 
S\left(r,z,t\right) &=& 
-\frac{\Sigma_0 \ddot{R}_r R_r^{2\gamma-1} R_z^{\gamma-1}}{\kappa_3}  
\label{entropy_uniform_entropy_ds_ds},
\end{eqnarray}
where $\Pi_0 < 0$ and $\Sigma_0 > 0$ are arbitrary constants. For the DS-DS type solution provided in Sec.~\ref{sec:DS-DS}, $\kappa_3 = 1$ as appearing in Eqs.~(\ref{density_uniform_entropy_ds_ds})-(\ref{entropy_uniform_entropy_ds_ds}), and the numerical representations of $R_r$ and $R_z$ are depicted in Fig.~\ref{DS_DS_plot}. 

Equations~(\ref{ur_assump}), (\ref{uz_assump}), (\ref{density_uniform_entropy_ds_ds}), (\ref{pressure_uniform_entropy_ds_ds}), and (\ref{sie_uniform_entropy_ds_ds}) are depicted in Figs.~\ref{isentropic_DS_DS_velocity_plot} and \ref{isentropic_DS_DS_plot}, for the example parameterization $\Pi_0 = -1$, $\Sigma_0 = 1$, and $\gamma = 5/3$, and featuring the $\gamma = 5/3$ DS-DS numerical solution depicted in Fig.~\ref{DS_DS_plot}. The associated time-dependent eccentricity of all ellipsoidal level surfaces in the density, pressure, and SIE state variables is given by Eq.~(\ref{eccentricity_R}), and is provided in Fig.~\ref{spatial_solution_set_plots}. Finally, the associated Eq.~(\ref{entropy_uniform_entropy_ds_ds}) is independent of both $r$ and $z$ by construction (i.e., it assumes the same time-dependent value at every spatial point within the solution field); therefore only its time-dependence is depicted in Fig.~\ref{isentropic_DS_DS_entropy_plot}. 

Figure~\ref{isentropic_DS_DS_velocity_plot} depicts the total velocity vector field associated with the DS-DS type solution [as Eqs.~(\ref{ur_assump}) and (\ref{uz_assump}) hold whether the spatial portion of the associated Nemchinov-Dyson solution is of uniform entropy type or not] featured in Sec.~\ref{sec:DS-DS}, including the appropriate, conjoined linear behavior in both $r$ and $z$. The directions of the various velocity vectors appearing in Fig.~\ref{isentropic_DS_DS_velocity_plot} are again directly proportional to the slopes of the $R_r$ and $R_z$ curves appearing in Fig.~\ref{DS_DS_plot}, as otherwise explicitly revealed by Eqs.~(\ref{ur_assump}) and (\ref{uz_assump}). For the DS-DS example depicted in Figs.~\ref{DS_DS_plot} and \ref{isentropic_DS_DS_velocity_plot}, the global motion of the associated uniform entropy solution is therefore largely dominated by motion in $z$ for $t \leq 0$, and becomes increasingly dominated by motion in $r$ at later times. 

Figure~\ref{isentropic_DS_DS_plot} depicts the presence of an ellipsoidal ``cavity'' in the solution field surrounding $r = z = 0$; the density and pressure given by Eqs. (\ref{density_uniform_entropy_ds_ds}) and (\ref{pressure_uniform_entropy_ds_ds}) are not real-valued in that region, though the SIE and entropy given by Eqs. (\ref{sie_uniform_entropy_ds_ds}) and (\ref{entropy_uniform_entropy_ds_ds}) are defined there. Figure~\ref{isentropic_DS_DS_plot} further indicates that all density, pressure, and SIE level surfaces exterior to the cavity surface are indeed ellipsoidal in shape, and (as depicted) continuously deform from oblate to prolate with increasing time, while also rarefying, depressurizing, and cooling. Furthermore, Fig.~\ref{isentropic_DS_DS_eccentricity_plot} shows that these level surfaces vary rapidly between extremely high eccentricity $\left(e > 0.9\right)$ to perfect sphericity $\left( e = 0.0\right)$ in the neighborhood of $t=0$ [by design, in light of Eqs.~(\ref{Rr_zero}) and (\ref{Rz_zero})]. Otherwise, the $r$ and $z$ variation of the density, pressure, and SIE solutions proceeds according to variously sharp power-law distributions, as also indicated by Eqs.~(\ref{density_uniform_entropy_ds_ds}), (\ref{pressure_uniform_entropy_ds_ds}), and (\ref{sie_uniform_entropy_ds_ds}). 

Finally, Fig.~\ref{isentropic_DS_DS_entropy_plot} shows that the spatially constant entropy is also constant in time, indicating that the solution is purely homentropic.
%
%%%%%%%%%%%%%%%%%%%%%%%%%%%%%%%%%%%%%%%%%%%%%%%%%%
%%%%%%%%%%%%%%%%%%%%%%%%%%%%%%%%%%%%%%%%%%%%%%%%%%
\subsection{Diffuse Surface DR-DS Solution}
\label{subsubsec:woodsaxon_DR_DS}
%%%%%%%%%%%%%%%%%%%%%%%%%%%%%%%%%%%%%%%%%%%%%%%%%%
%%%%%%%%%%%%%%%%%%%%%%%%%%%%%%%%%%%%%%%%%%%%%%%%%%
%
%
%
%
\begin{figure*}[]
        \includegraphics[scale=0.6]{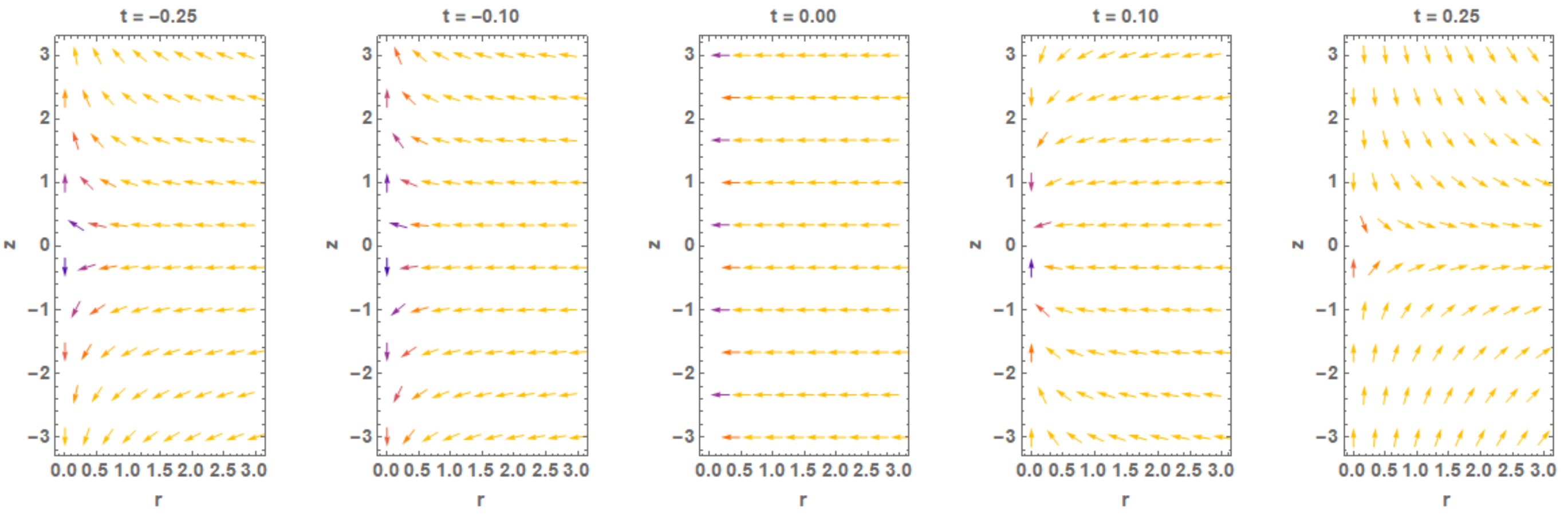}
        \caption{Equations~(\ref{ur_assump}) and (\ref{uz_assump}) evaluated at various times, featuring the $\gamma = 5/3$ DR-DS numerical solution depicted in Fig.~\ref{DR_DS_plot}.}
\label{woodsaxon_DR_DS_velocity_plot}
\end{figure*}
\begin{figure*}[]
        \begin{subfigure}[t]{1.1\textwidth}
      %  \centering
        \includegraphics[scale=0.5]{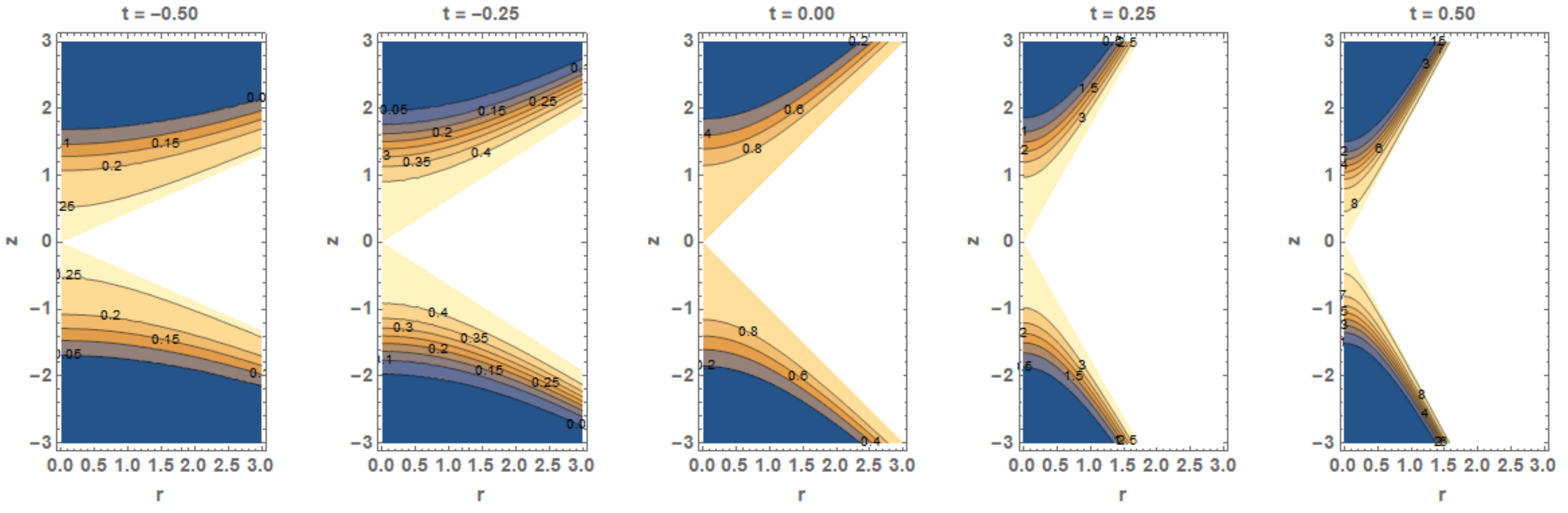}
        \caption{}\label{woodsaxon_dr_ds_density_plot}
    \end{subfigure} 
                \begin{subfigure}[t]{1.1\textwidth}
       % \centering
        \includegraphics[scale=0.5]{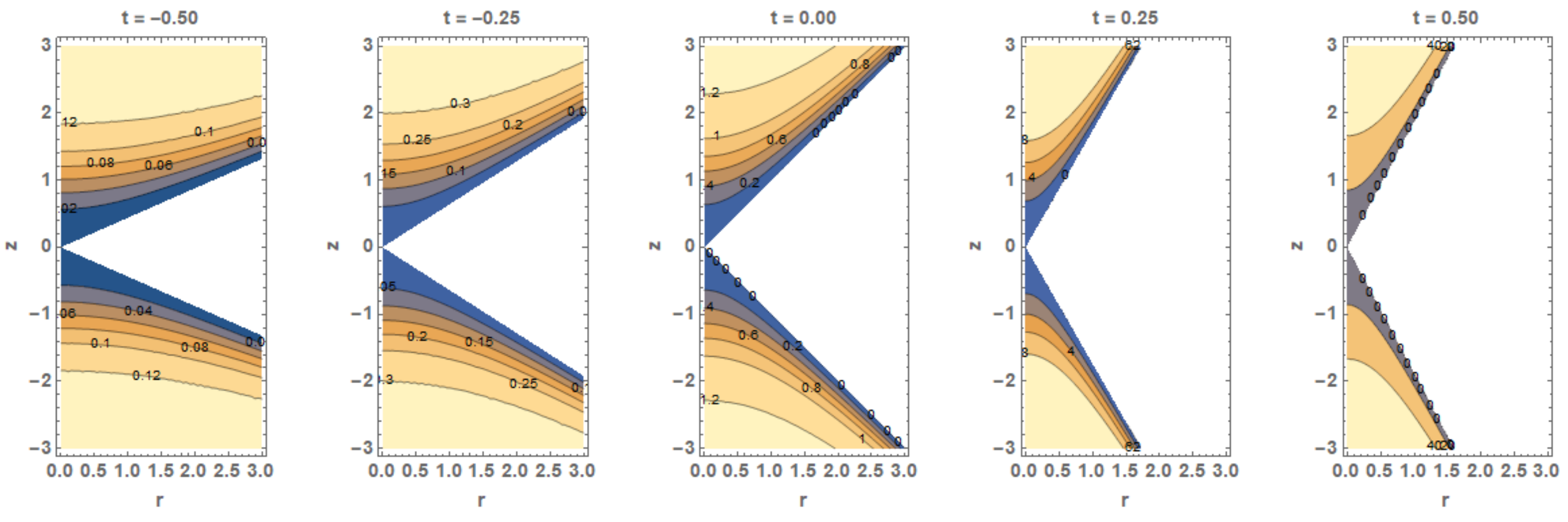}
        \caption{}\label{woodsaxon_dr_ds_pressure_plot}
    \end{subfigure} 
            \begin{subfigure}[t]{1.1\textwidth}
 %       \centering
        \includegraphics[scale=0.5]{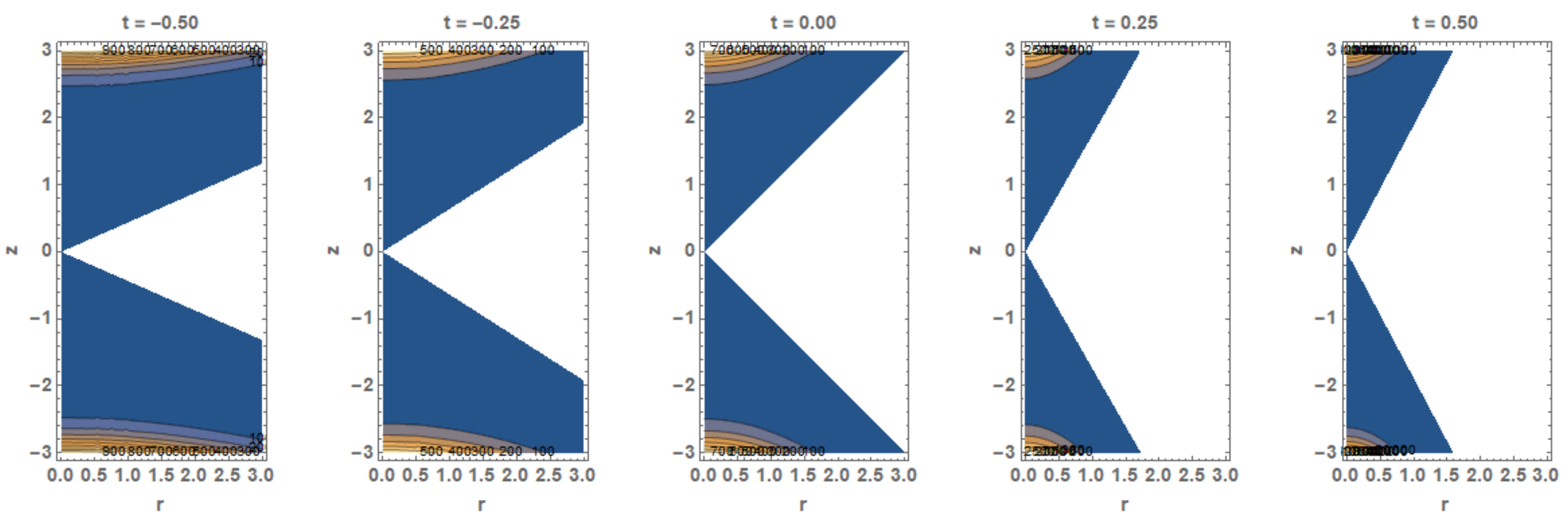}
        \caption{}\label{woodsaxon_dr_ds_sie_plot}
    \end{subfigure} 
            \begin{subfigure}[t]{1.1\textwidth}
 %       \centering
        \includegraphics[scale=0.5]{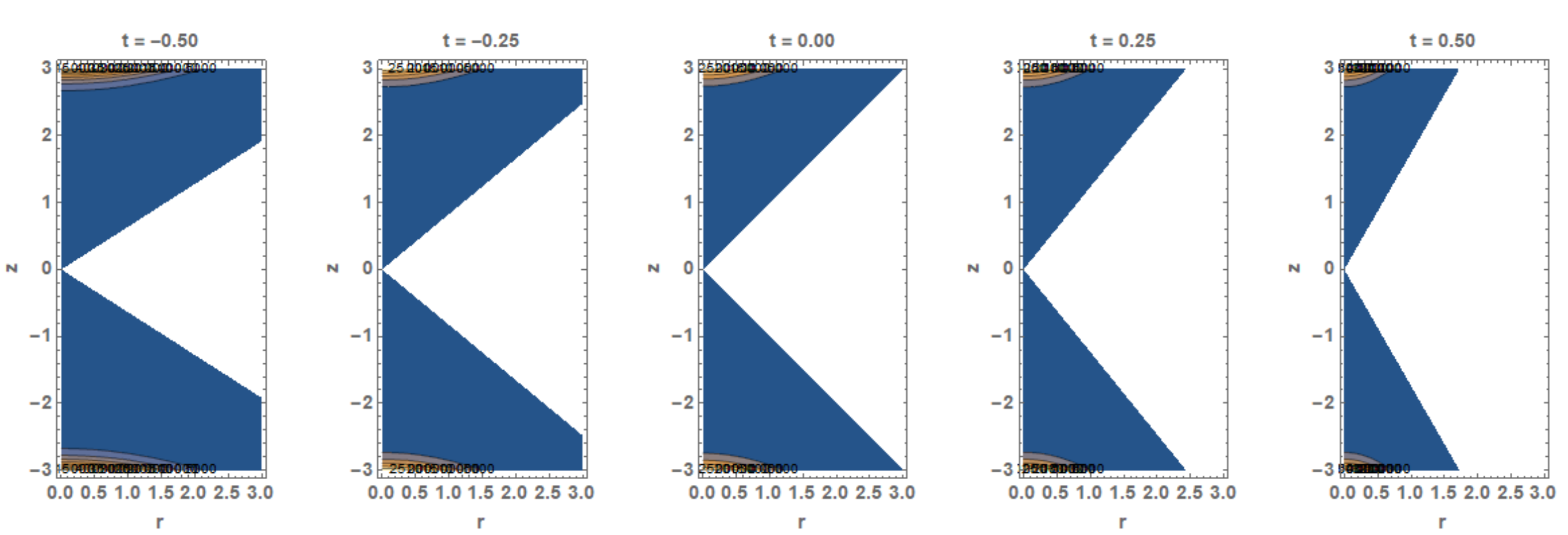}
        \caption{}\label{woodsaxon_dr_ds_entropy_plot}
    \end{subfigure} 
        \caption{Equations~(\ref{density_woodsaxon_dr_ds}), (\ref{pressure_woodsaxon_dr_ds}), (\ref{sie_woodsaxon_dr_ds}), and (\ref{entropy_woodsaxon_dr_ds}) ((a), (b), (c), and (d), respectively) evaluated at various times, under the example parameterization $\Pi_0 = 1$, $\Gamma_0$ given by Eq.~(\ref{gamma_0_woodsaxon}), $\zeta_0 = 1/4$, $\zeta_1 = 6$, and $\gamma = 5/3$, and featuring the $\gamma = 5/3$, $\kappa_3=-1$ DR-DS numerical solution depicted in Fig.~\ref{DR_DS_plot}.}
\label{woodsaxon_DR_DS_plot}
\end{figure*}
\begin{figure}[h!]
        \includegraphics[scale=0.4]{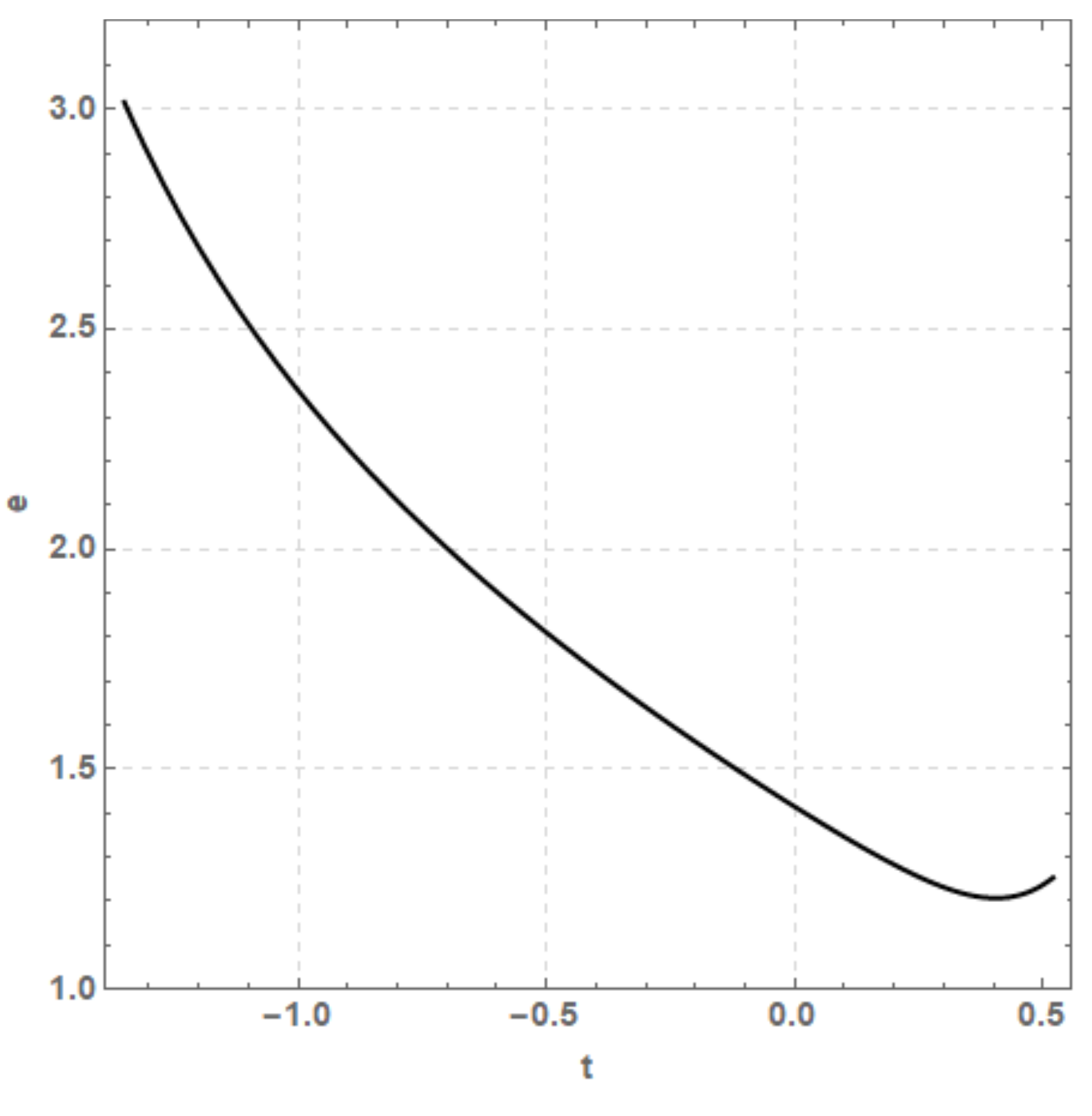}
        \caption{Equation~(\ref{eccentricity_hyper_R}) evaluated at various times, featuring the $\gamma = 5/3$, $\kappa_3 =-1$ DR-DS numerical solution depicted in Fig.~\ref{DR_DS_plot}.}
\label{woodsaxon_DR_DS_eccentricity_plot}
\end{figure}

As a final example, a ``diffuse surface DR-DS'' axisymmetric Nemchinov-Dyson solution follows from combining the results appearing in Secs.~\ref{sec:DR-DS} and \ref{subsubsec:woodsaxon}. This solution is again given by Eqs.~(\ref{ur_assump}) and (\ref{uz_assump}) for $u_r\left(r,z,t\right)$ and $u_z\left(r,z,t\right)$, respectively, and, with Eqs.~(\ref{xi_sim}), (\ref{eta_sim}), (\ref{zeta_def}), (\ref{rho_dyson_solution_zeta}), (\ref{p_dyson_solution_zeta}), (\ref{sie_dyson_solution_zeta}), (\ref{entropy_dyson_solution_zeta}), (\ref{pi_woodsaxon}), (\ref{gamma_woodsaxon}), (\ref{upsilon_woodsaxon}), and (\ref{sigma_woodsaxon}),
\begin{eqnarray}
\rho\left(r,z,t\right) &=& 
\frac{\Pi_0}{R_r^2 R_z} 
\frac{1}{1+\rm{exp}\left( \sqrt{ \frac{\kappa_3 r^2}{\zeta_0^2 R_r^2} + \frac{z^2}{\zeta_0^2 R_z^2} } -\zeta_1 \right) }
\label{density_woodsaxon_dr_ds}, \\ 
P\left(r,z,t\right) &=& 
-\frac{R_r \ddot{R}_r}{\kappa_3 R_r^2 R_z}\nonumber\\
&&\times\left\{\frac{\Pi_0}{4}\zeta_0\ln\left(\frac{1}{1+\mathrm{exp}\left(\sqrt{ \frac{\kappa_3 r^2}{\zeta_0^2 R_r^2} + \frac{z^2}{\zeta_0^2 R_z^2} } -\zeta_1 \right)}\right)\right.\nonumber\\
&&\left.-\frac{\Pi_0}{16}\mathrm{Li}_2\left[-\mathrm{exp}\left(\zeta_1-\sqrt{ \frac{\kappa_3 r^2}{\zeta_0^2 R_r^2} + \frac{z^2}{\zeta_0^2 R_z^2} }\right)\right]\right.\nonumber\\
&&\left.+-\frac{R_r \ddot{R}_r}{\kappa_3 R_r^2 R_z}\Gamma_0 \right\}
\label{pressure_woodsaxon_dr_ds},
\end{eqnarray}
\begin{eqnarray}
I\left(r,z,t\right) &=& 
\frac{R_r \ddot{R}_r}{\kappa_3 \left(\gamma-1\right)}\left\{ 
\zeta_0\left[1+\mathrm{exp}\left(\sqrt{ \frac{\kappa_3 r^2}{\zeta_0^2 R_r^2} + \frac{z^2}{\zeta_0^2 R_z^2} } -\zeta_1 \right)\right]\right.\nonumber\\
&&\times\bigg[\zeta\ln\left(\frac{1}{1+\mathrm{exp}\left(\sqrt{ \frac{\kappa_3 r^2}{\zeta_0^2 R_r^2} + \frac{z^2}{\zeta_0^2 R_z^2} } -\zeta_1 \right)}\right)\nonumber\\
&&+\zeta_0 \mathrm{Li}_2\left(-\mathrm{exp}\left(\zeta_1-\sqrt{ \frac{\kappa_3 r^2}{\zeta_0^2 R_r^2} + \frac{z^2}{\zeta_0^2 R_z^2} }\right)\right)\bigg]\nonumber\\
&&\left.+ \frac{1+\mathrm{exp}\left(\sqrt{ \frac{\kappa_3 r^2}{\zeta_0^2 R_r^2} + \frac{z^2}{\zeta_0^2 R_z^2} } -\zeta_1 \right)}{\Pi_0}\Gamma_0\right\}
\label{sie_woodsaxon_dr_ds},
\end{eqnarray}
\begin{eqnarray}
S\left(r,z,t\right) &=& 
-\frac{\ddot{R}_r R_r^{2\gamma-1} R_z^{\gamma-1}}{\kappa_3}\zeta_0\Pi_0^{1-\gamma}\nonumber\\
&&\times\left\{\left[1+\mathrm{exp}\left(\sqrt{ \frac{\kappa_3 r^2}{\zeta_0^2 R_r^2} + \frac{z^2}{\zeta_0^2 R_z^2} } -\zeta_1\right)\right]^\gamma\right.\nonumber\\
&&\times\bigg[\zeta \ln\left(\frac{1}{1+\mathrm{exp}\left(\sqrt{ \frac{\kappa_3 r^2}{\zeta_0^2 R_r^2} + \frac{z^2}{\zeta_0^2 R_z^2} } -\zeta_1\right)}\right)\nonumber\\
&&+\zeta_0 \mathrm{Li}_2\left(-\mathrm{exp}\left(\zeta_1-\sqrt{ \frac{\kappa_3 r^2}{\zeta_0^2 R_r^2} + \frac{z^2}{\zeta_0^2 R_z^2} } \right)\right)\bigg]\nonumber\\
&&\left.+\frac{\left[1+\mathrm{exp}\left(\sqrt{ \frac{\kappa_3 r^2}{\zeta_0^2 R_r^2} + \frac{z^2}{\zeta_0^2 R_z^2} } -\zeta_1\right)\right]^\gamma}{\Pi^{\gamma}}\Gamma_0\right\} \nonumber\\ 
\label{entropy_woodsaxon_dr_ds},
\end{eqnarray}
where $\Pi_0 > 0$, $\Gamma_0$, $\zeta_0 > 0$, and $\zeta_1 > 0$ are otherwise arbitrary constants. For the DR-DS type solution provided in Sec.~\ref{sec:DR-DS}, $\kappa_3 = -1$ as appearing in Eqs.~(\ref{density_woodsaxon_dr_ds})-(\ref{entropy_woodsaxon_dr_ds}), and the numerical representations of $R_r$ and $R_z$ are depicted in Fig.~\ref{DR_DS_plot}. 

Equations~(\ref{ur_assump}), (\ref{uz_assump}), and (\ref{density_woodsaxon_dr_ds})-(\ref{entropy_woodsaxon_dr_ds}) are depicted in Figs.~\ref{woodsaxon_DR_DS_velocity_plot} and \ref{woodsaxon_DR_DS_plot}, for the example parameterization $\Pi_0 = 1$, $\Gamma_0$ given by Eq.~(\ref{gamma_0_woodsaxon}), $\zeta_0 = 1/4$, $\zeta_1 = 6$, and $\gamma = 5/3$, and featuring the $\gamma = 5/3$ DR-DS numerical solution depicted in Fig.~\ref{DR_DS_plot}. The associated time-dependent eccentricity of all hyperbolic level surfaces in the density, pressure, SIE, and entropy state variables is given by Eq.~(\ref{eccentricity_hyper_R}), and is provided in Fig.~\ref{spatial_solution_set_plots}. 

Figure~\ref{woodsaxon_DR_DS_velocity_plot} depicts the total velocity vector field associated with the DR-DS type solution [as Eqs.~(\ref{ur_assump}) and (\ref{uz_assump}) hold whether the spatial portion of the associated Nemchinov-Dyson solution is of diffuse surface type or not] featured in Sec.~\ref{sec:DR-DS}, including the appropriate, conjoined linear behavior in both $r$ and $z$. The directions of the various velocity vectors appearing in Fig.~\ref{woodsaxon_DR_DS_velocity_plot} are again directly proportional to the slopes of the $R_r$ and $R_z$ curves appearing in Fig.~\ref{DR_DS_plot}, as otherwise explicitly revealed by Eqs.~(\ref{ur_assump}) and (\ref{uz_assump}). For the DR-DS example depicted in Figs.~\ref{DR_DS_plot} and \ref{woodsaxon_DR_DS_velocity_plot}, the global motion of the associated diffuse surface solution is therefore largely dominated for most times by motion in $r$, and and exhibits a sign reversal at later times. 

Figure~\ref{woodsaxon_DR_DS_plot} indicates the presence of a ``funnel''-like or double-conical ``outer surface'' in the diffuse surface DR-DS solution field; all state variables are not real-valued beyond that surface. Moreover, Fig.~\ref{woodsaxon_DR_DS_plot} indicates that with increasing time the outer surface ``opens up,'' while the material within simultaneously compresses, pressurizes, and heats. Figure~\ref{woodsaxon_DR_DS_plot} also shows that all hyperbolic state variable level surfaces diminish in eccentricity with increasing time. Otherwise, the $r$ and $z$ variation of the density, pressure, SIE, and entropy solutions proceeds according to highly non-trivial distributions, as also indicated by Eqs.~(\ref{density_woodsaxon_dr_ds})-(\ref{entropy_woodsaxon_dr_ds}). 
%
%%%%%%%%%%%%%%%%%%%%%%%%%%%%%%%%%%%%%%%%%%%%%%%%%%
%%%%%%%%%%%%%%%%%%%%%%%%%%%%%%%%%%%%%%%%%%%%%%%%%%
\section{Discussion and Conclusion}
\label{sec:conclusion}
%%%%%%%%%%%%%%%%%%%%%%%%%%%%%%%%%%%%%%%%%%%%%%%%%%
%%%%%%%%%%%%%%%%%%%%%%%%%%%%%%%%%%%%%%%%%%%%%%%%%%
%
In the spirit of and similar to the wealth of analogous results appearing within the existing literature, the results of Sec.~\ref{sec:dyson} define a general procedure for the construction of an infinite variety of Nemchinov-Dyson solutions of the 2D axisymmetric inviscid Euler equations, coupled to an ideal gas EOS. The solutions derived using this recipe share several common features, including:
\begin{itemize}
\item Space-time separability in each component of the velocity field. Moreover, this separable form is constrained to be linear in each associated direction, thus yielding a class of uniformly expanding or contracting solutions.
\item Self-consistent but otherwise arbitrary state variable distributions that depend solely on a 2D axial representation of the spherical radial coordinate. That the functional forms of these state variables in this coordinate are arbitrary owes to the absence of dissipation or other ancillary mechanisms within the attendant formulation of the inviscid Euler equations. 
\item The non-trivial level surfaces for all state variables (when they exist) are either elliptical or hyperbolic surfaces in $\left(r,z\right)$ space, and otherwise form surfaces of revolution about the 2D axisymmetric $z$-axis. Both the character and dynamical behavior of these level surfaces is associated with the sign of the solution's conjoined acceleration field (i.e., whether the acceleration field is positive or negative in each direction for all times).
\end{itemize}

Three example axisymmetric Nemchinov-Dyson solutions are provided in Secs.~\ref{subsubsec:isothermal_DR_DR}-\ref{subsubsec:woodsaxon_DR_DS}. The uniform SIE DR-DR solution appearing in Sec.~\ref{subsubsec:isothermal_DR_DR} may be regarded as a ``classical'' solution in that it features ellipsoidal state variable level surfaces that first contract and then expand in both the $r$ and $z$ directions. The uniform entropy DS-DS solution appearing in Sec.~\ref{subsubsec:isentropic_DS_DS} is in a sense the ``inverse'' of similar DR-DR solutions as disseminated by Nemchinov~\cite{nemchinov_1965}, in that instead of featuring a discrete object that either expands or contracts, it features a distending ellipsoidal cavity embedded within an otherwise infinite expanse of fluid. The diffuse surface DR-DS solution appearing in Sec.~\ref{subsubsec:woodsaxon_DR_DS} is entirely new and separate from the others, in that it features a distending, funnel-shaped figure with hyperbolic level surfaces in all interior state variables.

Without a doubt, and with little additional mathematical complexity, analogous solution behaviors are also extractible from the more extensive existing models featuring 3D Cartesian geometries. Even so, the absence from the existing literature of some of the more exotic solutions derived herein is likely due to historical, application-driven realities: conical or hyperbolic linear velocity solutions, for example, appear to have less immediately recognizable relevance to practical scenarios rooted in astrophysics or elsewhere, implications for solar flares or related processes notwithstanding. On the other hand, the expanding and contracting cavity solutions derived in this work (or implicitly contained in this work's broader result set) may have some utility in the fields of bubble collapse or cavitation, as otherwise discussed by Boyd et al~\cite{boyd2019collapsing}. 

Finally, and as discussed in Sec.~\ref{sec:intro}, any of the solutions derived as part of this program of study are expected to be of direct use in quantitative code verification or model qualification studies associated with inviscid Euler codes designed for the numerical solution of, for example, Eqs.~(\ref{cons_mass})-(\ref{cons_enegy_E}). In this sense, some of the more exotic solutions presented as part of Sec.~\ref{subsec:examples} - or their near neighbors - may find broader use beyond their limited physical implications, should they eventually come to serve as especially challenging or otherwise unique test problems or model solutions. 
%%%%%%%%%%%%%%%%%%%%%%%%%%%%%%%%%%%%%%%%%%%%%%%%%%
%%%%%%%%%%%%%%%%%%%%%%%%%%%%%%%%%%%%%%%%%%%%%%%%%%
\subsection{Recommendations for Future Study}
\label{sec:recommendations}
%%%%%%%%%%%%%%%%%%%%%%%%%%%%%%%%%%%%%%%%%%%%%%%%%%
%%%%%%%%%%%%%%%%%%%%%%%%%%%%%%%%%%%%%%%%%%%%%%%%%%
%
From a purely theoretical standpoint, the results of Sec.~\ref{sec:pi_sols} will no doubt prove readily extensible to a limitless variety of potential counterparts. A few examples of concrete physical significance (and with pedigree as established within the voluminous literature on solutions of the inviscid Euler equations featuring linear velocity assumptions) are provided in Secs.~\ref{subsubsec:uniform_density}-\ref{subsubsec:woodsaxon}, but others arising from analogous application-based motivations may of course be devised and coupled to any of the dynamical behaviors examined in Sec.~\ref{sec:scale_radius_sols}. This potential program of study represents perhaps the most straightforward path for further extension of the results appearing in Secs.~\ref{subsubsec:uniform_density}-\ref{subsubsec:woodsaxon}.

With additional relevance to Sec.~\ref{sec:scale_radius_sols}, and inspired by the analytical considerations of Anisimov and Lysikov~\cite{anisimov_1970}, Hunter and London~\cite{hunter1988multidimensional}, Gaffet~\cite{gaffet_1996,gaffet_1999}, Rozanova and Turzynsky~\cite{rozanova2016nonlinear,rozanova2017systems}, and Irtegov and Titorenko~\cite{irtegov2018qualitative} among many others, the dynamical system given by Eqs.~(\ref{ode_1_gen}) and (\ref{ode_2}) has been demonstrated in special cases to be amenable to analytical solution, or at the very least expression in terms of quadratures or special functions (e.g., elliptic integrals). Further detailed analytical studies of Eqs.~(\ref{ode_1_gen}) and (\ref{ode_2}) thus perhaps appear to be in order, potentially using the same techniques as rooted in symmetry analysis theory, and as previously employed in the context of the inviscid Euler equations by Coggeshall~\cite{coggeshall1986lie,coggeshall1992group,coggeshall1991analytic}. When they exist, the symmetry properties of any analytical or even semi-analytical solutions obtained through such means will prove invaluable for a more comprehensive understanding of the physical and mathematical properties of any affiliated axisymmetric Nemchinov-Dyson solutions. 

In addition, the symmetry analysis framework very likely represents a potential path toward further investigating some of the matters discussed in Sec.~\ref{sec:asymptotic}. In particular, a matter requiring further investigation is the construction of asymptotic scale ratio formulas in the style of Eq.~(\ref{Lambda_inf_HL}) for cases where the adiabatic index $\gamma \neq 5/3$. The establishment of any such results (should they exist) is expected to follow closely from considerations in Hamiltonian dynamics, integrals of motion, and perhaps even Noether's Theorem through introduction into the problem formulation of the symmetry analysis formalism. In turn, these rigorous considerations may help shed light on the on the still-outstanding matter of reconciling the Anisimov and Lysikov~\cite{anisimov_1970} formulas with those of Hunter and London~\cite{hunter1988multidimensional} (and the substantiating numerical evidence appearing in Sec.~\ref{sec:asymptotic}), or Gaffet~\cite{gaffet_1996,gaffet_1999}. The fit in all of these considerations of both the Dyson~\cite{dyson_original} and Nemchinov~\cite{nemchinov_1965} numerical tables -- and in the latter case, whether they are correct or not -- will also stand to be clarified though a future, more detailed study along these lines.

More broadly, the symmetry analysis formalism that may be brought to bear on Eqs.~(\ref{ode_1_gen}) and (\ref{ode_2}) may also be be used to better categorize and understand the physical implications of the axisymmetric Nemchinov-Dyson solutions themselves. For example, McHardy et al.~\cite{mchardy2019group} recently applied this technique in the context of linear velocity solutions of the 1D inviscid Euler equations, and discovered that not all such solutions necessarily share (or, more appropriately, are generated by) the same underlying symmetry properties. In addition to potentially yielding similar benefits in the context of axisymmetric Nemchinov-Dyson solutions, the symmetry analysis formalism also represents the best way to explicitly connect the results of this and related work to that of, for example, Coggeshall~\cite{coggeshall1986lie,coggeshall1992group,coggeshall1991analytic}. 

Otherwise, numerous modeling generalizations of this work are also possible, including but not limited to:
\begin{itemize}
\item Use of non-ideal alternatives to Eq.~(\ref{ideal_gas_eos}), such as the stiffened gas or Mie-Gruneisen EOS forms as disseminated by Harlow and Amsden~\cite{harlow1971fluid}, or various other forms depending on a practical application of interest). Some solutions along these lines have been developed with potential applications to an incredibly wide ranging set of circumstances including but not limited to quantum mechanics/superfluidity~\cite{kuznetsov2020expansion} and cosmology~\cite{bogoyavlenskymethods}. The mathematical framework for potentially constructing any additional solutions -- or identifying the most general possible class of solutions -- also exists and is based on the outcomes of symmetry analysis of the inviscid Euler equations~\cite{ovsyannikov1,ovsyannikov2,holm1976symmetry,axford1998solutions,ramsey2017symmetries}.
\item Coupling of Eqs.~(\ref{cons_mass})-(\ref{cons_enegy_E}) to various additional physical mechanisms (some of which are dissipative) such as viscosity, heat transport, charge transport, elasticity, plasticity, and/or electromagnetism, and in the style of many existing studies and solutions featuring gravitation as an ancillary process. In addition to the foundational developments by Grad~\cite{grad1949kinetic}, Chandrasekhar~\cite{chandrasekhar_1967}, and Bogoyavlensky~\cite{bogoyavlenskymethods} (and references therein), additional relevant examples featuring graviational processes owe to Borisov et al.~\cite{borisov2009hamiltonian}, Ragazzo and Ruiz~\cite{ragazzo2015dynamics}, Bizyaev et al.~\cite{bizyaev2015figures}, and Guo et al~\cite{guo2020continued}.
\item Investigation of Eqs.~(\ref{cons_mass})-(\ref{cons_enegy_E}) as written in other 2D and 3D geometries such as ellipsoidal, parabolic, or torodial coordinates where the Lam\'{e} coefficients are nontrivial~\cite{spiegel1959schaum}. 
\item Incorporation of rotational motion into the underlying mathematical framework, in the style of many existing solutions as disseminated by, for example, Ovsyannikov~\cite{ovsyannikov1,ovsyannikov2}, Dyson~\cite{dyson_spinning}, Bogoyavlensky~\cite{bogoyavlenskymethods}, and their many successors.
\end{itemize}
In conjunction with the linear velocity assumption as formulated within a given coordinate system, all of these generalizations may be combined in various ways to yield an ever-growing family of Nemchinov-Dyson solutions of increasing physical fidelity or application relevance.

Beyond even these non-trivial generalizations, further expansion of this work may be affected by either somehow enhancing or dispensing with the linear velocity assumption itself. Perhaps the lowest order generalization along these lines involves retaining space-time separability in each velocity field component (regardless of the underlying geometry), and subsequently investigating various non-uniform spatial velocity profiles. Several examples of non-linear velocity profiles also appear in, for example, the work of Coggeshall~\cite{coggeshall1986lie,coggeshall1992group,coggeshall1991analytic}.

In addition to this potentially extensive program of purely theoretical study, the rigorous exercise of 2D or 3D Nemchinov-Dyson solutions for the purposes of quantitative code verification or model qualification also remains as relatively unbroken ground. As noted in Sec.~\ref{sec:intro}, some quantitative code verification studies along these lines have been performed in the context of the 1D linear velocity solutions, but even the existing 2D and 3D solutions featuring the same underlying assumptions appear to have been much less utilized in this manner. Naturally, perhaps the most practical future use of the many results derived in this work (and the body of existing work upon which our results are founded) will be in the quantitative code verification context, or for use as diagnostic tools for computational simulations of more complicated physical processes. Indeed, and in closing, as appropriately noted by Sachdev~\cite{sachdev2001self},
\begin{quote}
``...understanding the validity and place of exact/approximate analytical solution[s] in the general context can be greatly enhanced by numerical simulation. In short, there must be a continuous interplay of analysis and computation if a ... problem is to be successfully tackled.''
\end{quote} 
%%%%%%%%%%%%%%%%%%%%%%%%%%%%%%%%%%%%%%%%%%%%%%%%%%
%%%%%%%%%%%%%%%%%%%%%%%%%%%%%%%%%%%%%%%%%%%%%%%%%%
\begin{acknowledgments}
The authors dedicate this work to the late Freeman J. Dyson and also the late Paul P. Whalen (who introduced SDR to this interesting topic). 

This work was supported by the U.S. Department of Energy (DOE) through the Los Alamos National Laboratory. Los Alamos National Laboratory is operated by Triad National Security, LLC, for the National Nuclear Security Administration of the DOE (contract number 89233218CNA000001). JFG was partially funded through supported by the National Science Foundation (NSF) under Grant No. PHY-1803912. The authors thank E. J. Albright, D. B. Garcia, P. J. Jaegers, J. D. McHardy, E. M. Schmidt, J. H. Schmidt, and B. A. Temple for their valuable insights on these topics. JFG thanks Prof. Richard F. Lebed (Arizona State University) for his insight and support.
\end{acknowledgments}
\section*{Data Availability}
The data that support the findings of this study are available from the corresponding author upon reasonable request.
%\clearpage
%\newpage
\bibliographystyle{apsrev4-1}
\bibliography{Dyson}{}

\begin{thebibliography}{73}%
\makeatletter
\providecommand \@ifxundefined [1]{%
 \@ifx{#1\undefined}
}%
\providecommand \@ifnum [1]{%
 \ifnum #1\expandafter \@firstoftwo
 \else \expandafter \@secondoftwo
 \fi
}%
\providecommand \@ifx [1]{%
 \ifx #1\expandafter \@firstoftwo
 \else \expandafter \@secondoftwo
 \fi
}%
\providecommand \natexlab [1]{#1}%
\providecommand \enquote  [1]{``#1''}%
\providecommand \bibnamefont  [1]{#1}%
\providecommand \bibfnamefont [1]{#1}%
\providecommand \citenamefont [1]{#1}%
\providecommand \href@noop [0]{\@secondoftwo}%
\providecommand \href [0]{\begingroup \@sanitize@url \@href}%
\providecommand \@href[1]{\@@startlink{#1}\@@href}%
\providecommand \@@href[1]{\endgroup#1\@@endlink}%
\providecommand \@sanitize@url [0]{\catcode `\\12\catcode `\$12\catcode
  `\&12\catcode `\#12\catcode `\^12\catcode `\_12\catcode `\%12\relax}%
\providecommand \@@startlink[1]{}%
\providecommand \@@endlink[0]{}%
\providecommand \url  [0]{\begingroup\@sanitize@url \@url }%
\providecommand \@url [1]{\endgroup\@href {#1}{\urlprefix }}%
\providecommand \urlprefix  [0]{URL }%
\providecommand \Eprint [0]{\href }%
\providecommand \doibase [0]{http://dx.doi.org/}%
\providecommand \selectlanguage [0]{\@gobble}%
\providecommand \bibinfo  [0]{\@secondoftwo}%
\providecommand \bibfield  [0]{\@secondoftwo}%
\providecommand \translation [1]{[#1]}%
\providecommand \BibitemOpen [0]{}%
\providecommand \bibitemStop [0]{}%
\providecommand \bibitemNoStop [0]{.\EOS\space}%
\providecommand \EOS [0]{\spacefactor3000\relax}%
\providecommand \BibitemShut  [1]{\csname bibitem#1\endcsname}%
\let\auto@bib@innerbib\@empty
%</preamble>
\bibitem [{\citenamefont {Sedov}(2018)}]{sedov}%
  \BibitemOpen
  \bibfield  {author} {\bibinfo {author} {\bibfnamefont {L.~I.}\ \bibnamefont
  {Sedov}},\ }\href@noop {} {\emph {\bibinfo {title} {Similarity and
  dimensional methods in mechanics}}}\ (\bibinfo  {publisher} {CRC press},\
  \bibinfo {year} {2018})\BibitemShut {NoStop}%
\bibitem [{\citenamefont {Zel’dovich}\ and\ \citenamefont
  {Raizer}(2012)}]{zeldovich_raizer}%
  \BibitemOpen
  \bibfield  {author} {\bibinfo {author} {\bibfnamefont {Y.}~\bibnamefont
  {Zel’dovich}}\ and\ \bibinfo {author} {\bibfnamefont {Y.}~\bibnamefont
  {Raizer}},\ }\href {https://books.google.com/books?id=PULCAgAAQBAJ} {\emph
  {\bibinfo {title} {Physics of Shock Waves and High-Temperature Hydrodynamic
  Phenomena}}},\ Dover Books on Physics\ (\bibinfo  {publisher} {Dover
  Publications},\ \bibinfo {year} {2012})\BibitemShut {NoStop}%
\bibitem [{\citenamefont {Cantwell}(2002)}]{cantwell}%
  \BibitemOpen
  \bibfield  {author} {\bibinfo {author} {\bibfnamefont {B.}~\bibnamefont
  {Cantwell}},\ }\href {https://books.google.com/books?id=76RS2ZQ0UyUC} {\emph
  {\bibinfo {title} {Introduction to Symmetry Analysis}}},\ Cambridge Texts in
  Applied Mathematics\ (\bibinfo  {publisher} {Cambridge University Press},\
  \bibinfo {year} {2002})\BibitemShut {NoStop}%
\bibitem [{\citenamefont {Atzeni}\ and\ \citenamefont {Meyer-ter
  Vehn}(2004)}]{atzeni2004physics}%
  \BibitemOpen
  \bibfield  {author} {\bibinfo {author} {\bibfnamefont {S.}~\bibnamefont
  {Atzeni}}\ and\ \bibinfo {author} {\bibfnamefont {J.}~\bibnamefont {Meyer-ter
  Vehn}},\ }\href@noop {} {\emph {\bibinfo {title} {The physics of inertial
  fusion: beam plasma interaction, hydrodynamics, hot dense matter}}},\ Vol.\
  \bibinfo {volume} {125}\ (\bibinfo  {publisher} {OUP Oxford},\ \bibinfo
  {year} {2004})\BibitemShut {NoStop}%
\bibitem [{\citenamefont {Stanyukovich}(2016)}]{stanyukovich2016unsteady}%
  \BibitemOpen
  \bibfield  {author} {\bibinfo {author} {\bibfnamefont {K.~P.}\ \bibnamefont
  {Stanyukovich}},\ }\href@noop {} {\emph {\bibinfo {title} {Unsteady motion of
  continuous media}}}\ (\bibinfo  {publisher} {Elsevier},\ \bibinfo {year}
  {2016})\BibitemShut {NoStop}%
\bibitem [{\citenamefont {Sachdev}(2016)}]{sachdev2016shock}%
  \BibitemOpen
  \bibfield  {author} {\bibinfo {author} {\bibfnamefont {P.}~\bibnamefont
  {Sachdev}},\ }\href@noop {} {\emph {\bibinfo {title} {Shock waves \&
  explosions}}}\ (\bibinfo  {publisher} {CRC Press},\ \bibinfo {year}
  {2016})\BibitemShut {NoStop}%
\bibitem [{\citenamefont {Motz}(1979)}]{motz1979physics}%
  \BibitemOpen
  \bibfield  {author} {\bibinfo {author} {\bibfnamefont {H.}~\bibnamefont
  {Motz}},\ }\href@noop {} {\bibfield  {journal} {\bibinfo  {journal} {London
  and New York, Academic Press, 1979, 299 p.}\ } (\bibinfo {year}
  {1979})}\BibitemShut {NoStop}%
\bibitem [{\citenamefont {Pert}(1980)}]{pert_1980}%
  \BibitemOpen
  \bibfield  {author} {\bibinfo {author} {\bibfnamefont {G.~J.}\ \bibnamefont
  {Pert}},\ }\href {\doibase 10.1017/S0022112080001140} {\bibfield  {journal}
  {\bibinfo  {journal} {Journal of Fluid Mechanics}\ }\textbf {\bibinfo
  {volume} {100}},\ \bibinfo {pages} {257–277} (\bibinfo {year}
  {1980})}\BibitemShut {NoStop}%
\bibitem [{\citenamefont {Pert}(1987)}]{pert1987use}%
  \BibitemOpen
  \bibfield  {author} {\bibinfo {author} {\bibfnamefont {G.}~\bibnamefont
  {Pert}},\ }\href@noop {} {\bibfield  {journal} {\bibinfo  {journal} {Laser
  and Particle beams}\ }\textbf {\bibinfo {volume} {5}},\ \bibinfo {pages}
  {643} (\bibinfo {year} {1987})}\BibitemShut {NoStop}%
\bibitem [{\citenamefont {Pert}(1989)}]{pert_1989}%
  \BibitemOpen
  \bibfield  {author} {\bibinfo {author} {\bibfnamefont {G.~J.}\ \bibnamefont
  {Pert}},\ }\href {\doibase 10.1017/S0022377800013854} {\bibfield  {journal}
  {\bibinfo  {journal} {Journal of Plasma Physics}\ }\textbf {\bibinfo {volume}
  {41}},\ \bibinfo {pages} {263–280} (\bibinfo {year} {1989})}\BibitemShut
  {NoStop}%
\bibitem [{\citenamefont {Hunter~Jr}\ and\ \citenamefont
  {London}(1988)}]{hunter1988multidimensional}%
  \BibitemOpen
  \bibfield  {author} {\bibinfo {author} {\bibfnamefont {J.~H.}\ \bibnamefont
  {Hunter~Jr}}\ and\ \bibinfo {author} {\bibfnamefont {R.~A.}\ \bibnamefont
  {London}},\ }\href@noop {} {\bibfield  {journal} {\bibinfo  {journal} {The
  Physics of fluids}\ }\textbf {\bibinfo {volume} {31}},\ \bibinfo {pages}
  {3102} (\bibinfo {year} {1988})}\BibitemShut {NoStop}%
\bibitem [{\citenamefont {Anisimov}\ \emph {et~al.}(1993)\citenamefont
  {Anisimov}, \citenamefont {B{\"a}uerle},\ and\ \citenamefont
  {Luk’Yanchuk}}]{anisimov1993gas}%
  \BibitemOpen
  \bibfield  {author} {\bibinfo {author} {\bibfnamefont {S.}~\bibnamefont
  {Anisimov}}, \bibinfo {author} {\bibfnamefont {D.}~\bibnamefont
  {B{\"a}uerle}}, \ and\ \bibinfo {author} {\bibfnamefont {B.}~\bibnamefont
  {Luk’Yanchuk}},\ }\href@noop {} {\bibfield  {journal} {\bibinfo  {journal}
  {Physical Review B}\ }\textbf {\bibinfo {volume} {48}},\ \bibinfo {pages}
  {12076} (\bibinfo {year} {1993})}\BibitemShut {NoStop}%
\bibitem [{\citenamefont {Anisimov}\ \emph {et~al.}(1996)\citenamefont
  {Anisimov}, \citenamefont {Luk'Yanchuk},\ and\ \citenamefont
  {Luches}}]{anisimov1996analytical}%
  \BibitemOpen
  \bibfield  {author} {\bibinfo {author} {\bibfnamefont {S.}~\bibnamefont
  {Anisimov}}, \bibinfo {author} {\bibfnamefont {B.}~\bibnamefont
  {Luk'Yanchuk}}, \ and\ \bibinfo {author} {\bibfnamefont {A.}~\bibnamefont
  {Luches}},\ }\href@noop {} {\bibfield  {journal} {\bibinfo  {journal}
  {Applied surface science}\ }\textbf {\bibinfo {volume} {96}},\ \bibinfo
  {pages} {24} (\bibinfo {year} {1996})}\BibitemShut {NoStop}%
\bibitem [{\citenamefont {Guo}\ \emph {et~al.}(2020)\citenamefont {Guo},
  \citenamefont {Had{\v{z}}i{\'c}},\ and\ \citenamefont
  {Jang}}]{guo2020continued}%
  \BibitemOpen
  \bibfield  {author} {\bibinfo {author} {\bibfnamefont {Y.}~\bibnamefont
  {Guo}}, \bibinfo {author} {\bibfnamefont {M.}~\bibnamefont
  {Had{\v{z}}i{\'c}}}, \ and\ \bibinfo {author} {\bibfnamefont
  {J.}~\bibnamefont {Jang}},\ }\href@noop {} {\bibfield  {journal} {\bibinfo
  {journal} {Archive for Rational Mechanics and Analysis}\ ,\ \bibinfo {pages}
  {1}} (\bibinfo {year} {2020})}\BibitemShut {NoStop}%
\bibitem [{\citenamefont {Kuznetsov}\ \emph {et~al.}(2020)\citenamefont
  {Kuznetsov}, \citenamefont {Kagan},\ and\ \citenamefont
  {Turlapov}}]{kuznetsov2020expansion}%
  \BibitemOpen
  \bibfield  {author} {\bibinfo {author} {\bibfnamefont {E.}~\bibnamefont
  {Kuznetsov}}, \bibinfo {author} {\bibfnamefont {M.~Y.}\ \bibnamefont
  {Kagan}}, \ and\ \bibinfo {author} {\bibfnamefont {A.}~\bibnamefont
  {Turlapov}},\ }\href@noop {} {\bibfield  {journal} {\bibinfo  {journal}
  {Physical Review A}\ }\textbf {\bibinfo {volume} {101}},\ \bibinfo {pages}
  {043612} (\bibinfo {year} {2020})}\BibitemShut {NoStop}%
\bibitem [{\citenamefont
  {Kidder}(1974{\natexlab{a}})}]{Kidder_1974_isentropic}%
  \BibitemOpen
  \bibfield  {author} {\bibinfo {author} {\bibfnamefont {R.}~\bibnamefont
  {Kidder}},\ }\href {\doibase 10.1088/0029-5515/14/1/008} {\bibfield
  {journal} {\bibinfo  {journal} {Nuclear Fusion}\ }\textbf {\bibinfo {volume}
  {14}},\ \bibinfo {pages} {53} (\bibinfo {year}
  {1974}{\natexlab{a}})}\BibitemShut {NoStop}%
\bibitem [{\citenamefont {Kidder}(1974{\natexlab{b}})}]{Kidder_1974_laser}%
  \BibitemOpen
  \bibfield  {author} {\bibinfo {author} {\bibfnamefont {R.}~\bibnamefont
  {Kidder}},\ }\href {\doibase 10.1088/0029-5515/14/6/005} {\bibfield
  {journal} {\bibinfo  {journal} {Nuclear Fusion}\ }\textbf {\bibinfo {volume}
  {14}},\ \bibinfo {pages} {797} (\bibinfo {year}
  {1974}{\natexlab{b}})}\BibitemShut {NoStop}%
\bibitem [{\citenamefont {Kidder}(1976)}]{Kidder_1976}%
  \BibitemOpen
  \bibfield  {author} {\bibinfo {author} {\bibfnamefont {R.}~\bibnamefont
  {Kidder}},\ }\href {\doibase 10.1088/0029-5515/16/1/001} {\bibfield
  {journal} {\bibinfo  {journal} {Nuclear Fusion}\ }\textbf {\bibinfo {volume}
  {16}},\ \bibinfo {pages} {3} (\bibinfo {year} {1976})}\BibitemShut {NoStop}%
\bibitem [{\citenamefont {Hora}(1971)}]{hora}%
  \BibitemOpen
  \bibfield  {author} {\bibinfo {author} {\bibfnamefont {H.}~\bibnamefont
  {Hora}},\ }in\ \href@noop {} {\emph {\bibinfo {booktitle} {Laser Interaction
  and Related Plasma Phenomena}}},\ \bibinfo {editor} {edited by\ \bibinfo
  {editor} {\bibfnamefont {H.~J.}\ \bibnamefont {Schwarz}}\ and\ \bibinfo
  {editor} {\bibfnamefont {H.}~\bibnamefont {Hora}}}\ (\bibinfo  {publisher}
  {Springer US},\ \bibinfo {address} {Boston, MA},\ \bibinfo {year} {1971})\
  pp.\ \bibinfo {pages} {365--382}\BibitemShut {NoStop}%
\bibitem [{\citenamefont {Hora}\ and\ \citenamefont
  {Pfirsch}(1972)}]{hora_pfirsch}%
  \BibitemOpen
  \bibfield  {author} {\bibinfo {author} {\bibfnamefont {H.}~\bibnamefont
  {Hora}}\ and\ \bibinfo {author} {\bibfnamefont {D.}~\bibnamefont {Pfirsch}},\
  }in\ \href@noop {} {\emph {\bibinfo {booktitle} {Laser Interaction and
  Related Plasma Phenomena}}},\ \bibinfo {editor} {edited by\ \bibinfo {editor}
  {\bibfnamefont {H.~J.}\ \bibnamefont {Schwarz}}\ and\ \bibinfo {editor}
  {\bibfnamefont {H.}~\bibnamefont {Hora}}}\ (\bibinfo  {publisher} {Springer
  US},\ \bibinfo {address} {Boston, MA},\ \bibinfo {year} {1972})\ pp.\
  \bibinfo {pages} {515--526}\BibitemShut {NoStop}%
\bibitem [{\citenamefont {Coggeshall}\ and\ \citenamefont
  {Axford}(1986)}]{coggeshall1986lie}%
  \BibitemOpen
  \bibfield  {author} {\bibinfo {author} {\bibfnamefont {S.~V.}\ \bibnamefont
  {Coggeshall}}\ and\ \bibinfo {author} {\bibfnamefont {R.~A.}\ \bibnamefont
  {Axford}},\ }\href@noop {} {\bibfield  {journal} {\bibinfo  {journal} {Phys.
  Fluids (1958-1988)}\ }\textbf {\bibinfo {volume} {29}},\ \bibinfo {pages}
  {2398} (\bibinfo {year} {1986})}\BibitemShut {NoStop}%
\bibitem [{\citenamefont {Coggeshall}\ and\ \citenamefont {Meyer-ter
  Vehn}(1992)}]{coggeshall1992group}%
  \BibitemOpen
  \bibfield  {author} {\bibinfo {author} {\bibfnamefont {S.~V.}\ \bibnamefont
  {Coggeshall}}\ and\ \bibinfo {author} {\bibfnamefont {J.}~\bibnamefont
  {Meyer-ter Vehn}},\ }\href@noop {} {\bibfield  {journal} {\bibinfo  {journal}
  {J. Math. Phys.}\ }\textbf {\bibinfo {volume} {33}},\ \bibinfo {pages} {3585}
  (\bibinfo {year} {1992})}\BibitemShut {NoStop}%
\bibitem [{\citenamefont {Coggeshall}(1991)}]{coggeshall1991analytic}%
  \BibitemOpen
  \bibfield  {author} {\bibinfo {author} {\bibfnamefont {S.~V.}\ \bibnamefont
  {Coggeshall}},\ }\href@noop {} {\bibfield  {journal} {\bibinfo  {journal}
  {Phys. Fluids A: Fluid Dynamics (1989-1993)}\ }\textbf {\bibinfo {volume}
  {3}},\ \bibinfo {pages} {757} (\bibinfo {year} {1991})}\BibitemShut {NoStop}%
\bibitem [{\citenamefont {Krauser}\ \emph {et~al.}(1996)\citenamefont {Krauser}
  \emph {et~al.}}]{krauser}%
  \BibitemOpen
  \bibfield  {author} {\bibinfo {author} {\bibfnamefont {W.~J.}\ \bibnamefont
  {Krauser}} \emph {et~al.},\ }\href {\doibase 10.1063/1.872006} {\bibfield
  {journal} {\bibinfo  {journal} {Physics of Plasmas}\ }\textbf {\bibinfo
  {volume} {3}},\ \bibinfo {pages} {2084} (\bibinfo {year} {1996})}\BibitemShut
  {NoStop}%
\bibitem [{\citenamefont {Morgan}\ \emph {et~al.}(2014)\citenamefont {Morgan}
  \emph {et~al.}}]{morgan2014lagrangian}%
  \BibitemOpen
  \bibfield  {author} {\bibinfo {author} {\bibfnamefont {N.~R.}\ \bibnamefont
  {Morgan}} \emph {et~al.},\ }\href@noop {} {\bibfield  {journal} {\bibinfo
  {journal} {Journal of Computational Physics}\ }\textbf {\bibinfo {volume}
  {259}},\ \bibinfo {pages} {568} (\bibinfo {year} {2014})}\BibitemShut
  {NoStop}%
\bibitem [{\citenamefont {Morgan}\ \emph {et~al.}(2015)\citenamefont {Morgan}
  \emph {et~al.}}]{morgan2015point}%
  \BibitemOpen
  \bibfield  {author} {\bibinfo {author} {\bibfnamefont {N.~R.}\ \bibnamefont
  {Morgan}} \emph {et~al.},\ }\href@noop {} {\bibfield  {journal} {\bibinfo
  {journal} {Journal of Computational Physics}\ }\textbf {\bibinfo {volume}
  {290}},\ \bibinfo {pages} {239} (\bibinfo {year} {2015})}\BibitemShut
  {NoStop}%
\bibitem [{\citenamefont {Burton}\ \emph {et~al.}(2015)\citenamefont {Burton}
  \emph {et~al.}}]{burton2015reduction}%
  \BibitemOpen
  \bibfield  {author} {\bibinfo {author} {\bibfnamefont {D.~E.}\ \bibnamefont
  {Burton}} \emph {et~al.},\ }\href@noop {} {\bibfield  {journal} {\bibinfo
  {journal} {Journal of Computational Physics}\ }\textbf {\bibinfo {volume}
  {299}},\ \bibinfo {pages} {229} (\bibinfo {year} {2015})}\BibitemShut
  {NoStop}%
\bibitem [{\citenamefont {Burton}\ \emph {et~al.}(2018)\citenamefont {Burton}
  \emph {et~al.}}]{burton2018compatible}%
  \BibitemOpen
  \bibfield  {author} {\bibinfo {author} {\bibfnamefont {D.~E.}\ \bibnamefont
  {Burton}} \emph {et~al.},\ }\href@noop {} {\bibfield  {journal} {\bibinfo
  {journal} {Journal of Computational Physics}\ }\textbf {\bibinfo {volume}
  {355}},\ \bibinfo {pages} {492} (\bibinfo {year} {2018})}\BibitemShut
  {NoStop}%
\bibitem [{\citenamefont {Oberkampf}\ \emph {et~al.}(2004)\citenamefont
  {Oberkampf} \emph {et~al.}}]{oberkampf}%
  \BibitemOpen
  \bibfield  {author} {\bibinfo {author} {\bibfnamefont {W.~L.}\ \bibnamefont
  {Oberkampf}} \emph {et~al.},\ }\href {\doibase 10.1115/1.1767847} {\bibfield
  {journal} {\bibinfo  {journal} {Applied Mechanics Reviews}\ }\textbf
  {\bibinfo {volume} {57}},\ \bibinfo {pages} {345} (\bibinfo {year}
  {2004})}\BibitemShut {NoStop}%
\bibitem [{\citenamefont {Chandrasekhar}(1967)}]{chandrasekhar_1967}%
  \BibitemOpen
  \bibfield  {author} {\bibinfo {author} {\bibfnamefont {S.}~\bibnamefont
  {Chandrasekhar}},\ }\href {\doibase 10.1002/cpa.3160200203} {\bibfield
  {journal} {\bibinfo  {journal} {Communications on Pure and Applied
  Mathematics}\ }\textbf {\bibinfo {volume} {20}},\ \bibinfo {pages} {251}
  (\bibinfo {year} {1967})}\BibitemShut {NoStop}%
\bibitem [{\citenamefont {Ovsyannikov}(1956)}]{ovsyannikovhydro}%
  \BibitemOpen
  \bibfield  {author} {\bibinfo {author} {\bibfnamefont {L.}~\bibnamefont
  {Ovsyannikov}},\ }\href@noop {} {\bibfield  {journal} {\bibinfo  {journal}
  {Dokl. Akad. Nauk}\ }\textbf {\bibinfo {volume} {111}},\ \bibinfo {pages}
  {47} (\bibinfo {year} {1956})}\BibitemShut {NoStop}%
\bibitem [{\citenamefont {Nemchinov}(1965)}]{nemchinov_1965}%
  \BibitemOpen
  \bibfield  {author} {\bibinfo {author} {\bibfnamefont {I.}~\bibnamefont
  {Nemchinov}},\ }\href {\doibase https://doi.org/10.1016/0021-8928(65)90158-9}
  {\bibfield  {journal} {\bibinfo  {journal} {Journal of Applied Mathematics
  and Mechanics}\ }\textbf {\bibinfo {volume} {29}},\ \bibinfo {pages} {143 }
  (\bibinfo {year} {1965})}\BibitemShut {NoStop}%
\bibitem [{\citenamefont {Anisimov}\ and\ \citenamefont
  {Lysikov}(1970)}]{anisimov_1970}%
  \BibitemOpen
  \bibfield  {author} {\bibinfo {author} {\bibfnamefont {S.}~\bibnamefont
  {Anisimov}}\ and\ \bibinfo {author} {\bibfnamefont {I.}~\bibnamefont
  {Lysikov}},\ }\href {\doibase 10.1016/0021-8928(70)90070-5} {\bibfield
  {journal} {\bibinfo  {journal} {Journal of Applied Mathematics and
  Mechanics}\ }\textbf {\bibinfo {volume} {34}},\ \bibinfo {pages} {882 }
  (\bibinfo {year} {1970})}\BibitemShut {NoStop}%
\bibitem [{\citenamefont {Dyson}(1958)}]{dyson_original}%
  \BibitemOpen
  \bibfield  {author} {\bibinfo {author} {\bibfnamefont {F.~J.}\ \bibnamefont
  {Dyson}},\ }\href@noop {} {\emph {\bibinfo {title} {Free Expansion of a Gas
  (II) Gaussian Model}}},\ \bibinfo {type} {Tech. Rep.}\ (\bibinfo {year}
  {1958})\BibitemShut {NoStop}%
\bibitem [{\citenamefont {Dyson}(1968)}]{dyson_spinning}%
  \BibitemOpen
  \bibfield  {author} {\bibinfo {author} {\bibfnamefont {F.~J.}\ \bibnamefont
  {Dyson}},\ }\href@noop {} {\bibfield  {journal} {\bibinfo  {journal} {Journal
  of Mathematics and Mechanics}\ }\textbf {\bibinfo {volume} {18}},\ \bibinfo
  {pages} {91} (\bibinfo {year} {1968})}\BibitemShut {NoStop}%
\bibitem [{\citenamefont {Hara}\ \emph {et~al.}(1973)\citenamefont {Hara},
  \citenamefont {Matsuda},\ and\ \citenamefont {Nakazawa}}]{hara}%
  \BibitemOpen
  \bibfield  {author} {\bibinfo {author} {\bibfnamefont {T.}~\bibnamefont
  {Hara}}, \bibinfo {author} {\bibfnamefont {T.}~\bibnamefont {Matsuda}}, \
  and\ \bibinfo {author} {\bibfnamefont {K.}~\bibnamefont {Nakazawa}},\ }\href
  {\doibase 10.1143/PTP.49.460} {\bibfield  {journal} {\bibinfo  {journal}
  {Progress of Theoretical Physics}\ }\textbf {\bibinfo {volume} {49}},\
  \bibinfo {pages} {460} (\bibinfo {year} {1973})}\BibitemShut {NoStop}%
\bibitem [{\citenamefont {Tarasova}(2010)}]{tarasova}%
  \BibitemOpen
  \bibfield  {author} {\bibinfo {author} {\bibfnamefont {Y.~V.}\ \bibnamefont
  {Tarasova}},\ }\href {\doibase 10.1134/S1990478910040125} {\bibfield
  {journal} {\bibinfo  {journal} {Journal of Applied and Industrial
  Mathematics}\ }\textbf {\bibinfo {volume} {4}},\ \bibinfo {pages} {570}
  (\bibinfo {year} {2010})}\BibitemShut {NoStop}%
\bibitem [{\citenamefont {Shieh}(1983)}]{shieh}%
  \BibitemOpen
  \bibfield  {author} {\bibinfo {author} {\bibfnamefont {S.~Y.}\ \bibnamefont
  {Shieh}},\ }\href {\doibase 10.1063/1.525625} {\bibfield  {journal} {\bibinfo
   {journal} {Journal of Mathematical Physics}\ }\textbf {\bibinfo {volume}
  {24}},\ \bibinfo {pages} {2438} (\bibinfo {year} {1983})}\BibitemShut
  {NoStop}%
\bibitem [{\citenamefont {Rogers}\ and\ \citenamefont
  {Schief}(2011)}]{Rogers_2011}%
  \BibitemOpen
  \bibfield  {author} {\bibinfo {author} {\bibfnamefont {C.}~\bibnamefont
  {Rogers}}\ and\ \bibinfo {author} {\bibfnamefont {W.~K.}\ \bibnamefont
  {Schief}},\ }\href {\doibase 10.1088/0951-7715/24/11/009} {\bibfield
  {journal} {\bibinfo  {journal} {Nonlinearity}\ }\textbf {\bibinfo {volume}
  {24}},\ \bibinfo {pages} {3165} (\bibinfo {year} {2011})}\BibitemShut
  {NoStop}%
\bibitem [{\citenamefont {Gaffet}(1996)}]{gaffet_1996}%
  \BibitemOpen
  \bibfield  {author} {\bibinfo {author} {\bibfnamefont {B.}~\bibnamefont
  {Gaffet}},\ }\href {\doibase 10.1017/S0022112096008051} {\bibfield  {journal}
  {\bibinfo  {journal} {Journal of Fluid Mechanics}\ }\textbf {\bibinfo
  {volume} {325}},\ \bibinfo {pages} {113–144} (\bibinfo {year}
  {1996})}\BibitemShut {NoStop}%
\bibitem [{\citenamefont {Gaffet}(1999)}]{gaffet_1999}%
  \BibitemOpen
  \bibfield  {author} {\bibinfo {author} {\bibfnamefont {B.}~\bibnamefont
  {Gaffet}},\ }\href {\doibase https://doi.org/10.1016/S0167-2789(99)00038-X}
  {\bibfield  {journal} {\bibinfo  {journal} {Physica D: Nonlinear Phenomena}\
  }\textbf {\bibinfo {volume} {132}},\ \bibinfo {pages} {233 } (\bibinfo {year}
  {1999})}\BibitemShut {NoStop}%
\bibitem [{\citenamefont {Gaffet}(2000)}]{gaffet_2000}%
  \BibitemOpen
  \bibfield  {author} {\bibinfo {author} {\bibfnamefont {B.}~\bibnamefont
  {Gaffet}},\ }\href {\doibase 10.1088/0305-4470/33/21/306} {\bibfield
  {journal} {\bibinfo  {journal} {Journal of Physics A: Mathematical and
  General}\ }\textbf {\bibinfo {volume} {33}},\ \bibinfo {pages} {3929}
  (\bibinfo {year} {2000})}\BibitemShut {NoStop}%
\bibitem [{\citenamefont {Gaffet}(2001{\natexlab{a}})}]{gaffet_2001_axis}%
  \BibitemOpen
  \bibfield  {author} {\bibinfo {author} {\bibfnamefont {B.}~\bibnamefont
  {Gaffet}},\ }\href {\doibase 10.1088/0305-4470/34/43/307} {\bibfield
  {journal} {\bibinfo  {journal} {Journal of Physics A: Mathematical and
  General}\ }\textbf {\bibinfo {volume} {34}},\ \bibinfo {pages} {9195}
  (\bibinfo {year} {2001}{\natexlab{a}})}\BibitemShut {NoStop}%
\bibitem [{\citenamefont {Gaffet}(2001{\natexlab{b}})}]{gaffet_2001_liouville}%
  \BibitemOpen
  \bibfield  {author} {\bibinfo {author} {\bibfnamefont {B.}~\bibnamefont
  {Gaffet}},\ }\href {\doibase 10.1088/0305-4470/34/11/303} {\bibfield
  {journal} {\bibinfo  {journal} {Journal of Physics A: Mathematical and
  General}\ }\textbf {\bibinfo {volume} {34}},\ \bibinfo {pages} {2097}
  (\bibinfo {year} {2001}{\natexlab{b}})}\BibitemShut {NoStop}%
\bibitem [{\citenamefont {Gaffet}(2005)}]{gaffet_2005}%
  \BibitemOpen
  \bibfield  {author} {\bibinfo {author} {\bibfnamefont {B.}~\bibnamefont
  {Gaffet}},\ }\href {\doibase 10.1088/0305-4470/39/1/008} {\bibfield
  {journal} {\bibinfo  {journal} {Journal of Physics A: Mathematical and
  General}\ }\textbf {\bibinfo {volume} {39}},\ \bibinfo {pages} {99} (\bibinfo
  {year} {2005})}\BibitemShut {NoStop}%
\bibitem [{\citenamefont {Gaffet}(2010)}]{gaffet_2010}%
  \BibitemOpen
  \bibfield  {author} {\bibinfo {author} {\bibfnamefont {B.}~\bibnamefont
  {Gaffet}},\ }\href {\doibase 10.1088/1751-8113/43/16/165207} {\bibfield
  {journal} {\bibinfo  {journal} {Journal of Physics A: Mathematical and
  Theoretical}\ }\textbf {\bibinfo {volume} {43}},\ \bibinfo {pages} {165207}
  (\bibinfo {year} {2010})}\BibitemShut {NoStop}%
\bibitem [{\citenamefont {Bogoyavlensky}()}]{bogoyavlenskymethods}%
  \BibitemOpen
  \bibfield  {author} {\bibinfo {author} {\bibfnamefont {O.}~\bibnamefont
  {Bogoyavlensky}},\ }\href@noop {} {\emph {\bibinfo {title} {Methods in the
  Qualitative Theory of Dynamical Systems in Astrophysics and Gas Dynamics,
  1985}}}\ (\bibinfo  {publisher} {Springer-Verlag: Berlin})\BibitemShut
  {NoStop}%
\bibitem [{\citenamefont {Coggeshall}(1994)}]{coggeshall_94_hydro}%
  \BibitemOpen
  \bibfield  {author} {\bibinfo {author} {\bibfnamefont {S.~V.}\ \bibnamefont
  {Coggeshall}},\ }\href@noop {} {\emph {\bibinfo {title} {Group-invariant
  Solutions of Hydrodynamics}}},\ \bibinfo {type} {Tech. Rep.}\ \bibinfo
  {number} {{LA}-UR 94-1277}\ (\bibinfo  {institution} {Los Alamos National
  Laboratory},\ \bibinfo {year} {1994})\BibitemShut {NoStop}%
\bibitem [{\citenamefont {McHardy}\ \emph {et~al.}(2019)\citenamefont {McHardy}
  \emph {et~al.}}]{mchardy2019group}%
  \BibitemOpen
  \bibfield  {author} {\bibinfo {author} {\bibfnamefont {J.}~\bibnamefont
  {McHardy}} \emph {et~al.},\ }\href@noop {} {\bibfield  {journal} {\bibinfo
  {journal} {AIP Advances}\ }\textbf {\bibinfo {volume} {9}},\ \bibinfo {pages}
  {085113} (\bibinfo {year} {2019})}\BibitemShut {NoStop}%
\bibitem [{\citenamefont {Harlow}\ and\ \citenamefont
  {Amsden}(1971)}]{harlow1971fluid}%
  \BibitemOpen
  \bibfield  {author} {\bibinfo {author} {\bibfnamefont {F.~H.}\ \bibnamefont
  {Harlow}}\ and\ \bibinfo {author} {\bibfnamefont {A.~A.}\ \bibnamefont
  {Amsden}},\ }\href@noop {} {\emph {\bibinfo {title} {Fluid dynamics: a {LASL}
  monograph (Mathematical solutions for problems in fluid dynamics)}}},\
  \bibinfo {type} {Tech. Rep.}\ \bibinfo {number} {{LA} 4700}\ (\bibinfo
  {institution} {Los Alamos National Laboratory},\ \bibinfo {year}
  {1971})\BibitemShut {NoStop}%
\bibitem [{\citenamefont {Mandl}(1988)}]{mandl1988statistical}%
  \BibitemOpen
  \bibfield  {author} {\bibinfo {author} {\bibfnamefont {F.}~\bibnamefont
  {Mandl}},\ }\href {https://books.google.com/books?id=0IEpAQAAMAAJ} {\emph
  {\bibinfo {title} {Statistical Physics}}},\ CIBA Foundation Symposium\
  (\bibinfo  {publisher} {Wiley},\ \bibinfo {year} {1988})\BibitemShut
  {NoStop}%
\bibitem [{\citenamefont {Bowley}\ and\ \citenamefont
  {S{\'a}nchez}(1999)}]{bowley1999introductory}%
  \BibitemOpen
  \bibfield  {author} {\bibinfo {author} {\bibfnamefont {R.}~\bibnamefont
  {Bowley}}\ and\ \bibinfo {author} {\bibfnamefont {M.}~\bibnamefont
  {S{\'a}nchez}},\ }\href {https://books.google.com/books?id=o6zvAAAAMAAJ}
  {\emph {\bibinfo {title} {Introductory Statistical Mechanics}}},\ Oxford
  science publications\ (\bibinfo  {publisher} {Clarendon Press},\ \bibinfo
  {year} {1999})\BibitemShut {NoStop}%
\bibitem [{\citenamefont {Adkins}\ and\ \citenamefont
  {Adkins}(1983)}]{adkins1983equilibrium}%
  \BibitemOpen
  \bibfield  {author} {\bibinfo {author} {\bibfnamefont {C.}~\bibnamefont
  {Adkins}}\ and\ \bibinfo {author} {\bibfnamefont {C.}~\bibnamefont
  {Adkins}},\ }\href {https://books.google.com/books?id=FW4Oz48TWwQC} {\emph
  {\bibinfo {title} {Equilibrium Thermodynamics}}}\ (\bibinfo  {publisher}
  {Cambridge University Press},\ \bibinfo {year} {1983})\BibitemShut {NoStop}%
\bibitem [{\citenamefont {Landau}\ and\ \citenamefont
  {Lifshitz}(2013)}]{landau2013statistical}%
  \BibitemOpen
  \bibfield  {author} {\bibinfo {author} {\bibfnamefont {L.}~\bibnamefont
  {Landau}}\ and\ \bibinfo {author} {\bibfnamefont {E.}~\bibnamefont
  {Lifshitz}},\ }\href {https://books.google.com/books?id=VzgJN-XPTRsC} {\emph
  {\bibinfo {title} {Statistical Physics}}},\ \bibinfo {number} {v. 5}\
  (\bibinfo  {publisher} {Elsevier Science},\ \bibinfo {year}
  {2013})\BibitemShut {NoStop}%
\bibitem [{\citenamefont {Zemansky}\ \emph {et~al.}(1966)\citenamefont
  {Zemansky}, \citenamefont {Abbott},\ and\ \citenamefont
  {Van~Ness}}]{zemansky1966basic}%
  \BibitemOpen
  \bibfield  {author} {\bibinfo {author} {\bibfnamefont {M.}~\bibnamefont
  {Zemansky}}, \bibinfo {author} {\bibfnamefont {M.}~\bibnamefont {Abbott}}, \
  and\ \bibinfo {author} {\bibfnamefont {H.}~\bibnamefont {Van~Ness}},\ }\href
  {https://books.google.com/books?id=b9tSAAAAMAAJ} {\emph {\bibinfo {title}
  {Basic engineering thermodynamics}}},\ International student edition\
  (\bibinfo  {publisher} {McGraw-Hill},\ \bibinfo {year} {1966})\BibitemShut
  {NoStop}%
\bibitem [{\citenamefont {Axford}(2000)}]{axford}%
  \BibitemOpen
  \bibfield  {author} {\bibinfo {author} {\bibfnamefont {R.~A.}\ \bibnamefont
  {Axford}},\ }\href@noop {} {\bibfield  {journal} {\bibinfo  {journal} {Laser
  and Particle Beams}\ }\textbf {\bibinfo {volume} {18}},\ \bibinfo {pages}
  {93} (\bibinfo {year} {2000})}\BibitemShut {NoStop}%
\bibitem [{\citenamefont {Tsinopoulos}\ \emph {et~al.}(1999)\citenamefont
  {Tsinopoulos} \emph {et~al.}}]{TsinopoulosStephanosV1999Aabe}%
  \BibitemOpen
  \bibfield  {author} {\bibinfo {author} {\bibfnamefont {S.~V.}\ \bibnamefont
  {Tsinopoulos}} \emph {et~al.},\ }\href@noop {} {\bibfield  {journal}
  {\bibinfo  {journal} {Journal of the Acoustical Society of America}\ }\textbf
  {\bibinfo {volume} {105}},\ \bibinfo {pages} {1517} (\bibinfo {year}
  {1999})}\BibitemShut {NoStop}%
\bibitem [{\citenamefont {Spiegel}(1959)}]{spiegel1959schaum}%
  \BibitemOpen
  \bibfield  {author} {\bibinfo {author} {\bibfnamefont {M.}~\bibnamefont
  {Spiegel}},\ }\href@noop {} {\emph {\bibinfo {title} {Schaum's Outline of
  Theory and Problems of Vector Analysis and an Introduction to Tensor
  Analysis}}},\ Schaum's outline series of theory and problems of vector
  analysis\ (\bibinfo  {publisher} {Schaum Publishing Company},\ \bibinfo
  {year} {1959})\BibitemShut {NoStop}%
\bibitem [{\citenamefont {Hendon}\ and\ \citenamefont
  {Ramsey}(2012)}]{hendon2012radiation}%
  \BibitemOpen
  \bibfield  {author} {\bibinfo {author} {\bibfnamefont {R.~C.}\ \bibnamefont
  {Hendon}}\ and\ \bibinfo {author} {\bibfnamefont {S.~D.}\ \bibnamefont
  {Ramsey}},\ }\href@noop {} {\emph {\bibinfo {title} {Radiation Hydrodynamics
  Test Problems with Linear Velocity Profiles}}},\ \bibinfo {type} {Tech.
  Rep.}\ (\bibinfo  {institution} {Los Alamos National Lab.(LANL), Los Alamos,
  NM (United States)},\ \bibinfo {year} {2012})\BibitemShut {NoStop}%
\bibitem [{\citenamefont {Boyd}\ \emph {et~al.}(2019)\citenamefont {Boyd},
  \citenamefont {Schmidt}, \citenamefont {Ramsey},\ and\ \citenamefont
  {Baty}}]{boyd2019collapsing}%
  \BibitemOpen
  \bibfield  {author} {\bibinfo {author} {\bibfnamefont {Z.~M.}\ \bibnamefont
  {Boyd}}, \bibinfo {author} {\bibfnamefont {E.~M.}\ \bibnamefont {Schmidt}},
  \bibinfo {author} {\bibfnamefont {S.~D.}\ \bibnamefont {Ramsey}}, \ and\
  \bibinfo {author} {\bibfnamefont {R.~S.}\ \bibnamefont {Baty}},\ }\href@noop
  {} {\bibfield  {journal} {\bibinfo  {journal} {The Quarterly Journal of
  Mechanics and Applied Mathematics}\ }\textbf {\bibinfo {volume} {72}},\
  \bibinfo {pages} {501} (\bibinfo {year} {2019})}\BibitemShut {NoStop}%
\bibitem [{\citenamefont {Rozanova}\ and\ \citenamefont
  {Turzynsky}(2016)}]{rozanova2016nonlinear}%
  \BibitemOpen
  \bibfield  {author} {\bibinfo {author} {\bibfnamefont {O.~S.}\ \bibnamefont
  {Rozanova}}\ and\ \bibinfo {author} {\bibfnamefont {M.~K.}\ \bibnamefont
  {Turzynsky}},\ }in\ \href@noop {} {\emph {\bibinfo {booktitle} {XVI
  International Conference on Hyperbolic Problems: Theory, Numerics,
  Applications}}}\ (\bibinfo {organization} {Springer},\ \bibinfo {year}
  {2016})\ pp.\ \bibinfo {pages} {549--561}\BibitemShut {NoStop}%
\bibitem [{\citenamefont {Rozanova}\ and\ \citenamefont
  {Turzynski}(2017)}]{rozanova2017systems}%
  \BibitemOpen
  \bibfield  {author} {\bibinfo {author} {\bibfnamefont {O.~S.}\ \bibnamefont
  {Rozanova}}\ and\ \bibinfo {author} {\bibfnamefont {M.~K.}\ \bibnamefont
  {Turzynski}},\ }in\ \href@noop {} {\emph {\bibinfo {booktitle} {International
  Conference on Differential \& Difference Equations and Applications}}}\
  (\bibinfo {organization} {Springer},\ \bibinfo {year} {2017})\ pp.\ \bibinfo
  {pages} {387--398}\BibitemShut {NoStop}%
\bibitem [{\citenamefont {Irtegov}\ and\ \citenamefont
  {Titorenko}(2018)}]{irtegov2018qualitative}%
  \BibitemOpen
  \bibfield  {author} {\bibinfo {author} {\bibfnamefont {V.}~\bibnamefont
  {Irtegov}}\ and\ \bibinfo {author} {\bibfnamefont {T.}~\bibnamefont
  {Titorenko}},\ }in\ \href@noop {} {\emph {\bibinfo {booktitle} {International
  Workshop on Computer Algebra in Scientific Computing}}}\ (\bibinfo
  {organization} {Springer},\ \bibinfo {year} {2018})\ pp.\ \bibinfo {pages}
  {254--271}\BibitemShut {NoStop}%
\bibitem [{\citenamefont {Ovsyannikov}(2014)}]{ovsyannikov1}%
  \BibitemOpen
  \bibfield  {author} {\bibinfo {author} {\bibfnamefont {L.}~\bibnamefont
  {Ovsyannikov}},\ }\href {https://books.google.com/books?id=YpPiBQAAQBAJ}
  {\emph {\bibinfo {title} {Group Analysis of Differential Equations}}}\
  (\bibinfo  {publisher} {Elsevier Science},\ \bibinfo {year}
  {2014})\BibitemShut {NoStop}%
\bibitem [{\citenamefont {Ovsyannikov}(2013)}]{ovsyannikov2}%
  \BibitemOpen
  \bibfield  {author} {\bibinfo {author} {\bibfnamefont {L.}~\bibnamefont
  {Ovsyannikov}},\ }\href {https://books.google.com/books?id=CSk8DQAAQBAJ}
  {\emph {\bibinfo {title} {Lectures on the Theory of Group Properties of
  Differential Equations}}}\ (\bibinfo  {publisher} {World Scientific
  Publishing Company},\ \bibinfo {year} {2013})\BibitemShut {NoStop}%
\bibitem [{\citenamefont {Holm}(1976)}]{holm1976symmetry}%
  \BibitemOpen
  \bibfield  {author} {\bibinfo {author} {\bibfnamefont {D.~D.}\ \bibnamefont
  {Holm}},\ }\href@noop {} {\emph {\bibinfo {title} {Symmetry breaking in fluid
  dynamics: Lie group reducible motions for real fluids}}},\ \bibinfo {type}
  {Tech. Rep.}\ (\bibinfo  {institution} {Los Alamos Scientific Lab., N.
  Mex.(USA)},\ \bibinfo {year} {1976})\BibitemShut {NoStop}%
\bibitem [{\citenamefont {Axford}(1998)}]{axford1998solutions}%
  \BibitemOpen
  \bibfield  {author} {\bibinfo {author} {\bibfnamefont {R.~A.}\ \bibnamefont
  {Axford}},\ }\href@noop {} {\emph {\bibinfo {title} {Solutions of the Noh
  problem for various equations of state using Lie groups}}},\ \bibinfo {type}
  {Tech. Rep.}\ (\bibinfo  {institution} {Los Alamos National Lab., NM (US)},\
  \bibinfo {year} {1998})\BibitemShut {NoStop}%
\bibitem [{\citenamefont {Ramsey}\ and\ \citenamefont
  {Baty}(2017)}]{ramsey2017symmetries}%
  \BibitemOpen
  \bibfield  {author} {\bibinfo {author} {\bibfnamefont {S.~D.}\ \bibnamefont
  {Ramsey}}\ and\ \bibinfo {author} {\bibfnamefont {R.~S.}\ \bibnamefont
  {Baty}},\ }\href@noop {} {\bibfield  {journal} {\bibinfo  {journal} {Journal
  of Mathematical Physics}\ }\textbf {\bibinfo {volume} {58}},\ \bibinfo
  {pages} {111506} (\bibinfo {year} {2017})}\BibitemShut {NoStop}%
\bibitem [{\citenamefont {Grad}(1949)}]{grad1949kinetic}%
  \BibitemOpen
  \bibfield  {author} {\bibinfo {author} {\bibfnamefont {H.}~\bibnamefont
  {Grad}},\ }\href@noop {} {\bibfield  {journal} {\bibinfo  {journal}
  {Communications on pure and applied mathematics}\ }\textbf {\bibinfo {volume}
  {2}},\ \bibinfo {pages} {331} (\bibinfo {year} {1949})}\BibitemShut {NoStop}%
\bibitem [{\citenamefont {Borisov}\ \emph {et~al.}(2009)\citenamefont
  {Borisov}, \citenamefont {Kilin},\ and\ \citenamefont
  {Mamaev}}]{borisov2009hamiltonian}%
  \BibitemOpen
  \bibfield  {author} {\bibinfo {author} {\bibfnamefont {A.~V.}\ \bibnamefont
  {Borisov}}, \bibinfo {author} {\bibfnamefont {A.}~\bibnamefont {Kilin}}, \
  and\ \bibinfo {author} {\bibfnamefont {I.}~\bibnamefont {Mamaev}},\
  }\href@noop {} {\bibfield  {journal} {\bibinfo  {journal} {Regular and
  Chaotic Dynamics}\ }\textbf {\bibinfo {volume} {14}},\ \bibinfo {pages} {179}
  (\bibinfo {year} {2009})}\BibitemShut {NoStop}%
\bibitem [{\citenamefont {Ragazzo}\ and\ \citenamefont
  {Ruiz}(2015)}]{ragazzo2015dynamics}%
  \BibitemOpen
  \bibfield  {author} {\bibinfo {author} {\bibfnamefont {C.}~\bibnamefont
  {Ragazzo}}\ and\ \bibinfo {author} {\bibfnamefont {L.}~\bibnamefont {Ruiz}},\
  }\href@noop {} {\bibfield  {journal} {\bibinfo  {journal} {Celestial
  Mechanics and Dynamical Astronomy}\ }\textbf {\bibinfo {volume} {122}},\
  \bibinfo {pages} {303} (\bibinfo {year} {2015})}\BibitemShut {NoStop}%
\bibitem [{\citenamefont {Bizyaev}\ \emph {et~al.}(2015)\citenamefont
  {Bizyaev}, \citenamefont {Borisov},\ and\ \citenamefont
  {Mamaev}}]{bizyaev2015figures}%
  \BibitemOpen
  \bibfield  {author} {\bibinfo {author} {\bibfnamefont {I.~A.}\ \bibnamefont
  {Bizyaev}}, \bibinfo {author} {\bibfnamefont {A.~V.}\ \bibnamefont
  {Borisov}}, \ and\ \bibinfo {author} {\bibfnamefont {I.~S.}\ \bibnamefont
  {Mamaev}},\ }\href@noop {} {\bibfield  {journal} {\bibinfo  {journal}
  {Celestial Mechanics and Dynamical Astronomy}\ }\textbf {\bibinfo {volume}
  {122}},\ \bibinfo {pages} {1} (\bibinfo {year} {2015})}\BibitemShut {NoStop}%
\bibitem [{\citenamefont {Sachdev}\ and\ \citenamefont
  {Janna}(2001)}]{sachdev2001self}%
  \BibitemOpen
  \bibfield  {author} {\bibinfo {author} {\bibfnamefont {P.}~\bibnamefont
  {Sachdev}}\ and\ \bibinfo {author} {\bibfnamefont {W.}~\bibnamefont
  {Janna}},\ }\href@noop {} {\bibfield  {journal} {\bibinfo  {journal} {Appl.
  Mech. Rev.}\ }\textbf {\bibinfo {volume} {54}},\ \bibinfo {pages} {B108}
  (\bibinfo {year} {2001})}\BibitemShut {NoStop}%
\end{thebibliography}%
\end{document}